\documentclass[11pt,a4paper]{article}
\DeclareUnicodeCharacter{2212}{\textminus}
\usepackage{jheppub}
\usepackage{enumitem}
\usepackage{booktabs}
\usepackage{microtype}
\usepackage{amsfonts}
\usepackage{mathrsfs}
\usepackage{amssymb}
\usepackage{amsmath}
\usepackage{braket}
\usepackage{hyperref} 
\usepackage{xcolor}
\usepackage{cases}
\usepackage{tikz}
\usepackage{bm}
\usepackage{physics}
\usetikzlibrary{
    arrows.meta,
    decorations.pathreplacing,
    decorations.markings,
    decorations.pathmorphing,
    calc
}

\newcommand{\cO}{\mathcal{O}}

\newcommand{\cM}{\mathcal{M}}
\newcommand{\cH}{\mathcal{H}}

\newcommand{\hDelta}{\widehat\Delta}

\renewcommand{\Tr}{\text{Tr}}

\newcommand{\SU}{\mathrm{SU}}
\newcommand{\PSU}{\mathrm{PSU}}
\newcommand{\U}{\mathrm{U}}
\renewcommand{\O}{\mathrm{O}}
\newcommand{\ZZ}{\mathbb{Z}}

\renewcommand{\i}{\text{i}}
\renewcommand{\d}{\mathrm{d}}

\newcommand{\R}{\mathbb{R}}
\newcommand{\Z}{\mathbb{Z}}
\newcommand{\Adj}{\mathbf{Adj}}
\newcommand{\one}{\mathbf{1}}

\newcommand{\cG}{\mathcal G}

\newcommand{\ii}{\mathrm i}

\newcommand{\Li}{\operatorname{Li}}

\newcommand{\hatlambda}{\widehat\lambda}

\usepackage[normalem]{ulem}

\title{Bootstrapping Wilson lines in Critical Bosonic QED$_3$ Part 1: Large-$N$ endpoint correlators and CFT data}
\author[a]{Jun Maeda,}
\author[b]{Toshiki Onagi}
\author[c]{and Soichiro Shimamori}

\affiliation[a]{Department of Physics, Kyoto University, Kyoto, 606-8502, Japan}
\affiliation[b]{Yukawa Institute for Theoretical Physics, Kyoto University, Kitashirakawa Oiwakecho, Sakyo-ku, Kyoto 606-8502 Japan}
\affiliation[c]{SISSA, Via Bonomea 265, 34136 Trieste, Italy}

\preprint{KUNS-3136, YITP-26-143, SISSA\,22/2026/FISI}

\abstract{We study the conformal Wilson line with general charge $q$ in critical bosonic QED$_3$ with $N$ complex scalar flavors by combining diagrammatic four-point calculations with endpoint crossing.
We formulate the endpoint crossing equations with the full $\SU(N)$ flavor structure, keeping positivity manifest. For arbitrary positive integer charge $q$ held fixed at large $N$, we obtain closed-form expressions for four-endpoint correlators and mixed correlators involving two bulk Hubbard--Stratonovich fields
through first subleading order in $1/N$. These results determine the endpoint dimension, the defect
$g$-function, and the dimensions and endpoint OPE coefficients of the lowest non-singlet Wilson-line primaries. Moreover, combining the diagrammatic correlator with endpoint crossing, somewhat surprisingly, we find agreement between the bulk and Wilson-line spectra in every non-singlet flavor channel through order $1/N$. At degenerate levels, this agreement concerns the OPE-weighted averages of anomalous dimensions probed by the endpoint correlator. For unit charge, we independently reproduce these averages from the known bulk higher-spin spectrum.
}

\begin{document}
\maketitle
\flushbottom

\section{Introduction}

Conformal field theory (CFT) provides a universal description of continuous
phase transitions and strongly coupled fixed points.  Its local
dynamics is encoded in the spectrum of primary operators and their
operator-product-expansion (OPE) coefficients, while associativity of
the OPE imposes exact crossing equations on these data.  The conformal
bootstrap turns this observation into a nonperturbative method that
does not require a Lagrangian or a small coupling.  Its modern numerical
form has led to precise determinations of CFT data in the three-dimensional
Ising and $\O(N)$ universality classes
\cite{Rattazzi:2008pe,Poland:2010wg,El-Showk:2012cjh,
El-Showk:2014dwa,Kos:2014bka,Kos:2016ysd}, and its numerical and
analytic developments have been reviewed in
\cite{Simmons-Duffin:2016gjk,Poland:2018epd,Chester:2019wfx,
Rychkov:2023wsd}.  In parallel, lightcone methods, inversion formulas,
and analytic functionals have exposed detailed structures that are
difficult to see numerically
\cite{Fitzpatrick:2012yx,Komargodski:2012ek,Alday:2016njk,
Caron-Huot:2017vep,Simmons-Duffin:2017nub,Karateev:2017jgd,
Mazac:2016qev,Mazac:2018mdx,Mazac:2018ycv}.

Extended objects are equally fundamental observables of a quantum
field theory.  Boundaries, interfaces, impurities, and line operators
occur in statistical systems, quantum many-body problems, and gauge
theories.  At a bulk critical point they can themselves flow to
scale-invariant defects and support localized operators with their own
spectrum and OPE.  Boundary conformal field theory and the Kondo
problem supplied early and influential examples of this structure
\cite{Cardy:1989ir,Ishibashi:1988kg,Affleck:1990by,Affleck:1991tk,
Affleck:1995ge,Vasseur:2013rda,McAvity:1995zd,Diehl:1996kd,
DiFrancesco:1997nk}.  In two dimensions, conformal interfaces and
boundary flows also admit many exact constructions
\cite{Oshikawa:1996dj,Bachas:2001vj,Fredenhagen:2002qn,
Recknagel:2002qq,Bachas:2004sy,Quella:2006de}; see
\cite{Billo:2016cpy,Andrei:2018die} for broader accounts of defect CFT.
Besides local defect data, one would like to determine bulk-to-defect
couplings, one-point functions, and intrinsic observables such as the
defect $g$-function.  The latter generalizes boundary entropy and is
sensitive to defect renormalization-group flows
\cite{Friedan:2003yc,Jensen:2015swa,Kobayashi:2018lil,
Cuomo:2021rkm, Wang:2021mdq, Zhou:2023fqu}.  Developing systematic tools for extracting these
defect data is therefore an essential extension of the ordinary bulk
bootstrap.

Several bootstrap equations for boundaries and defects are already
known.  Two-point functions of bulk operators in the presence of a
defect equate a bulk OPE to a defect OPE, whereas correlators of
operators confined to a defect obey an ordinary lower-dimensional
crossing equation
\cite{Liendo:2012hy,Gaiotto:2013nva,Gliozzi:2015qsa,Gadde:2016fbj,
Fukuda:2017cup,Lauria:2017wav,Lauria:2018klo}.  These approaches have
been explored from the numerical and analytic viewpoints~\cite{Bissi:2018mcq,Mazac:2018biw,Kaviraj:2018tfd,
Gimenez-Grau:2019hez,Gimenez-Grau:2022czc,Dey:2024ilw,
Padayasi:2021sik,Nishioka:2022qmj}, and defect inversion formulas have also been developed~\cite{Liendo:2019jpu}. 
There is, however, a basic tension between the
two setups.  In the bulk channel of the usual bulk-to-defect equation,
the coefficients are products of bulk OPE coefficients and defect
one-point functions and need not have a definite sign.  Correlators of
operators living on a fixed defect retain positivity, but by themselves
contain little information identifying the bulk CFT.  Consequently,
neither bootstrap equation alone generally provides a positive, theory-specific
bootstrap, and in practice it has been difficult to obtain sharp
constraints on the CFT data of a specified defect.
\begin{figure}[t]
\centering
\begin{tikzpicture}[
    x=0.85cm,
    y=0.85cm,
    wilson/.style={
        draw=black,
        double=white,
        double distance=2.6pt,
        line width=0.65pt,
        line cap=butt,
        line join=round
    }
]
    \draw[wilson]
        (0,0)
        .. controls (0,1.50) and (0.90,1.65) ..
        (1.80,1.90)
        .. controls (2.65,2.15) and (2.85,2.05) ..
        (3.15,2.85)
        .. controls (3.30,3.25) and (3.40,3.60) ..
        (3.45,3.95);

    \fill[red!85!black] (0,0) circle[radius=3.8pt];

    \node[
        anchor=north west,
        inner sep=0pt,
        text=black,
        font=\Large
    ] at (0.24,-0.08) {$E^{(q)}$};
\end{tikzpicture}
\caption{A charge-$q$ Wilson line ending at the defect-changing operator $E^{(q)}$.}
\label{fig:intro-wilson-endpoint}
\end{figure}
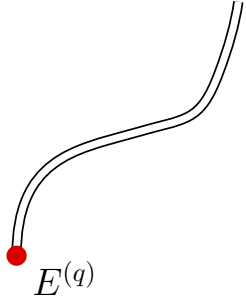
For a line defect that can end, this difficulty can be overcome by treating
its endpoints as external defect-changing operators.  Their four-point
function admits one OPE channel containing bulk local operators and
another containing local operators on the line.  Reflection positivity
acts in both Hilbert spaces, so the resulting crossing relation directly
connects positive bulk and defect spectra.  This endpoint bootstrap was
recently formulated and implemented both analytically and numerically for the pinning-field line defect of the three-dimensional Ising CFT
\cite{Lanzetta:2025xfw}.  It opens a route to importing known bulk CFT
data into a rigorous defect bootstrap. In this work, we take the
endpoint-bootstrap program one step further by formulating and
analytically solving it for charged Wilson lines in an interacting
critical gauge theory at large $N$.

Wilson lines are natural probes of electric charge and are central to
the distinction between confinement, screening, and Higgs behavior
\cite{Wilson:1974sk,Polyakov:1976fu,Fradkin:1978dv}.  They also furnish
some of the best studied examples of conformal line defects, from
holographic Wilson loops to supersymmetric and integrable defect CFTs
\cite{Rey:1998ik,Maldacena:1998im,Erickson:2000af,Drukker:2000rr,
Polchinski:2011im,Correa:2012at,Bianchi:2018zpb,Giombi:2018qox,
Liendo:2018ukf,Agmon:2020pde,Gabai:2023lax}.  In a gauge theory with dynamical matter,
an open Wilson line can end on a charged field.  Its endpoint is a
gauge-invariant description of charged matter together with its electric
dressing, and joins the bulk and defect Hilbert spaces
\cite{Aharony:2023amq,Iqbal:2025yop}.  We study this structure in
critical bosonic QED$_3$, equivalently the critical
$\mathbb{CP}^{N-1}$ model.  This theory is a canonical example of a
three-dimensional conformal gauge theory and is closely connected to
deconfined quantum criticality in two-dimensional quantum magnets
\cite{Senthil:2003eed,Kaul:2008xw,Senthil:2023vqd}.  Its large-$N$
expansion gives analytic access to otherwise strongly coupled bulk and
defect observables
\cite{Metlitski_2008,Dyer:2015zha,Benvenuti:2019ujm,
DiPietro:2015taa,Borokhov:2002ib,Pufu:2013vpa,Chester:2017vdh,
Chai:2021uhv}, while
three-dimensional dualities place it in a broader network of critical
gauge theories \cite{Karch:2016sxi,Seiberg:2016gmd,Ihrig:2018ojl}. 

The conformal bootstrap has also been applied to bosonic and fermionic QED$_3$, including the deconfined critical point at $N=2$, using correlators of local gauge-invariant operators such as monopole operators, flavor-adjoint scalars, and conserved currents~\cite{Chester:2016wrc,Nakayama:2016jhq,Li:2018lyb,He:2020azu, Manenti:2021elk, He:2021sto,Albayrak:2021xtd, He:2023ewx,Chester:2023njo,Chester:2025uxb,Nakayama:2025mrm,Li:2026cpu}. 
Endpoint correlators of Wilson lines give crossing equations in which both channels are expanded with squared OPE coefficients, for each gauge charge $q$ of the line.\footnote{Throughout this paper, we limit ourself to $q=O(N^0)$. Our fixed-$q$ and large-$N$ analysis probes the small-$q/N$ regime of critical bosonic QED$_3$, where conformal Wilson lines are expected to be stable against screening~\cite{Aharony:2023amq}.}

An important aim of this work is to formulate endpoint crossing in the presence of nontrivial flavor symmetry. Building on the endpoint-bootstrap framework of~\cite{Lanzetta:2025xfw}, we consider Wilson-line
endpoints carrying nontrivial $\SU(N)$ representations. We derive the crossing equations relating their flavor channels and make the positivity of the OPE weights explicit.

We then compute endpoint four-point functions in critical bosonic QED$_3$ and show how these equations can be used to extract conformal data of the Wilson line and its
endpoints. We obtain closed-form results through first subleading order in $1/N$ for arbitrary positive integer charge $q$, held fixed in the large-$N$ limit. Specifically, we study
\begin{align*}
    \left\langle
        \bar E_{(q)} E^{(q)}
        \bar E_{(q)} E^{(q)}
    \right\rangle,
    \quad
    \left\langle
        \bar E_{(q)} E^{(q)}\sigma\sigma
    \right\rangle,
\end{align*}
where $E^{(q)}$ and $\bar E_{(q)}$ are the Wilson-line endpoints and $\sigma$ is the bulk Hubbard--Stratonovich field. Flavor indices and insertion points are suppressed here.

Combining the four-endpoint correlator with crossing, we determine the endpoint dimension, the defect $g$-function, and the dimensions and endpoint OPE coefficients of the lowest Wilson-line primaries in every non-singlet flavor channel. The results give their full charge dependence
at this order. We also derive a nonperturbative interpretation of the $g$-function in terms of the relative high-dimension OPE weight in the bulk and Wilson-line sectors.

We also show that the correlator reveals a further connection between bulk and defect dynamics. Their full spectral functions obey a proportionality relation through order $1/N$. For the fundamental endpoint, $q=1$, we explore this
relation beyond the lowest operators. The correlator determines OPE-weighted anomalous dimensions at every
one-dimensional conformal level, which we independently reconstruct from the known bulk higher-spin spectrum and its decomposition into line primaries \cite{Maldacena:2011jn,Maldacena:2012sf,
Giombi:2016hkj,Zhou:2022pah}. This identifies how currents of different bulk spins
contribute to the spectral information measured by Wilson-line endpoints.

Finally, the mixed correlator accesses new defect-changing operators produced when $\sigma$
approaches an endpoint. Its crossing relation yields their OPE-weighted dimensions and summed squared OPE
coefficients for general $q$. Thus, changing the bulk operators inserted alongside the endpoints opens access to additional sectors of the defect theory. Together, these results demonstrate how explicit
four-point calculations and endpoint crossing can systematically uncover the conformal data of
non-supersymmetric Wilson lines.

The paper is organized as follows.  Section~\ref{sec:general-strategy}
introduces the charge-$q$ Wilson line, its endpoint operators, and the
general crossing equations.  Section~\ref{sec:analytical-bootstrap}
computes and solves the four-endpoint correlator at large $N$.
Section~\ref{sec:higher-spin-reconstruction} develops the all-level
higher-spin reconstruction for $q=1$.  Section~\ref{sec:mixed-endpoint-correlators-general-q}
applies endpoint crossing to the mixed endpoint--bulk correlator.  We
conclude with a summary and open directions.  The appendices contain details of the flavor tensors
and diagrammatic calculations, together with an independent reconstruction of the $q=1$ adjoint data by alpha-space inversion in appendix~\ref{sec:q-one-alpha-space-inversion}.

\section{General strategy of endpoint bootstrap of Wilson lines in bosonic QED$_3$}
\label{sec:general-strategy}

\subsection{Bulk theory: critical $N$-flavor bosonic QED$_3$}

We work with $N$ complex scalars of unit gauge charge. A convenient large-$N$ representation of the critical fixed point is
\begin{align}\label{eq:CP-action}
S
=
\int_{\R^3}\d^3x\,
\left[
\sum_{i=1}^N
\abs{(\partial_\mu+\i A_\mu)\Phi_i}^2
+
\lambda\left(
\sum_{i=1}^N\abs{\Phi_i}^2-\frac{N}{g_c}
\right)
\right]\, , 
\end{align}
where we have dropped the Maxwell term, which is irrelevant at the infrared fixed point, and $\lambda$ is the Lagrange multiplier imposing the constraint, which plays the role of the Hubbard--Stratonovich field
at the critical point.\footnote{For later use, we normalize the fundamental generators of $\SU(N)$ as
\begin{align}\label{eq:SU-N-generator-normalization}
    \Tr(T^aT^b)
    =
    \frac12\delta^{ab}\, .
\end{align}} Throughout this paper, we use the unit-normalized primary 
\begin{align}\label{eq:general-q-mixed-sigma-definition}
    \sigma(x)
    \equiv
    \frac{\pi\sqrt N}{2\sqrt2}\lambda(x)\, , \quad
    \left\langle\sigma(x)\sigma(0)\right\rangle_0
    =
    \frac{1}{|x|^4}\, , 
\end{align}
where $\langle\cdots\rangle_0$ denotes the vacuum expectation value at $N=\infty$. This is the standard large-$N$ formulation of the critical
$\mathbb{CP}^{N-1}$ model, used in the classic analyses
\cite{Read:1989zza,Murthy:1989ps} and reviewed in
\cite{Moshe:2003xn}. We follow the conventions of
\cite{Iqbal:2025yop}. Related large-$N$ calculations of the local and
monopole spectra appear in~\cite{Metlitski_2008,Dyer:2015zha,Benvenuti:2019ujm}. The faithful zero-form global symmetry is
\begin{align}\label{eq:global-group}
G_0
=
\left(\PSU(N)\times \U(1)_{\rm m}\right)\rtimes\Z_2^{\rm C}\, , \qquad \PSU(N) = \SU(N)/\Z_N\,.
\end{align}
Its relation to the projective flavor quantum numbers carried by Wilson defects is made
explicit in \cite{Aharony:2023amq,Iqbal:2025yop}. The quotient follows because the generator of the center
\begin{align}
z_{\rm c}
=
\exp\left(\frac{2\pi\ii}{N}\right)\mathbf 1_N
\end{align}
of $\SU(N)$ acts on $\Phi_i$ in exactly the same way as a constant $\U(1)$ gauge transformation. Hence the center acts trivially on the physical local Hilbert space.

We now introduce the line defects studied in this paper.  For an
oriented curve $C$ and an integer electric charge $q$, we define
\begin{align}\label{eq:charge-q-Wilson-line-definition}
    W_q(C)
    \equiv
    \exp\!\left(
        \i q\int_C A_\mu\,\d x^\mu
    \right)\, ,
    \quad 
    q\in\mathbb Z\, .
\end{align}
We assume that a straight Wilson line flows to a conformal line defect $\mathcal D_q$, with $\mathcal D_0$ denoting the trivial defect~\cite{Cuomo:2021rkm,Cuomo:2022xgw,Aharony:2022ntz,Aharony:2023amq}. 
In the large-$N$ regime studied in this paper, this assumption is supported by the absence of nontrivial relevant flavor-singlet operators on the defect (see also section~\ref{subsec:general-q-bootstrap-extraction-main}), and by the absence of a logarithmic divergence in the defect $g$-function (appendix~\ref{app:defect-g-function-check}).
An open Wilson line is not gauge invariant by itself:
under a gauge transformation it acquires phases at its two endpoints.
These phases must be canceled by charged local insertions.  Since the
unit-charge field $\Phi_i$ is a flavor fundamental, gauge
invariance relates a jump in the Wilson charge to the projective
flavor representation carried by the endpoint.  This is the defect
counterpart of the quotient in~\eqref{eq:global-group}.  We next make
this relation precise by introducing defect-changing operators and
deriving their charge-selection rule.

\subsection{Defect-changing operators and the charge selection rule}
\label{subsec:defect-changing-operators}

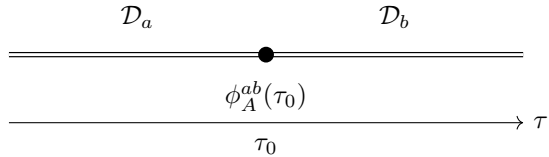
\begin{figure}[t]
\centering
\begin{tikzpicture}[
    wilson/.style={
        double,
        double distance=1.1pt,
        line width=0.45pt
    },
    insertion/.style={circle,fill=black,inner sep=2.1pt},
    every node/.style={font=\small}
]
    \draw[wilson] (-3.4,0)--(0,0);
    \draw[wilson] (0,0)--(3.4,0);
    \node[insertion] at (0,0) {};

    \node[above=6pt] at (-1.7,0) {$\mathcal D_a$};
    \node[above=6pt] at (1.7,0) {$\mathcal D_b$};
    \node[below=6pt] at (0, 0) {$\phi_A^{ab}(\tau_0)$};

    \draw[->,thin] (-3.4,-0.9)--(3.4,-0.9)
        node[right] {$\tau$};
    \node[below=2pt] at (0,-0.9) {$\tau_0$};
\end{tikzpicture}
\caption{A defect-changing operator inserted on a straight Wilson
line.  With $\tau$ increasing from left to right,
$\phi_A^{ab}(\tau_0)$ changes the defect from $\mathcal D_a$ to
$\mathcal D_b$.  The double lines denote the two Wilson defects and
the black dot denotes the local insertion.}
\label{fig:defect-changing-operator-convention}
\end{figure}

We define a defect-changing operator by
\begin{align}\label{eq:defect-changing-convention}
\phi_A^{ab}(\tau_0):
\quad
\mathcal D_a
\quad (\tau<\tau_0)
\longrightarrow
\mathcal D_b
\quad (\tau>\tau_0)\, .
\end{align}
Here $\mathcal D_a$ and $\mathcal D_b$ denote Wilson defects of
electric charges $q_a$ and $q_b$, respectively, while $A$ labels
different conformal primaries in the same defect-changing sector. 
The convention is illustrated in
figure~\ref{fig:defect-changing-operator-convention}.
We denote by $\cH^{ab}$ the Hilbert space of local operators that
interpolate from $\mathcal D_a$ to $\mathcal D_b$,
\begin{align}\label{eq:hilbert_space}
\cH^{ab}
\equiv
\left\{
\ket{\phi_A^{ab}}:
\mathcal D_a
\longrightarrow
\mathcal D_b
\right\}\, .
\end{align}
The diagonal sector $\cH^{aa}$ contains genuine local operators living
on the defect $\mathcal D_a$, whereas for $a\neq b$ the space
$\cH^{ab}$ consists of genuine defect-changing operators interpolating between
two distinct Wilson defects. 

We assume that the bulk vacuum is unique and that the Wilson defect $\mathcal D_q$ is simple, meaning that it has no dimension-zero defect local operators other than the identity. Thus the dimension-zero subspaces of $\mathcal H^{00}$ and $\mathcal H^{qq}$ are each one-dimensional, spanned by the bulk identity $I^{00}$ and defect identity $I^{qq}$, respectively. This assumption is
consistent with our perturbative large-$N$ results.

We first derive the gauge charge required at a junction between a left Wilson line with charge $q_a$ and a right Wilson line with charge $q_b$ as depicted in figure~\ref{fig:defect-changing-operator-convention}. Placing the junction at $\tau=0$, the right half-line
transforms under the gauge transformation as
\begin{align}\label{eq:upper-Wilson-halfline-gauge-transformation}
\exp\left(
\i q_b
\int_0^{\infty}
A_\tau \d \tau
\right)
\longrightarrow
e^{+\i q_b\alpha(0)}
\exp\left(
\i q_b
\int_0^{\infty}
A_\tau \d \tau
\right)\, ,
\end{align}
where $\alpha(\infty)=0$.  Similarly, the left half-line transforms as
\begin{align}\label{eq:lower-Wilson-halfline-gauge-transformation}
\exp\left(
\i q_a
\int_{-\infty}^{0}
A_\tau \d \tau
\right)
\longrightarrow
e^{-\i q_a\alpha(0)}
\exp\left(
\i q_a
\int_{-\infty}^{0}
A_\tau \d \tau
\right)\, .
\end{align}
Thus the two Wilson factors together carry the junction phase $(q_b-q_a)\alpha(0)$. If the defect-changing operator $\phi_A^{ab}$ carries gauge charge
$Q_{ab}$, gauge invariance therefore requires
\begin{align}\label{eq:junction-charge}
Q_{ab}
=
q_a-q_b \, .
\end{align}

We now translate this gauge-charge constraint into a flavor selection
rule.  For an operator $\cO$ transforming in an irreducible
$\SU(N)$ representation, we define its $N$-ality
$\nu(\cO)\in\ZZ_N$ by the action of the center generator $z_{\rm c}$
introduced above,
\begin{align}\label{eq:Nality-definition}
z_{\rm c} \, \cO
=
\exp\left(
\frac{2\pi\i}{N}
\nu(\cO)
\right)
\cO\, .
\end{align}
Equivalently, for an irreducible representation $R$ described by a
Young diagram,
\begin{align}\label{eq:Nality-Young-diagram}
\nu(R)
=
\# \, (\text{boxes of }R)
\pmod N\, .
\end{align}
In particular, the fundamental and anti-fundamental scalars carry
\begin{align}\label{eq:fundamental-scalar-Nality}
\nu(\Phi_i)
=
1
\pmod N\, ,
\qquad
\nu(\Phi^{\dagger i})
=
-1
\pmod N\, .
\end{align}

As explained above, the action of $z_{\rm c}$ on the fundamental fields is
identical to a constant gauge transformation with parameter
$2\pi/N$.  It follows that, in the electric sector considered here, a
gauge-covariant local operator $\cO_Q$ of gauge charge $Q$ transforms
as
\begin{align}\label{eq:center-action-on-charged-operator}
z_{\rm c}\, \cO_Q
=
\exp\left(
\frac{2\pi\ii Q}{N}
\right)
\cO_Q\, .
\end{align}
Comparing this equation with \eqref{eq:Nality-definition}, we obtain
\begin{align}\label{eq:gauge-charge-Nality}
\nu(\cO_Q)
=
Q
\pmod N \, .
\end{align}
Applying this relation to the defect-changing operator $\phi_A^{ab}$ and using
\eqref{eq:junction-charge} give the following selection rule
\begin{align}\label{eq:junction-Nality}
\nu(\phi_A^{ab})
=
q_a - q_b
\pmod N\, .
\end{align}
Thus the difference of Wilson-line charges fixes the center charge of the flavor representation carried by the junction.\footnote{Since the faithful flavor symmetry is $\PSU(N)$ rather than
$\SU(N)$, a defect-changing operator with nonzero $N$-ality
transforms projectively under the physical flavor group.  In
particular, the projective class carried by the junction is determined
by
\begin{align}\label{eq:junction-projective-class}
[\omega_{ab}]
=
q_a-q_b
\pmod N\, .
\end{align}
This is the symmetry-fractionalization data associated with the
Wilson-line junction \cite{Aharony:2023amq,Iqbal:2025yop}.}

\subsection{OPE structures of endpoint operators}
\begin{figure}[t]
\centering
\begin{tikzpicture}[
    wilson/.style={
        double,
        double distance=1.1pt,
        line width=0.45pt
    },
    insertion/.style={circle,fill=black,inner sep=2.1pt},
    every node/.style={font=\small}
]
    \draw[wilson] (-5.2,0)--(-3.8,0);
    \draw[wilson] (-3.8,0)--(-2.2,0);
    \draw[wilson] (-2.2,0)--(-0.8,0);
    \node[insertion] at (-3.8,0) {};
    \node[insertion] at (-2.2,0) {};

    \node[above=6pt] at (-4.5,0) {$\mathcal D_a$};
    \node[above=6pt] at (-3.0,0) {$\mathcal D_b$};
    \node[above=6pt] at (-1.5,0) {$\mathcal D_c$};
    \node[below=6pt] at (-3.8,0) {$\phi_A^{ab}(0)$};
    \node[below=6pt] at (-2.2,0) {$\phi_B^{bc}(\tau)$};

    \draw[->,thick] (-0.45,0)--(0.45,0);

    \node at (0.95,0) {$\displaystyle\sum_C$};
    \draw[wilson] (1.55,0)--(3.25,0);
    \draw[wilson] (3.25,0)--(4.95,0);
    \node[insertion] at (3.25,0) {};

    \node[above=6pt] at (2.4,0) {$\mathcal D_a$};
    \node[above=6pt] at (4.1,0) {$\mathcal D_c$};
    \node[below=6pt] at (3.25,0) {$\phi_C^{ac}(0)$};
\end{tikzpicture}
\caption{Schematic OPE of two compatible defect-changing operators.
The adjacent sectors $\mathcal D_a\to\mathcal D_b\to\mathcal D_c$
combine into a single defect-changing insertion from $\mathcal D_a$
to $\mathcal D_c$.  The OPE coefficients and the powers of $\tau$ in
\eqref{eq:defect-changing-operator-OPE} are suppressed.}
\label{fig:defect-changing-operator-OPE}
\end{figure}
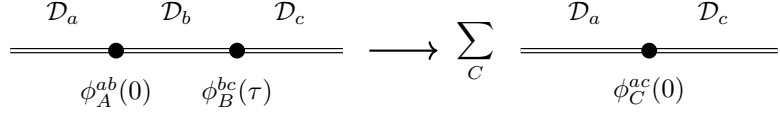

We next make the OPE structure more precise.  Consider two compatible
defect-changing operators,
\begin{align}\label{eq:compatible-defect-changing-operators}
\phi_A^{ab}:
\mathcal D_a
\longrightarrow
\mathcal D_b\, ,
\quad
\phi_B^{bc}:
\mathcal D_b
\longrightarrow
\mathcal D_c \, .
\end{align}
For $\tau>0$, their OPE takes the form
\begin{align}\label{eq:defect-changing-operator-OPE}
\phi_A^{ab}(0)
\phi_B^{bc}(\tau)
\sim
\sum_C
\lambda_{AB}{}^{C}\, 
|\tau|^{
\Delta_C^{ac}
-
\Delta_A^{ab}
-
\Delta_B^{bc}
}
\phi_C^{ac}(0)
\, .
\end{align}
Here and below, contributions from descendant operators are omitted. The corresponding composition of Wilson defects is illustrated in
figure~\ref{fig:defect-changing-operator-OPE}.
Thus the defect labels compose according to
\begin{align}\label{eq:defect-changing-sector-composition}
\mathcal H^{ab}
\times
\mathcal H^{bc}
\longrightarrow
\mathcal H^{ac}\, .
\end{align}
If the three operators transform in irreducible representations
$R_A^{ab}$, $R_B^{bc}$, and $R_C^{ac}$, a non-vanishing OPE
coefficient requires
\begin{align}\label{eq:junction-flavor-selection-rule}
R_C^{ac}
\subset
R_A^{ab}
\otimes
R_B^{bc}\, .
\end{align}
The $N$-ality rule is automatically compatible with this OPE,
since
\begin{align}\label{eq:junction-Nality-composition}
\begin{aligned}
\nu(\phi_A^{ab})
+
\nu(\phi_B^{bc})
=
\nu(\phi_C^{ac})
\pmod N\, .
\end{aligned}
\end{align}
Thus junction composition realizes the additive composition law of
the projective $\PSU(N)$ charges.

Although the discussion above applies to general junctions between Wilson defects of different
charges, in this work we will mainly focus on this special class of
defect-changing operators interpolating between the
trivial defect $\mathcal D_0$ and a charge-$q$ Wilson defect
$\mathcal D_q$. For a positive charge-$q$ Wilson line, we denote the lowest derivative-free endpoint and its conjugate by
\begin{align}
E^{(q)}_I
\equiv
\phi^{q0}_I \, ,
\quad
\bar E_{(q)}^{I}
\equiv
\phi^{0q,I} \, ,
\end{align}
where $I$ denotes an index in the rank-$q$ symmetric representation\footnote{Remark that the index $I$ used here should not be confused with the primary label $A$ in \eqref{eq:defect-changing-convention}.}\footnote{See \eqref{eq:general-q-endpoint-infinite-N} for the concrete expressions of these endpoint operators at $N=\infty$.}
\begin{align}
E^{(q)}
\in
S_q
\equiv
\mathrm{Sym}^q\mathbf N \, ,
\quad
\bar E_{(q)}
\in
\overline{S_q} \, .
\end{align}
Since the collective index $I\in S_q$ becomes cumbersome for general $q$, it is convenient to suppress these indices by introducing commuting flavor polarizations $u^i$ and
$\bar u_i$:
\begin{align}\label{eq:general-q-collective-index-polarization}
    E^{(q)}(u)
    \equiv
    u^{i_1}\cdots u^{i_q}
    E^{(q)}_{i_1\cdots i_q}\, , \quad 
    \bar E_{(q)}(\bar u)
    \equiv
    \bar u_{i_1}\cdots\bar u_{i_q}
    \bar E_{(q)}^{i_1\cdots i_q}\, .
\end{align}
The operators with explicit
flavor indices are recovered by
\begin{align}\label{eq:general-q-inverse-polarization-map}
    E^{(q)}_{i_1\cdots i_q}
    =
    \left.
    \frac{1}{q!}
    \frac{\partial}{\partial u^{i_1}}
    \cdots
    \frac{\partial}{\partial u^{i_q}}
    E^{(q)}(u)
    \right|_{u=0}\, , \quad 
    \bar E_{(q)}^{i_1\cdots i_q}
    =
    \left.
    \frac{1}{q!}
    \frac{\partial}{\partial\bar u_{i_1}}
    \cdots
    \frac{\partial}{\partial\bar u_{i_q}}
    \bar E_{(q)}(\bar u)
    \right|_{\bar u=0}\, .
\end{align}
Their $N$-alities are
\begin{align}
\nu \left(E^{(q)}\right)
=
q
\, ,
\quad
\nu \left(\bar E_{(q)}\right)
=
-q
\pmod N \, .
\end{align}
The composition law of defect-changing operators gives two distinct
OPEs depending on the ordering of the endpoint pair. Fusing the
anti-endpoint with the endpoint removes a finite charge-$q$ Wilson
segment and leaves the trivial defect on both sides of the fused
insertion,
\begin{align}
\bar E_{(q)}^{I}
\times
E^{(q)}_J
\subset
\cH^{00}\, .
\end{align}
By contrast, the opposite ordering removes a finite trivial segment
and joins the two neighboring charge-$q$ Wilson segments into a
single defect,
\begin{align}
E^{(q)}_J
\times
\bar E_{(q)}^{I}
\subset
\cH^{qq}\, .
\end{align}
The two OPEs therefore probe different Hilbert spaces. 

Indeed, both endpoint products carry vanishing total $N$-ality, and
their flavor content is determined by
\begin{align}\label{eq:decomposition}
\overline{S_q}
\otimes
S_q
=
\bigoplus_{k=0}^{q}
R_k \, ,
\quad
R_k
\equiv
[k,0,\cdots,0,k]\, .
\end{align}
Here $[a_1, a_2, \cdots,a_{N-1}]$ denotes the irreducible
$\SU(N)$ representation with Dynkin labels
$(a_1,\cdots,a_{N-1})$. Also, the representation dimension of $R_k$ is given by
\begin{align}\label{eq:dimension-Rk}
    d_k
    \equiv
    \dim R_k
    =
    \frac{2k+N-1}{N-1}
    \left[
        \frac{(k+N-2)!}{k!(N-2)!}
    \right]^2\, .
\end{align}
This decomposition is multiplicity free. In particular,
\begin{align}
R_0
=
\one \, ,
\quad
R_1
=
\Adj \, ,
\end{align}
while for $q\geq2$ additional center-neutral representations appear.
For example,
\begin{align}
\overline{S_2}
\otimes
S_2
=
\one
\oplus
\Adj
\oplus
[2,0,\cdots,0,2] \, .
\end{align}
More explicitly, let
$\mathcal C^{(k)A,I}{}_{J}$ denote a Clebsch--Gordan tensor projecting
$\overline{S_q}\otimes S_q$ onto $R_k$, where $A$ is an index in
$R_k$. We choose the normalization
\begin{align}\label{eq:normalization}
\sum_{I,J}
\overline{\mathcal C}^{(k)}_{A,I}{}^{J}
\mathcal C^{(\ell)B,I}{}_{J}
= 
\delta_{k\ell}\delta^{AB}\, , 
\quad
\overline{\mathcal C}^{(k)}_{A,I}{}^{J}
\equiv
\left(
\mathcal C^{(k)A,I}{}_{J}
\right)^* \, .
\end{align}
An explicit tensor realization of
$\mathcal C^{(k)A,I}{}_{J}$ is given in
appendix~\ref{app:tensor_index_CG}. The bulk-channel OPE can then be written as
\begin{align}
\bar E_{(q)}^{I}(0)
E^{(q)}_J(\tau)
\sim
\sum_{k=0}^{q}
\sum_{\cO \in\cH^{00}_{R_k}}
\lambda^{00}_{q,k,\cO}
\tau^{
\Delta_{\cO}
-
2\Delta_{E_q}
}
\sum_{A=1}^{d_k}
\mathcal C^{(k)A,I}{}_{J} \,
\cO_A(0)\, .
\end{align}
The opposite ordering instead gives the Wilson-line OPE
\begin{align}
E^{(q)}_J(0)
\bar E_{(q)}^{I}(\tau)
\sim
\sum_{k=0}^{q}
\sum_{\widehat{\cO}\in\cH^{qq}_{R_k}}
\hatlambda^{qq}_{q,k,\widehat{\cO}}
\tau^{
\hDelta_{\widehat{\cO}}
-
2\Delta_{E_q}
}
\sum_{A=1}^{d_k}
\overline{\mathcal C}^{(k)}_{A,J}{}^{I}\,  
\widehat{\cO}_A(0) \, .
\end{align}
Thus the same set of flavor representations $R_k$ appears in both
channels, while the corresponding conformal primaries belong to the
distinct Hilbert spaces $\cH^{00}$ and $\cH^{qq}$. As we will see in the next subsection, the endpoint
crossing equation therefore relates ordinary bulk CFT data in
$\cH^{00}$ to local CFT data of the charge-$q$ Wilson line in
$\cH^{qq}$.

\subsection{Endpoint bootstrap equations}

\begin{figure}[t]
\centering
\begin{tikzpicture}[
    wilson/.style={
        double,
        double distance=1.1pt,
        line width=0.45pt
    },
    insertion/.style={circle,fill=black,inner sep=2.1pt},
    every node/.style={font=\small}
]
    \draw[wilson] (-3.8,4.0)--(-1.5,4.0);
    \draw[wilson] (1.5,4.0)--(3.8,4.0);
    \node[insertion] at (-3.8,4.0) {};
    \node[insertion] at (-1.5,4.0) {};
    \node[insertion] at (1.5,4.0) {};
    \node[insertion] at (3.8,4.0) {};

    \node[below=6pt] at (-3.8,4.0)
        {$\bar{E}_{(q)}^{I}(\tau_1)$};
    \node[below=6pt] at (-1.5,4.0)
        {$E^{(q)}_{J}(\tau_2)$};
    \node[below=6pt] at (1.5,4.0)
        {$\bar{E}_{(q)}^{K}(\tau_3)$};
    \node[below=6pt] at (3.8,4.0)
        {$E^{(q)}_{L}(\tau_4)$};

    \draw[->,thick] (-0.15,2.90)--(-2.20,0.85)
        node[midway,above,sloped] {$(12)(34)$};
    \draw[->,thick] (0.15,2.90)--(2.20,0.85)
        node[midway,above,sloped] {$(14)(32)$};

    \node at (-2.55,-1.55)
        {bulk channel: $\mathcal H^{00}$};
    \node at (-4.05,0) {$\displaystyle\sum_{\mathcal O}$};
    \node[insertion] at (-3.25,0) {};
    \node[insertion] at (-1.85,0) {};
    \node[below=5pt] at (-3.25,0) {$\mathcal O$};
    \node[below=5pt] at (-1.85,0) {$\bar{\mathcal O}$};

    \node at (2.55,-1.55)
        {defect channel: $\mathcal H^{qq}$};
    \node at (0.55,0) {$\displaystyle\sum_{\widehat{\mathcal O}}$};
    \draw[wilson] (1.15,0)--(3.95,0);
    \node[insertion] at (1.85,0) {};
    \node[insertion] at (3.25,0) {};
    \node[below=5pt] at (1.85,0) {$\widehat{\mathcal O}$};
    \node[below=5pt] at (3.25,0) {$\widehat{\bar{\mathcal O}}$};
\end{tikzpicture}
\caption{Configuration of four endpoint operators and the two distinct OPE channels of the ordered four-endpoint
correlator.  In the bulk channel, the two finite charge-$q$ Wilson
segments shrink and the exchanged operators belong to
$\mathcal H^{00}$. (The trivial defect is not drawn.)  In the defect
channel, the intervening trivial segments shrink and the two Wilson
segments join, so the exchanged operators are local operators on
$\mathcal D_q$ and belong to $\mathcal H^{qq}$.}
\label{fig:general-q-two-endpoint-OPE-channels}
\end{figure}
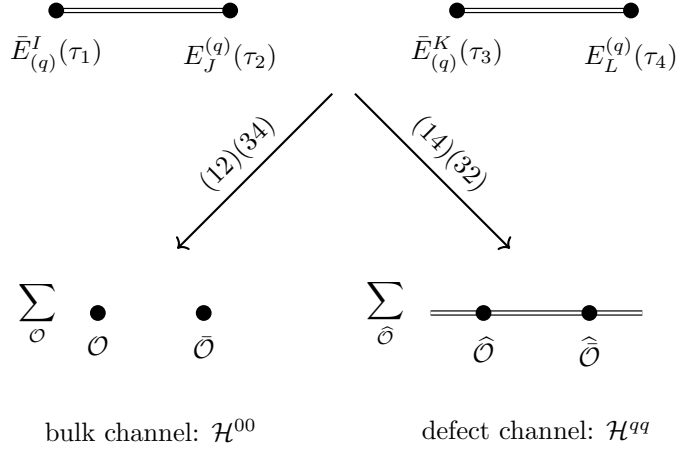

We now formulate the crossing equations associated to the endpoints of a
charge-$q$ Wilson line.  We take $q\in\mathbb Z_{>0}$ and the equations for
negative $q$ follow by charge conjugation.  We choose the relative phase
and normalization of the endpoint operators such that
\begin{align}\label{eq:general-q-endpoint-dagger-and-two-point}
    \left\langle
        \bar E_{(q)}^{I}(\tau_1)
        E^{(q)}_J(\tau_2)
    \right\rangle
    =
    \frac{\delta^I{}_J}{|\tau_{12}|^{2\Delta_{E_q}}}\, ,
    \quad
    \tau_1 < \tau_2\, ,
\end{align}
and
\begin{align}\label{eq:dimension-Sq}
    n_q
    \equiv
    \dim S_q
    =
    \binom{N+q-1}{q}\, .
\end{align}
Notably, this quantity is related to the dimensions $d_k$ of $R_k$ via
\begin{align}\label{eq:sumRk_nq}
    \sum_{k=0}^{q}d_k
    =
    n_q^2\, . 
\end{align}
Consider the ordered four-point function
\begin{align}\label{eq:general-q-ordered-four-point}
    &\left\langle
        \bar E_{(q)}^{I}(\tau_1)
        E^{(q)}_J(\tau_2)
        \bar E_{(q)}^{K}(\tau_3)
        E^{(q)}_L(\tau_4)
    \right\rangle
    =
    \frac{1}{(\tau_{12}\tau_{34})^{2\Delta_{E_q}}}
    \mathcal G^{I}{}_{J}{}^{K}{}_{L}(z)\, ,
    \quad
    z
    =
    \frac{\tau_{12}\tau_{34}}{\tau_{13}\tau_{24}}\, ,
    \quad
    0<z<1\, ,
\end{align}
where $\tau_1<\tau_2<\tau_3 < \tau_4$. This configuration is shown in
figure~\ref{fig:general-q-two-endpoint-OPE-channels}.
In this configuration, the $(12)$~OPE removes a finite charge-$q$ Wilson
segment and is expanded in $\mathcal H^{00}$, whereas the $(23)$ OPE
removes the intervening trivial segment and joins the two neighboring
Wilson segments.  The defect channel is therefore expanded in
$\mathcal H^{qq}$. The two operations are summarized in
figure~\ref{fig:general-q-two-endpoint-OPE-channels}.

As we have seen in the previous section, both channels resolve the multiplicity-free flavor decomposition given in \eqref{eq:decomposition}. Let $\mathbb T_k^{(\mathrm{s})}$ and $\mathbb T_k^{(\mathrm{t})}$ be the orthonormal flavor tensors associated with the $(12)(34)$ and $(14)(32)$ pairings, respectively, 
\begin{align}\label{eq:general-q-two-flavor-bases}
(\mathbb T_k^{(\mathrm{s})})^{\,I}{}_{J}{}^{K}{}_{L}
&\equiv
\frac{1}{\sqrt{d_k}}
\sum_{A=1}^{d_k}
\mathcal C^{(k)A,I}{}_{J}\,
\overline{\mathcal C}^{(k)}_{A,L}{}^{K}\, , \\
(\mathbb T_k^{(\mathrm{t})})^{I}{}_{J}{}^{K}{}_{L}
&\equiv
\frac{1}{\sqrt{d_k}}
\sum_{B=1}^{d_k}
\mathcal C^{(k)B,I}{}_{L}\, 
\overline{\mathcal C}^{(k)}_{B,J}{}^{K}\, .
\end{align}
These two bases are related by the linear transformation, 
\begin{align}\label{eq:general-q-Racah-crossing}
    \mathbb T_k^{(\mathrm{s})}
    =
    \sum_{\ell=0}^{q}
    F_{k\ell}^{(q,N)}
    \mathbb T_\ell^{(\mathrm{t})}\, ,
\end{align}
where $F_{k\ell}^{(q,N)}$ is the Racah matrix given by
\begin{align}\label{eq:general-q-Racah-six-j}
    F_{k\ell}^{(q,N)}
    =
    \sqrt{d_kd_\ell}\,
    \left\{
    \begin{matrix}
        S_q & \overline{S_q} & R_k\\
        S_q & \overline{S_q} & R_\ell
    \end{matrix}
    \right\}_{\SU(N)}\, .
\end{align}
Our convention for the $6j$ symbol is fixed by
\eqref{eq:general-q-Racah-crossing}.  Explicit expressions for these Racah coefficients for symmetric representations are known~\cite{Mironov:2017jgo}. With the real phase convention used here, $F^{(q,N)}$ is symmetric and orthogonal,
\begin{align}\label{eq:general-q-Racah-involution}
    (F^{(q,N)})^T 
    =
    F^{(q,N)}\, ,
    \quad
    (F^{(q,N)})^2
    =
    \mathbf 1_{q+1}\, .
\end{align}
The entries involving the singlet take the particularly simple form
\begin{align}\label{eq:general-q-Racah-singlet-row}
    F_{0k}^{(q,N)}
    =
    F_{k0}^{(q,N)}
    =
    \frac{\sqrt{d_k}}{n_q}\, .
\end{align}
Notice that the relation \eqref{eq:sumRk_nq} follows from the unit norm of the first row of $F^{(q,N)}$. In this basis, the two conformal-block decompositions of the reduced
correlator are
\begin{align}\label{eq:general-q-bulk-channel-decomposition}
    \mathcal G^{I}{}_{J}{}^{K}{}_{L}(z)
    =
    \sum_{k=0}^{q}
    \sqrt{d_k}\,
    (\mathbb T_k^{(\mathrm{s})})^{I}{}_{J}{}^{K}{}_{L}\,
    \mathcal G^{00}_{q,k}(z)\, ,
\end{align}
and
\begin{align}\label{eq:general-q-Wilson-channel-decomposition}
    \mathcal G^{I}{}_{J}{}^{K}{}_{L}(z)
    =
    \rho_q(z)
    \sum_{k=0}^{q}
    \sqrt{d_k}\,
    (\mathbb T_k^{(\mathrm{t})})^{I}{}_{J}{}^{K}{}_{L}\, 
    \mathcal G^{qq}_{q,k}(1-z)\, ,
\end{align}
where
\begin{align}\label{eq:general-q-rho}
    \rho_q(z)
    \equiv
    \left(\frac{z}{1-z}\right)^{2\Delta_{E_q}}\, .
\end{align}
The channel functions have positive block expansions
\begin{align}\label{eq:general-q-channel-block-sums}
    \mathcal G^{00}_{q,k}(z)
    =
    \sum_{\mathcal O\in\mathcal H^{00}_{R_k}}
    \left|\lambda^{00}_{q,k,\mathcal O}\right|^2
    g_{\Delta_{\mathcal O}}(z)\, , \quad 
    \mathcal G^{qq}_{q,k}(x)
    =
    \sum_{\widehat{\mathcal O}\in\mathcal H^{qq}_{R_k}}
    \left|\hatlambda^{qq}_{q,k,\widehat{\mathcal O}}\right|^2
    g_{\hDelta_{\widehat{\mathcal O}}}(x)\, ,
\end{align}
where $x \equiv 1-z$ is the cross ratio of the defect channel and $g_\Delta(z)$ is the $\mathrm{SL}(2, \mathbb{R})$ conformal block given by
\begin{align}\label{eq:general-q-SL2-block}
    g_\Delta(z)
    =
    z^\Delta\,{}_2F_1(\Delta,\Delta;2\Delta;z)\, .
\end{align}
The identity terms in \eqref{eq:general-q-channel-block-sums} require
some care because all correlators containing defect-changing operators
are normalized by the defect-free bulk partition function, rather than
by the partition function in each defect sector separately.  We place
the charge-$q$ Wilson defect $\mathcal D_q$ on a great circle of $S^3$ and define its
$g$-function by
\begin{align}\label{eq:general-q-g-function}
    g_q
    \equiv
    \frac{Z_{S^3}[\mathcal D_q]}{Z_{S^3}[\mathcal D_0]}
    =
    \left\langle I^{qq}\right\rangle\, .
\end{align}
Here $I^{qq}$ is the identity primary in $\mathcal H^{qq}$.  The last
equality means that insertion of the defect identity leaves the
charge-$q$ defect partition function, whose expectation value in our
bulk-normalized convention is $g_q$, rather than one.  In particular,
\begin{align}\label{eq:general-q-trivial-g-function}
    g_0=1\, ,
    \qquad
    \left\langle I^{00}\right\rangle=1\, .
\end{align}
The corresponding flat-space definition is obtained by conformally
mapping the circular defect to a straight line and using the induced
conformally invariant regulator.  The perimeter divergence is removed
by the usual defect-local cosmological counterterm.  Charge conjugation
maps $\mathcal D_q$ to $\mathcal D_{-q}$ and hence implies
$g_q=g_{-q}$, while reflection positivity requires $g_q>0$.

Non-identity primaries are normalized in the same bulk-normalized convention.
For $\widehat{\mathcal O} \in \mathcal H^{qq}$ we take $\langle \widehat{\mathcal O}(\tau)\widehat{\mathcal{O}}^\dagger(0)\rangle = |\tau|^{-2\hDelta_{\widehat{\mathcal{O}}}}$, where the correlator is divided by $Z_{S^3}[\mathcal D_0]$, so that the coefficients in \eqref{eq:general-q-channel-block-sums} are the squared OPE coefficients.

Our normalization also fixes the identity OPE coefficients as follows:
\begin{align}\label{eq:general-q-identity-OPEs}
    \bar E_{(q)}^{I}(0)E^{(q)}_J(\tau)
    \supset
    \delta^I{}_J\,
    |\tau|^{-2\Delta_{E_q}}I^{00}\, , \quad 
    E^{(q)}_J(0)\bar E_{(q)}^{I}(\tau)
    \supset
    \frac{\delta^I{}_J}{g_q}\,
    |\tau|^{-2\Delta_{E_q}}I^{qq}\, .
\end{align}
Since the normalized singlet Clebsch--Gordan coefficient is 
\begin{align}
    \mathcal C^{(0),I}{}_{J}=\frac{\delta^I{}_J}{\sqrt{n_q}}\, , 
\end{align}
the effective identity block coefficients in our orthonormal Clebsch--Gordan
convention are
\begin{align}\label{eq:general-q-identity-block-coefficients}
    \begin{aligned}
        \mathcal G^{00}_{q,0}(z)
        &=
        n_q
        +
        \sum_{\substack{
            \mathcal O\in\mathcal H^{00}_{\one}\\
            \mathcal O\neq I^{00}
        }}
        \left|\lambda^{00}_{q,0,\mathcal O}\right|^2
        g_{\Delta_{\mathcal O}}(z)\, ,
        \\
        \mathcal G^{qq}_{q,0}(x)
        &=
        \frac{n_q}{g_q}
        +
        \sum_{\substack{
            \widehat{\mathcal O}\in\mathcal H^{qq}_{\one}\\
            \widehat{\mathcal O}\neq I^{qq}
        }}
        \left|\hatlambda^{qq}_{q,0,\widehat{\mathcal O}}\right|^2
        g_{\hDelta_{\widehat{\mathcal O}}}(x)\, .
    \end{aligned}
\end{align}
For $k\geq1$, the sums in
\eqref{eq:general-q-channel-block-sums} contain no identity term.

As illustrated in figure~\ref{fig:general-q-two-endpoint-OPE-channels},
the two channel decompositions correspond geometrically to shrinking
the two finite Wilson segments or joining them into a single
charge-$q$ Wilson line. Equating~\eqref{eq:general-q-bulk-channel-decomposition} and~\eqref{eq:general-q-Wilson-channel-decomposition}, and using \eqref{eq:general-q-Racah-crossing}, gives the charge-$q$ endpoint crossing equation
\begin{align}\label{eq:general-q-endpoint-crossing}
    \sum_{k=0}^{q}
    \sqrt{d_k}\,
    F_{k\ell}^{(q,N)}\,
    \mathcal G^{00}_{q,k}(z)
    =
    \rho_q(z)\sqrt{d_\ell}\,
    \mathcal G^{qq}_{q,\ell}(1-z)\, ,
    \quad
    \ell=0, \cdots, q
    \, .
\end{align}
Thus crossing relates the bulk spectrum in $\mathcal H^{00}$ to the
local spectrum of the charge-$q$ Wilson line in $\mathcal H^{qq}$.  Introducing
\begin{align}
    \mathbf d^{1/2}
    \equiv
    \operatorname{diag}(\sqrt{d_0},\cdots,\sqrt{d_q})\, ,
\end{align}
the crossing relation~\eqref{eq:general-q-endpoint-crossing} can be written compactly
as
\begin{align}\label{eq:general-q-endpoint-crossing-matrix}
    (F^{(q,N)})^T \mathbf d^{1/2}
    \vec{\mathcal G}^{\,00}_q(z)
    =
    \rho_q(z)\mathbf d^{1/2}
    \vec{\mathcal G}^{\,qq}_q(1-z)\, .
\end{align}
It is useful to make this matrix equation explicit in the fundamental case.  For $q=1$, the matrices appearing in the crossing relation are
\begin{align}\label{eq:q-one-Racah-and-d-matrices}
    F^{(1,N)}
    =
    \frac{1}{N}
    \begin{pmatrix}
        1 & \sqrt{N^2-1}\\
        \sqrt{N^2-1} & -1
    \end{pmatrix}\, ,
    \quad
    \mathbf d^{1/2}
    =
    \begin{pmatrix}
        1 & 0\\
        0 & \sqrt{N^2-1}
    \end{pmatrix}\, .
\end{align}
Equation \eqref{eq:general-q-endpoint-crossing-matrix} therefore becomes
\begin{align}\label{eq:q-one-explicit-crossing-matrix}
    \frac{1}{N}
    \begin{pmatrix}
        1 & N^2-1\\
        \sqrt{N^2-1} & -\sqrt{N^2-1}
    \end{pmatrix}
    \begin{pmatrix}
        \mathcal G^{00}_{1,0}(z)\\
        \mathcal G^{00}_{1,1}(z)
    \end{pmatrix}
    =
    \rho_1(z)
    \begin{pmatrix}
        \mathcal G^{11}_{1,0}(1-z)\\
        \sqrt{N^2-1}\,\mathcal G^{11}_{1,1}(1-z)
    \end{pmatrix}\, .
\end{align}
Since the channel functions in \eqref{eq:general-q-channel-block-sums} have non-negative spectral coefficients and $g_q>0$, the general matrix equation retains the positivity required for the endpoint bootstrap.

\subsection{A spectral interpretation of the defect $g$-function}
\label{subsec:general-q-g-spectral-interpretation}

The endpoint crossing equations relate two different sets of local operators: bulk operators in $\mathcal H^{00}$ and defect operators in $\mathcal H^{qq}$.  Can they constrain the relation between these spectra without first computing the four-point function? We show that the answer is affirmative for their cumulative OPE weights.  In each flavor channel, the ratio of these weights, summed up to a large scaling dimension, approaches the defect $g$-function. This gives a spectral characterization of the partition-function ratio~\eqref{eq:general-q-g-function}, without using the large-$N$ expansion.

We assume $\Delta_{E_q}>0$, reflection positivity, and convergence of the block expansions for $0<z<1$.  We also assume that each of the sectors $\mathcal H^{00}$ and $\mathcal H^{qq}$ has a unique identity, separated from the other exchanged operators by a positive gap.
The identity coefficients in~\eqref{eq:general-q-identity-block-coefficients} then imply
\begin{align}\label{eq:general-q-identity-dominated-limits}
    \mathcal G^{00}_{q,k}(x)
    =n_q\delta_{k0}+o(1)\, , \quad
    \mathcal G^{qq}_{q,k}(x)
    =\frac{n_q}{g_q}\delta_{k0}+o(1)\, ,
    \quad x\to0^+\, .
\end{align}
Although the identity occurs only in the singlet, crossing makes it contribute to the leading singularity of every flavor channel in the opposite OPE. To see this explicitly, we invert~\eqref{eq:general-q-endpoint-crossing-matrix} using the orthogonality of $F^{(q,N)}$, 
    \begin{align}
    F^{(q,N)}(F^{(q,N)})^T
    =\mathbf 1_{q+1}\quad \Longleftrightarrow\quad 
    \sum_{\ell=0}^{q}
    F_{k\ell}^{(q,N)}F_{j\ell}^{(q,N)}
    =\delta_{kj}\, ,
\end{align}
which gives 
\begin{align}\label{eq:general-q-crossing-for-spectral-asymptotics}
    \mathcal G^{00}_{q,k}(z)
    =
    \frac{\rho_q(z)}{\sqrt{d_k}}
    \sum_{\ell=0}^{q}
    F_{k\ell}^{(q,N)}\sqrt{d_\ell}\,
    \mathcal G^{qq}_{q,\ell}(1-z)\, .
\end{align}
As $z\to1^-$, only the identity term with $\ell=0$ survives inside the sum.  Since $d_0=1$ and $F_{k0}^{(q,N)}=\sqrt{d_k}/n_q$, its coefficient is
\begin{align}
    \frac{F_{k0}^{(q,N)}}{\sqrt{d_k}}\,
    \frac{n_q}{g_q}
    =\frac{1}{g_q}\, .
\end{align}
Applying the same argument in the reverse direction gives
\begin{align}\label{eq:general-q-channel-leading-singularities}
    \mathcal G^{00}_{q,k}(z)
    \sim
    \frac{1}{g_q}(1-z)^{-2\Delta_{E_q}}\, , \quad
    \mathcal G^{qq}_{q,k}(z)
    \sim
    (1-z)^{-2\Delta_{E_q}}\, , 
    \quad z\to1^-\, .
\end{align}
These relations hold for every $k=0,\cdots,q$. Here and below, $\sim$ means that the ratio of the two expressions tends to one as $z\to1^-$.

To understand what these singularities imply for the OPE data, consider the contribution of a single exchanged operator. At fixed $\Delta>0$, its block behaves as
\begin{align}\label{eq:general-q-fixed-dimension-block-logarithm}
    g_\Delta(z)
    =
    \frac{\Gamma(2\Delta)}{\Gamma(\Delta)^2}
    \log\frac{1}{1-z}
    +O(1)\, ,
    \quad z\to1^-\, .
\end{align}
A finite sum of blocks therefore grows only logarithmically. It cannot reproduce the stronger power-law singularities $(1-z)^{-2\Delta_{E_q}}$ required by crossing. These singularities must instead arise from summing contributions of operators with arbitrarily large dimensions. Crossing thus constrains their combined OPE weight, even without determining the dimensions or OPE coefficients of individual operators~\cite{Qiao:2017xif}.

To quantify this combined contribution, let $H$ be a cutoff on the exchanged scaling dimension and define the cumulative OPE weights
\begin{align}\label{eq:general-q-cumulative-OPE-weights}
    \mathscr S^{00}_{q,k}(H)
    \equiv
    \sum_{\substack{
        \mathcal O\in\mathcal H^{00}_{R_k}\\
        0<\Delta_{\mathcal O}\leq H
    }}
    \frac{\Gamma(2\Delta_{\mathcal O})}
         {\Gamma(\Delta_{\mathcal O})^2}
    \left|\lambda^{00}_{q,k,\mathcal O}\right|^2\, , \quad
    \mathscr S^{qq}_{q,k}(H)
    \equiv
    \sum_{\substack{
        \widehat{\mathcal O}\in\mathcal H^{qq}_{R_k}\\
        0<\hDelta_{\widehat{\mathcal O}}\leq H
    }}
    \frac{\Gamma(2\hDelta_{\widehat{\mathcal O}})}
         {\Gamma(\hDelta_{\widehat{\mathcal O}})^2}
    \left|\hatlambda^{qq}_{q,k,\widehat{\mathcal O}}\right|^2\, .
\end{align}
The sums run over the $\mathrm{SL}(2,\mathbb R)$ primaries appearing in \eqref{eq:general-q-channel-block-sums}, with the identity omitted. Each squared OPE coefficient is weighted by the gamma-function factor appearing in~\eqref{eq:general-q-fixed-dimension-block-logarithm}.
Thus these sums measure the total OPE weight up to dimension $H$. 

We now determine how these sums grow as $H$ increases. As $z$ approaches one, contributions from increasingly large exchanged dimensions become important.  We therefore need an
approximation that captures large $\Delta$ and $z\to1$ simultaneously, rather than the behavior of each block at fixed $\Delta$. Setting $\epsilon\equiv 1-z$, the relevant block asymptotics is \cite{Qiao:2017xif}
\begin{align}\label{eq:general-q-large-dimension-block}
    g_\Delta(1-\epsilon)
    \sim
    2\frac{\Gamma(2\Delta)}{\Gamma(\Delta)^2}
    K_0(2\Delta\sqrt{\epsilon})\, ,
\end{align}
where $K_0$ is the modified Bessel function of the second kind, and the limit is $\Delta\to\infty$, $\epsilon\to0^+$ with $\Delta\sqrt{\epsilon}$ fixed and positive. For a positive block expansion, the Tauberian theorem~\cite{Qiao:2017xif} relates this limit of the correlator to the large-$H$ growth of its cumulative OPE weight. 
Applying it to the two singularities in \eqref{eq:general-q-channel-leading-singularities} gives\footnote{Positivity is essential here. The theorem constrains the cumulative sums without assuming that the spectrum is evenly spaced or that individual OPE coefficients vary smoothly.} 
\begin{align}\label{eq:general-q-cumulative-OPE-asymptotics}
    \mathscr S^{00}_{q,k}(H)
    \sim
    \frac{H^{4\Delta_{E_q}}}
         {2g_q\Delta_{E_q}\Gamma(2\Delta_{E_q})^2}\, , \quad
    \mathscr S^{qq}_{q,k}(H)
    \sim
    \frac{H^{4\Delta_{E_q}}}
         {2\Delta_{E_q}\Gamma(2\Delta_{E_q})^2}\, ,
    \quad H\to\infty\, .
\end{align}
The two Hilbert spaces need not have matching spectra or individual OPE coefficients.  Nevertheless, crossing fixes the ratio of their leading cumulative OPE weights in terms of the defect partition function:
\begin{align}\label{eq:general-q-g-function-spectral-ratio}
        g_q
        =
        \lim_{H\to\infty}
        \frac{\mathscr S^{qq}_{q,k}(H)}
             {\mathscr S^{00}_{q,k}(H)}
    \, ,
    \quad k=0,\cdots,q\, .
\end{align}
Equation~\eqref{eq:general-q-g-function-spectral-ratio} gives a nonperturbative interpretation of the defect $g$-function in terms of endpoint OPE data.

\section{Four-endpoint correlators from large-$N$ diagrammatics}
\label{sec:analytical-bootstrap}

We begin by relating the reduced correlator \eqref{eq:general-q-ordered-four-point} used in the bootstrap to the microscopic path-integral insertion.  For $x_r=(\tau_r,\bm 0_\perp)$ with
$\tau_1<\tau_2<\tau_3<\tau_4$, define
\begin{align}\label{eq:general-q-microscopic-four-endpoint-insertion}
    \mathcal E_q
    &\equiv
    \frac{1}{(q!)^2}
    \left(\bar u_1\cdot\Phi^\dagger(x_1)\right)^q
    W_q(C_{12})
    \left(u_2\cdot\Phi(x_2)\right)^q
    \left(\bar u_3\cdot\Phi^\dagger(x_3)\right)^q
    W_q(C_{34})
    \left(u_4\cdot\Phi(x_4)\right)^q\, ,
\end{align}
where
$W_q(C_{rs})=\exp\!\left(\i q\int_{\tau_r}^{\tau_s}
A_\tau\,\d\tau\right)$ for $(rs)=(12),(34)$. Also, $\bar u_1, u_2, \bar u_3, u_4$ are polarization vectors introduced in~\eqref{eq:general-q-collective-index-polarization}. The factor
$(q!)^{-1/2}$ at each endpoint implements the normalization of the
symmetric product of $q$ scalar fields. 

Let the index-free bare two-point function on $C_{rs}$ be
\begin{align}\label{eq:general-q-bare-two-point-amplitude}
    C^{\rm B}_{q,rs}
    \equiv
    \frac{1}{q!(\bar u_r\cdot u_s)^q}
    \left\langle
        \left(\bar u_r\cdot\Phi^\dagger(x_r)\right)^q
        W_q(C_{rs})
        \left(u_s\cdot\Phi(x_s)\right)^q
    \right\rangle_{\rm B}\, .
\end{align}
Here $\langle\cdots\rangle_{\rm B}$ denotes an expectation value in the
UV-regulated microscopic theory, evaluated with bare fields and the
bare Wilson operator before endpoint renormalization and subtraction
of defect-local counterterms. The index-free four-point function is then
\begin{align}\label{eq:general-q-microscopic-to-reduced-correlator}
    \begin{aligned}
    \cG_q(\bar u_1,u_2,\bar u_3,u_4;z)
    &\equiv
    \frac{
        \left\langle\mathcal E_q\right\rangle_{\rm B}
    }{
        C^{\rm B}_{q,12}C^{\rm B}_{q,34}
    }
    \\
    &=
    \bar u_{1,I}u_2^{J}\bar u_{3,K}u_4^{L}
    \mathcal G^{I}{}_{J}{}^{K}{}_{L}(z)\, .
    \end{aligned}
\end{align}
In the second line, a collective index denotes a symmetric
multi-index of rank $q$.

We compute the endpoint four-point function of a charge-$q$ Wilson line in the large-$N$ expansion, keeping $q$ fixed. Starting from \eqref{eq:general-q-microscopic-to-reduced-correlator}, we evaluate the index-free correlator at $N=\infty$ and through first subleading order in $1/N$. We then project the result onto irreducible flavor channels and use the crossing relation \eqref{eq:general-q-endpoint-crossing-matrix}
to obtain its defect-channel conformal-block expansion. Matching this expansion with the defect OPE allows us to extract various defect CFT data including the defect $g$-function, endpoint dimension, and endpoint OPE coefficients.

\subsection{The strict $N=\infty$ solution at general charge $q$}
\label{subsec:general-q-infinite-N}

At $N=\infty$, fluctuations of the
gauge field and the unnormalized Hubbard--Stratonovich field $\lambda$ are suppressed.  The
charge-$q$ endpoint therefore reduces, for the purpose of its
correlation functions, to the normalized symmetric product of $q$ free
fundamental scalars.  In the index-free notation introduced in
\eqref{eq:general-q-collective-index-polarization}, this becomes
\begin{align}\label{eq:general-q-endpoint-infinite-N}
    E^{(q)}(u,\tau)
    \longrightarrow
    \frac{(4\pi)^{q/2}}{\sqrt{q!}}
    :\!\left(u^i\Phi_i(\tau)\right)^q\!:\, , \quad 
    \bar E_{(q)}(\bar u,\tau)
    \longrightarrow
    \frac{(4\pi)^{q/2}}{\sqrt{q!}}
    :\!\left(\bar u_i\Phi^{\dagger i}(\tau)\right)^q\!:\, .
\end{align}
Wick contraction with the endpoint normalization~\eqref{eq:general-q-endpoint-dagger-and-two-point} gives the index-free two-point function.\footnote{Here, we normalize the two-point function of $\Phi_i$ and $\Phi^{\dagger j}$ as follows,
\begin{align}
    \left\langle
        \Phi_i(\tau)
        \Phi^{\dagger j}(\tau')
    \right\rangle_{0}
    =
    \frac{\delta_i^{\ j }}{4\pi}\frac{1}{|\tau - \tau'|}\, . 
\end{align}
}
\begin{align}\label{eq:polarized-general-q-two-point}
    \left\langle
        \bar E_{(q)}(\bar u_1,\tau_1)
        E^{(q)}(u_2,\tau_2)
    \right\rangle_{0}
    =
    \frac{(\bar u_1\cdot u_2)^q}
    {|\tau_{12}|^q}\, .
\end{align}
This immediately gives the leading endpoint dimension
$\Delta_{E_q}^{(0)}=q/2$ and $g_q^{(0)}=1$.

The leading four-point function again follows from Wick contractions.  Let
\begin{align}\label{eq:r-variable}
    r(z)
    \equiv
    \frac{\tau_{12}\tau_{34}}{\tau_{14}\tau_{23}}
    =
    \frac{z}{1-z}\, .
\end{align}
A contraction pattern is specified by the number $m$ of scalar lines
connecting the defect pairs $(14)$ and $(32)$.  The remaining $q-m$
lines connect the bulk pairs $(12)$ and $(34)$.  There are
$\binom{q}{m}^2$ normalized Wick contractions of this type, and hence
\begin{align}\label{eq:general-q-free-four-point-polarized}
    \cG_q^{(0)}
    (\bar u_1,u_2,\bar u_3,u_4;z)
    =
    \sum_{m=0}^{q}
    \binom{q}{m}^{\!2}
    r(z)^m
    \,\mathbb X_m \, ,
\end{align}
where $\mathbb X_m$ is the flavor monomial defined by
\begin{align}\label{eq:general-q-Xm-main}
    \mathbb X_m
    &\equiv
    (\bar u_1\!\cdot u_2)^{q-m}
    (\bar u_3\!\cdot u_4)^{q-m}
    (\bar u_1\!\cdot u_4)^m
    (\bar u_3\!\cdot u_2)^m\, .
\end{align}
Equation \eqref{eq:general-q-free-four-point-polarized} is the complete strict $N=\infty$ solution for the derivative-free charge-$q$ endpoint.
Setting $m\to q-m$ gives
\begin{align}\label{eq:general-q-free-crossing-check}
    \cG_q^{(0)}(\bar u_1,u_2,\bar u_3,u_4;z)
    =
    r(1-z)^{-q}
    \sum_{m=0}^{q}
    \binom{q}{m}^{\!2}
    r(1-z)^{m}
    \,\mathbb X_{q-m}\, .
\end{align}
Since $\mathbb X_{q-m}$ is $\mathbb X_m$ with $u_2\leftrightarrow u_4$, and $r(z)^q=\rho_q(z)$ at $N = \infty$, this reduces to
\begin{align}
    \begin{aligned}
    \cG_q^{(0)}
    (\bar u_1,u_2,\bar u_3,u_4;z)
    =
    \rho_q(z) \, \cG_q^{(0)}(\bar u_1,u_4,\bar u_3,u_2;1-z)\, . 
    \end{aligned}
\end{align}
The $N=\infty$ correlator therefore satisfies the crossing equation \eqref{eq:general-q-endpoint-crossing-matrix}. Also, this implies
\begin{align}
    \mathcal G^{qq}_{q,k}=\mathcal G^{00}_{q,k}\, , 
\end{align}
which is naturally expected because the Wilson line becomes trivial at strict $N=\infty$. 

\subsection{General-charge diagrammatics at first subleading order}
\label{subsec:general-q-first-subleading}

We now keep the Wilson charge $q=O(N^0)$ fixed and compute the
four-endpoint function through first subleading order.  We use the
same contraction sectors $m=0,\cdots,q$ introduced in
\eqref{eq:general-q-free-four-point-polarized}.  This is only a
classification of Wick contractions within the full correlator and
should not be confused with a choice of OPE channel.  After computing
all diagrams multiplying the flavor monomial $\mathbb X_m$ in
\eqref{eq:general-q-Xm-main}, we project their sum onto irreducible
$\SU(N)$ flavor representations.
After normalizing the endpoints by their charge-$q$ two-point
function~\eqref{eq:general-q-microscopic-to-reduced-correlator}, the most general result at this order is
\begin{align}\label{eq:general-q-correlator-main-result}
    \begin{aligned}
    \cG_q
    (\bar u_1,u_2,\bar u_3,u_4;z)
    &=
    \sum_{m=0}^{q}
    \binom qm^{\!2}
    r(z)^m\mathbb X_m
    \left[
        1+\frac{1}{N}\mathscr F_{q,m}(z)
    \right]
    +O(N^{-2})\, .
    \end{aligned}
\end{align}
The flavor-diagonal photon and Hubbard--Stratonovich vertices do not
alter which endpoints are joined by scalar propagators.  Consequently,
the $q+1$ tensors
$\mathbb X_m$ form a closed basis even at order $1/N$, and interactions only
correct their coefficient functions.

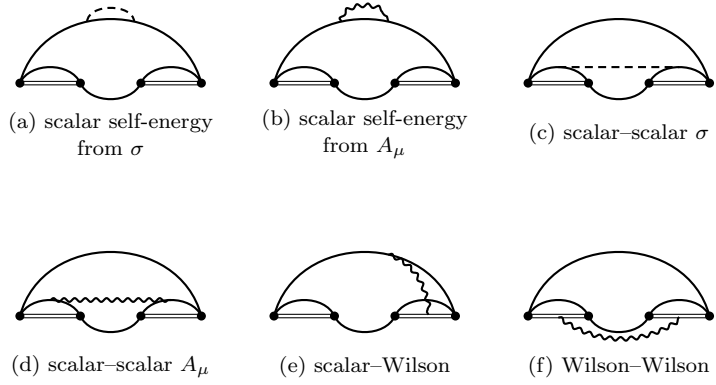
\begin{figure}[t]
\centering
\begin{tikzpicture}[
    x=0.80cm,
    y=0.70cm,
    scalar/.style={thick},
    photon/.style={thick,decorate,decoration={snake,amplitude=0.7pt,segment length=4pt}},
    hs/.style={thick,densely dashed},
    wilson/.style={double,double distance=0.9pt,line width=0.4pt},
    endpoint/.style={circle,fill,inner sep=1.2pt},
    every node/.style={font=\scriptsize}
]
\def\drawmtwosector#1{
  \draw[wilson] (0,0)--(1,0) (2,0)--(3,0);
  \draw[scalar]
    (0,0) to[out=65,in=115]
    node[pos=.5,coordinate](#1done){} (1,0);
  \draw[scalar]
    (2,0) to[out=65,in=115]
    node[pos=.5,coordinate](#1dtwo){} (3,0);
  \draw[scalar]
    (0,0) to[out=75,in=105,looseness=1.25]
    node[pos=.40,coordinate](#1xone){}
    node[pos=.60,coordinate](#1xtwo){} (3,0);
  \draw[scalar]
    (2,0) to[out=-115,in=-65] (1,0);
  \foreach \x in {0,1,2,3}
    \node[endpoint] at (\x,0){};
}

\begin{scope}[shift={(0,4.4)}]
  \drawmtwosector{a}
  \draw[hs] (axone) to[bend left=85] (axtwo);
  \node[align=center] at (1.5,-.95) {(a) scalar self-energy\\ from $\sigma$};
\end{scope}
\begin{scope}[shift={(4.2,4.4)}]
  \drawmtwosector{b}
  \draw[photon] (bxone) to[bend left=85] (bxtwo);
  \node[align=center] at (1.5,-.95) {(b) scalar self-energy\\ from $A_{\mu}$};
\end{scope}
\begin{scope}[shift={(8.4,4.4)}]
  \drawmtwosector{c}
  \draw[hs] (cdone)--(cdtwo);
  \node at (1.5,-.95) {(c) scalar--scalar $\sigma$};
\end{scope}
\begin{scope}[shift={(0,0)}]
  \drawmtwosector{d}
  \draw[photon] (ddone)--(ddtwo);
  \node at (1.5,-.95) {(d) scalar--scalar $A_\mu$};
\end{scope}
\begin{scope}[shift={(4.2,0)}]
  \drawmtwosector{e}
  \draw[photon] (extwo) to[out=-20,in=100] (2.55,.02);
  \node at (1.5,-.95) {(e) scalar--Wilson};
\end{scope}
\begin{scope}[shift={(8.4,0)}]
  \drawmtwosector{f}
  \draw[photon] (.5,-.02) to[out=-35,in=-145] (2.5,-.02);
  \node at (1.5,-.95) {(f) Wilson--Wilson};
\end{scope}
\end{tikzpicture}
\caption{The complete order-$1/N$ diagram classes, illustrated for
$q=2$ in the sector $m=1$.  The four black dots are ordered as
$(1,2,3,4)$, and one scalar propagator joins each of $(12)$, $(34)$,
$(14)$, and $(32)$.  A dashed line is the effective $\sigma$
propagator, a wavy line is the effective photon, and a double line is
a physical charge-two Wilson segment.  In each panel the interaction
is attached to every admissible scalar propagator or Wilson segment.
Seagull tadpoles and the critical-mass insertion are scaleless in
dimensional regularization; line-tension and endpoint counterterms
cancel in the normalized correlator.}
\label{fig:general-q-diagram-classes}
\end{figure}

We now compute the subleading correction $\mathscr F_{q,m}(z)$ in~\eqref{eq:general-q-correlator-main-result}. 
At leading order, the endpoint $E^{(q)}$ is represented by the
symmetrized product of $q$ fundamental scalar fields.  A Wick
contraction of the four endpoint insertions therefore contains $2q$
scalar propagators, exactly $q$ of which end at each insertion.  The
correlator also contains the two physical charge-$q$ Wilson segments
$C_{12}$ and $C_{34}$.  Both the photon exchange and the
$\sigma$ exchange are of order $1/N$, so every first-subleading
diagram contains exactly one of them.  The complete list is shown in
figure~\ref{fig:general-q-diagram-classes}: a self-energy insertion on
a scalar propagator, photon or $\sigma$ exchange between two scalar
propagators, photon exchange between a scalar propagator and a Wilson
segment, and photon exchange between Wilson segments.

For each $m$, we sum the diagrams in
figure~\ref{fig:general-q-diagram-classes} over all scalar
propagators and both Wilson segments, and then divide by the two
charge-$q$ endpoint two-point functions. After performing the diagram sum and the two-point subtraction, the resulting finite coefficient in \eqref{eq:general-q-correlator-main-result} organizes into four charge-independent functions:
\begin{align}\label{eq:general-q-four-master-functions-main}
    \mathscr F_{q,m}(z)
    =
    (q-m)^2\mathscr A(z)
    +m^2\mathscr B(z)
    +m(q-m)\mathscr C(z)
    +m\mathscr D(z)
    \, .
\end{align}
This form makes the general-$q$ content transparent. Here $q-m$ and $m$ are the numbers of scalar propagators in each
direct and crossed pairing, respectively.  The four prefactors in
\eqref{eq:general-q-four-master-functions-main} are the four
independent multiplicities produced by the diagram sum: three are
quadratic in these two occupation numbers, while the last is linear
in the number of crossed propagators.  We define
$\mathscr A,\mathscr B,\mathscr C$, and $\mathscr D$ as the
corresponding coefficient functions.  They depend on $z$, but not on
$q$ or $m$. Their diagrammatic definitions and the evaluation of the individual integrals are given in
appendix~\ref{app:general-q-diagrammatics_single}. 

The diagram integrals give
\begin{align}\label{eq:general-q-A-master-function-main}
    \mathscr A(z)
    =
    \left(\frac4z-1\right)\Lambda(z)
    +\frac{8}{\pi^2}\log(1-z)+4\, ,
\end{align}
where
\begin{align}\label{eq:general-q-Lambda-main}
    \Lambda(z)
    &\equiv
    -\frac{4z}{\pi^2\sqrt{1-z}}
    \Bigl[
        \Li_2(\sqrt{1-z})-\Li_2(-\sqrt{1-z})
        -\log(1-z)\operatorname{arctanh}\sqrt{1-z}
    \Bigr]\, ,
\end{align}
where $\Li_2(t)$ is the dilogarithm. The other three functions are
\begin{align}\label{eq:general-q-BCD-master-functions-main}
    \begin{aligned}
    \mathscr B(z)
    &=
    \frac{3+z}{1-z}\Lambda(1-z)
    -\frac{4}{\pi^2}\log z
    +\frac{12}{\pi^2}\log(1-z)
    -4\, ,
    \\
    \mathscr C(z)
    &=
    -8
    +\frac{16}{\pi^2}
    +\frac{24}{\pi^2}\log(1-z)\, ,
    \\
    \mathscr D(z)
    &=
    -\frac{24}{\pi^2}\log r(z)\, .
    \end{aligned}
\end{align}
The two channel limits of these functions are used repeatedly below. 
They follow from
\begin{align}\label{eq:general-q-Lambda-limits}
    \Lambda(z)
    &=
    -z+\Bigl(\frac{2}{\pi^2}-\frac12\Bigr)z^2
    +\Bigl(\frac{2}{\pi^2}-\frac38\Bigr)z^3+O(z^4)\, ,
    \\
    \Lambda(1-x)
    &=
    \frac{4}{\pi^2}\log x-\frac{8}{\pi^2}+O(x\log x)\, .
\end{align}
In the bulk channel limit $z\to0$,
\begin{align}\label{eq:general-q-master-small-z}
    \begin{aligned}
    \mathscr A(z)
    &=
    -z + \left(\frac{2}{\pi^2}-1\right)z^2 + O(z^3)\, ,
    &
    \mathscr B(z)
    &=
    \frac{8}{\pi^2}\log z
    -4-\frac{24}{\pi^2}
    +O(z\log z)\, ,
    \\
    \mathscr C(z)
    &=
    -8+\frac{16}{\pi^2}
    +O(z)\, ,
    &
    \mathscr D(z)
    &=
    -\frac{24}{\pi^2}\log z
    +O(z)\, ,
    \end{aligned}
\end{align}
while in the defect channel limit $z=1-x$ with $x\to0$,
\begin{align}\label{eq:general-q-master-small-x-for-defect-data}
    \begin{aligned}
    \mathscr A(1-x)
    &=
    \frac{20}{\pi^2}\log x
    +4-\frac{24}{\pi^2}
    +O(x\log x)\, ,
    &
    \mathscr B(1-x)
    &=
    \frac{12}{\pi^2}\log x
    -8
    +O(x\log x)\, ,
    \\
    \mathscr C(1-x)
    &=
    \frac{24}{\pi^2}\log x
    -8+\frac{16}{\pi^2}
    +O(x)\, ,
    &
    \mathscr D(1-x)
    &=
    \frac{24}{\pi^2}\log x
    +O(x)\, .
    \end{aligned}
\end{align}

\subsection{Bootstrapping bulk CFT data and consistency checks}
\label{subsec:bulk-CFT-data-bootstrap}

We now bootstrap bulk CFT data from the result of the diagrammatic calculation. This provides consistency checks against the known spectrum of bosonic QED$_3$. 

For each fixed-$m$ contraction pattern, we first combine all graphs that multiply the flavor monomial $\mathbb X_m$.  Only then do we convert the result to the irreducible flavor basis.  Recall that the polarization polynomial $\mathbb X_m$ retains the complete flavor-index information. 
Applying the inverse polarization map described in \eqref{eq:general-q-inverse-polarization-map}, we may regard it as an invariant tensor $(\mathbb X_m)^{I}{}_{J}{}^{K}{}_{L}$; below we use the same symbol for the tensor and its polarized form.

The $q+1$ tensors $\mathbb X_m$ and the $q+1$ orthonormal tensors $\mathbb T_k^{(\mathrm s)}$ are two bases for the same space of $\SU(N)$-invariant four-index tensors. 
There is therefore a unique change of basis
\begin{align}\label{eq:general-q-Xm-irrep-expansion-main}
    \mathbb X_m
    =
    \sum_{k=0}^{q}
    \sqrt{d_k}\,
    \Pi_{km}^{(\mathrm s)}
    \mathbb T_k^{(\mathrm s)}\, ,
\end{align}
where orthonormality of $\mathbb T_k^{(\mathrm s)}$ gives
\begin{align}\label{eq:general-q-Pi-definition-main}
    \Pi_{km}^{(\mathrm s)}
    \equiv
    \frac{1}{\sqrt{d_k}}
    \left(
        \mathbb T_k^{(\mathrm s)},
        \mathbb X_m
    \right)\, . 
\end{align}
Here the parentheses denote the Hilbert--Schmidt inner product. 
Note that $\Pi_{km}^{(\mathrm s)}$ is a purely group-theoretic, cross-ratio independent matrix. 
It records how much of the flavor structure $\mathbb X_m$ belongs to the irreducible bulk-channel representation $R_k$. 

Substituting~\eqref{eq:general-q-Xm-irrep-expansion-main} into the diagrammatic correlator~\eqref{eq:general-q-correlator-main-result} and comparing with~\eqref{eq:general-q-bulk-channel-decomposition}, we obtain
\begin{align}\label{eq:general-q-channel-projection-main}
    \mathcal G^{00}_{q,k}(z)
    =
    \sum_{m=0}^{q}
    \binom qm^{\!2}
    \Pi_{km}^{(\mathrm s)}r(z)^m
    \left[
        1+\frac1N\mathscr F_{q,m}(z)+O(N^{-2})
    \right]
    \, .
\end{align}
Through order $1/N$ only $m=k$ and $m=k+1$ contribute to \eqref{eq:general-q-channel-projection-main},\footnote{For $m<k$, the coefficients $\Pi^{(\mathrm s)}_{km}$ vanish.} 
\begin{align}\label{eq:general-q-channel-NLO}
    \mathcal{G}_{q,k}^{00}(z)
    =
    \binom{N+q+k-1}{q-k}r(z)^k
    \left[
    1
    + \frac{1}{N}\Big(\mathscr{F}_{q,k}(z)
    + (q-k)^2r(z)\Big) + O(N^{-2})
    \right]\, , 
\end{align}
where we used the following formulas, 
\begin{align}\label{eq:general-q-Pi-leading-NLO}
    \binom qk^{\!2}\Pi^{(\mathrm s)}_{kk}
    =\binom{N+q+k-1}{q-k}\, ,
    \quad
    \binom q{k+1}^{\!2}\Pi^{(\mathrm s)}_{k,k+1}
    =\binom qk^{\!2}\Pi^{(\mathrm s)}_{kk}\,
    \frac{(q-k)^2}{N+2k}\, .
\end{align}
We relegate their proofs to appendix~\ref{app:Pi_km}. 
The small-$z$ expansion of $\mathcal{G}_{q,k}^{00}(z)$ then determines the bulk CFT data.  Substituting the asymptotic form of the diagrammatic result~\eqref{eq:general-q-master-small-z}, we find
\begin{align}\label{eq:general-q-F-q-k-small-z}
    \mathscr F_{q,k}(z)
    =
    \frac{8k(k-3)}{\pi^2}\log z
    +4k^2-8qk
    +\frac{8k(2q-5k)}{\pi^2}
    +O(z\log z)\, .
\end{align} 
One can compare the result \eqref{eq:general-q-channel-NLO} with the general systematics derived from conformal symmetry. The small-$z$ behavior of a one-dimensional conformal block is 
\begin{align}\label{eq:general-q-small-z-block-main}
    g_{\Delta}(z)
    =
    z^{\Delta}\left[1+O(z)\right] \,  .
\end{align}
Consequently, the powers of $z$ determine the dimension of lowest-lying bulk operators, 
while their coefficients determine the squared OPE coefficients. 
At first subleading order, an anomalous dimension is identified through 
\begin{align}\label{eq:general-q-small-z-anomalous-log-main}
    z^{\Delta^{(0)}+\gamma/N}
    =
    z^{\Delta^{(0)}}
    \left[
        1+\frac{\gamma}{N}\log z
        +O(N^{-2})
    \right] \, .
\end{align}

For $k\neq0$, the corresponding bulk primary is the traceless symmetric product of $k$ fields $\Phi$ and $k$ fields $\Phi^\dagger$ at $N=\infty$, which we denote by $\mathcal{O}_{k,0}$.  It has canonical dimension $k$. Comparison with a single block, using \eqref{eq:general-q-small-z-anomalous-log-main}, gives
\begin{align}\label{eq:general-q-lowest-bulk-data}
    \Delta_{\mathcal O_{k,0}}
    &=
    k+\frac{8k(k-3)}{\pi^2N}+O(N^{-2})\, ,
    \\
    \left|\lambda^{00}_{q,k,\mathcal O_{k,0}}\right|^2
    &=
    \binom{N+q+k-1}{q-k}
    \left[
        1+\frac1N
        \left(
            4k^2-8qk+\frac{8k(2q-5k)}{\pi^2}
        \right)
    \right]
    +O\!\left(N^{q-k-2}\right)\, .
    \label{eq:general-q-lowest-squared-OPE}
\end{align}
Although the individual terms in \eqref{eq:general-q-F-q-k-small-z} depend on $q$, the dimension is independent of $q$, as expected.  For $k=1,2$, it reproduces the known results \cite[equations~(2.23) and~(2.27)]{Benvenuti:2019ujm}.\footnote{As far as we know, there is no literature which gives analytic anomalous dimensions for general $k$.}

The singlet channel provides a further check. 
Setting $k=0$ in \eqref{eq:general-q-channel-NLO} gives
\begin{align}\label{eq:general-q-bulk-singlet}
    \frac{1}{n_q}\mathcal{G}^{00}_{q,0}(z)
    =
    1+
    \frac{2q^2}{\pi^2 N }z^2
    +O(z^3, N^{-2})\, . 
\end{align}
The absence of an $O(z)$ term reflects that $\Phi^{\dagger i}\Phi_i$ is not an independent primary at the critical point. 
The leading singlet contribution is the $z^2$ term, which is the block of the Hubbard--Stratonovich field $\sigma$ with dimension 2. 
Because this coefficient is already of order $1/N$, the anomalous dimension $\gamma_\sigma$ would appear only at order $N^{-2}$ and cannot be extracted at the order $1/N$.

\subsection{Bootstrapping Wilson line data from crossing relation}
\label{subsec:general-q-bootstrap-extraction-main}
We organize the endpoint dimension and defect $g$-function as
\begin{align}\label{eq:general-q-large-N-expansions}
    \Delta_{E_q}
    &=
    \frac{q}{2}
    +
    \frac{\gamma_{q}}{N}
    +
    O(N^{-2})\, ,
    \\
    g_q
    &=
    1
    +
    \frac{g_q^{(1)}}{N}
    +
    O(N^{-2})\, ,
\end{align}
where $q=O(N^0)$ is held fixed. In this subsection, we extract these
large-$N$ corrections by combining the diagrammatic results obtained
above with the endpoint crossing relation. Furthermore, the
defect-channel conformal-block decomposition also determines the
dimensions of local operators
$\widehat{\mathcal O}\in\mathcal H^{qq}_{R_k}$ on the charge-$q$
Wilson line and the endpoint OPE coefficients
$\hatlambda^{qq}_{q,k,\widehat{\mathcal O}}$ that couple them to a pair
of defect-changing endpoints. 
We extract these additional defect CFT data below.

To extract data of operators living on the charge-$q$ Wilson line, we reconstruct the channel functions associated with $\mathcal{H}^{qq}$ from the bulk channel functions~\eqref{eq:general-q-channel-projection-main}. 
Setting $x = 1 - z$ and solving the crossing equation~\eqref{eq:general-q-endpoint-crossing-matrix} for these functions gives 
\begin{align}\label{eq:general-q-Wilson-channel-from-crossing-main}
    \vec{\mathcal G}^{\,qq}_q(x)
    =
    \frac{1}{\rho_q(1-x)}
    \mathbf d^{-1/2}
    (F^{(q,N)})^T
    \mathbf d^{1/2}
    \vec{\mathcal G}^{\,00}_q(1-x)\, .
\end{align}
Since the four-point function has been computed with its flavor structure, the channel functions in $\mathcal{H}^{00}$ and $\mathcal{H}^{qq}$ are two projections of the same known tensor function, and \eqref{eq:general-q-Wilson-channel-from-crossing-main} relates them by a change of flavor basis. 
It is simpler to bypass the Racah matrix and expand each monomial  $\mathbb X_m$ directly in the defect-channel tensors,
\begin{align}\label{eq:general-q-Xm-t-channel-expansion-main}
    \mathbb X_m
    =
    \sum_{k=0}^{q}
    \sqrt{d_k}\,
    \Pi_{km}^{(\mathrm t)}
    \mathbb T_k^{(\mathrm t)}\, .
\end{align}
Exchanging the flavor indices $J$ and $L$, or equivalently the
polarizations $u_2$ and $u_4$, maps $\mathbb X_m$ to $\mathbb X_{q-m}$
and $\mathbb T_k^{(\mathrm s)}$ to $\mathbb T_k^{(\mathrm t)}$, and
leaves the Hilbert--Schmidt inner product invariant. 
Hence 
\begin{align}\label{eq:general-q-t-from-s-relabeling}
    \Pi_{km}^{(\mathrm t)}
    =
    \Pi_{k,q-m}^{(\mathrm s)}\, ,
\end{align}
which is equivalent to the Racah form
\begin{align}\label{eq:general-q-s-to-t-projection-main}
    \Pi_{km}^{(\mathrm t)}
    =
    \frac{1}{\sqrt{d_k}}
    \sum_{j=0}^{q}
    \sqrt{d_j}\,
    F_{jk}^{(q,N)}
    \Pi_{jm}^{(\mathrm s)}
\end{align}
obtained by substituting \eqref{eq:general-q-Racah-crossing} into \eqref{eq:general-q-Xm-irrep-expansion-main}. 
Combining with \eqref{eq:general-q-channel-projection-main}, the defect-channel functions are therefore 
\begin{align}\label{eq:general-q-Wilson-channel-projection-main}
    \mathcal G^{qq}_{q,k}(x)
    =
    \frac{1}{\rho_q(1-x)}
    \sum_{m=0}^{q}
    \binom qm^{\!2}
    \Pi_{km}^{(\mathrm t)}
    r(1-x)^m
    \left[
        1+\frac1N\mathscr F_{q,m}(1-x)
        +O(N^{-2})
    \right]\, .
\end{align}
By \eqref{eq:general-q-t-from-s-relabeling} and \eqref{eq:general-q-Pi-leading-NLO}, only $m=q-k$ and $m=q-k-1$ contribute through order $1/N$, with coefficients $\binom{N+q+k-1}{q-k}$ and $\binom{N+q+k-1}{q-k}\frac{(q-k)^2}{N+2k}$. 
Using
\begin{align}\label{eq:general-q-fixed-m-Wilson-limit}
    \frac{r(1-x)^m}{\rho_q(1-x)}
    =
    r(x)^{q-m}
    \left[
        1+\frac{2\gamma_q}{N}\log r(x)
        +O(N^{-2})
    \right]\, ,
\end{align}
the defect channel functions become
\small
\begin{align}\label{eq:general-q-Wilson-channel-NLO}
    \mathcal{G}_{q,k}^{qq}(x) 
    =
    \binom{N+q+k-1}{q-k}r(x)^k
    \left[
    1
    + \frac{1}{N}\Big(\mathscr{F}_{q,q-k}(1-x)
    + (q-k)^2r(x)
    + 2\gamma_q\log r(x)\Big)
    + O(N^{-2})
    \right]\, .
\end{align}
\normalsize

The defect CFT data are read off from the small-$x$ expansion of~\eqref{eq:general-q-Wilson-channel-projection-main} in the same way as in the bulk channel,
\begin{align}\label{eq:general-q-small-x-block-main}
    g_{\widehat\Delta}(x)
    =
    x^{\widehat\Delta}\left[1+O(x)\right] \, ,
    \quad
    x^{\widehat\Delta^{(0)}+\widehat\gamma/N}
    =
    x^{\widehat\Delta^{(0)}}
    \left[
        1+\frac{\widehat\gamma}{N}\log x
        +O(N^{-2})
    \right] \, .
\end{align}
Thus the powers and logarithms in the defect channel functions contain the defect spectrum and OPE coefficients that we wish to determine.

\subsubsection{Endpoint dimension and defect $g$-function}

We first isolate the defect identity in the block expansion since its contribution determines $\Delta_{E_q}$ and $g_q$. The identity occurs only in the singlet channel $k=0$ and requires
\begin{align}\label{eq:general-q-Wilson-singlet-required-limit-main}
    \begin{aligned}
    \frac{1}{n_q}\mathcal G^{qq}_{q,0}(x)
    &=
    \frac{1}{g_q}
    +O\!\left(x^{\widehat\Delta_{\rm gap}}\right) \\
    &=
    1-\frac{g_q^{(1)}}{N}
    +O\!\left(x^{\widehat\Delta_{\rm gap}}, N^{-2}\right)
    \quad (x\to0)\,,
    \end{aligned}
\end{align}
with the normalization of
\eqref{eq:general-q-identity-block-coefficients}. Here $\widehat{\Delta}_{\rm gap}>0$ is the spectral gap above the defect identity, assumed in section~\ref{subsec:general-q-g-spectral-interpretation}. 
For $k=0$, \eqref{eq:general-q-Wilson-channel-NLO} gives 
\begin{align}\label{eq:general-q-Wilson-singlet-before-gamma}
    \frac{1}{n_q}\mathcal G^{qq}_{q,0}(x)
    =
    1+\frac1N
    \Bigl[
        \mathscr F_{q,q}(1-x)
        +q^2r(x)
        +2\gamma_q\log r(x)
    \Bigr]
    +O(N^{-2})\, ,
\end{align}
with $\mathscr F_{q,q}(1-x)=q^2\mathscr B(1-x)+q\mathscr D(1-x)$. 
The asymptotic forms~\eqref{eq:general-q-master-small-x-for-defect-data} then give 
\begin{align}\label{eq:general-q-Wilson-singlet-small-x-main}
    \frac{1}{n_q}\mathcal G^{qq}_{q,0}(x)
    &=
    1+\frac1N\left[
        \left(
            \frac{12q(q+2)}{\pi^2}
            +2\gamma_q
        \right)\log x
        -8q^2
        +O(x\log x)
    \right]
    +O(N^{-2})\, .
\end{align}
Comparing this result with the required identity behavior~\eqref{eq:general-q-Wilson-singlet-required-limit-main}, the absence of
an $x^0\log x$ term gives
\begin{align}\label{eq:general-q-endpoint-dimension-bootstrap-main}
    \gamma_q
    =-\frac{6q(q+2)}{\pi^2}
    \, , \quad
    \Delta_{E_q}
    =\frac q2-\frac{6q(q+2)}{\pi^2N}
    +O(N^{-2})\, .
\end{align}
After imposing \eqref{eq:general-q-endpoint-dimension-bootstrap-main},
comparison of the constant term with $1/g_q$ gives
\begin{align}\label{eq:general-q-g-function-bootstrap-main}
    g_q
    =1+\frac{8q^2}{N}+O(N^{-2})
    \, .
\end{align}
An independent calculation of the circular Wilson loop expectation value reproduces this result, as shown in appendix~\ref{app:defect-g-function-check}.  Here it has instead been obtained from the identity term of endpoint crossing.

\subsubsection{Defect spectrum and endpoint OPE coefficients}

We now turn to the non-identity spectrum.  For each
$k=1,\cdots,q$, the defect channel OPE contains a unique
derivative-free primary $\widehat{\mathcal O}_{k,0}$ in the flavor
representation $R_k$. Its conformal dimension is also expanded as
\begin{align}
    \widehat\Delta_{k,0}
    =
    \widehat\Delta_{k,0}^{(0)}+\frac{\widehat\Delta_{k,0}^{(1)}}{N} + O(N^{-2})\, . 
\end{align}
At $N=\infty$, the operator $\widehat{\mathcal O}_{k,0}$ is the traceless symmetric
product of $k$ fields $\Phi$ and $k$ fields $\Phi^\dagger$ inserted on
the charge-$q$ Wilson line, and hence has dimension
\begin{align}\label{eq:general-q-lowest-defect-leading-dimension}
    \widehat\Delta_{k,0}^{(0)}
    =
    k\, .
\end{align}
The remaining $q-k$ constituent fields in each endpoint are contracted
across the OPE.  Combining this Wick contraction with the
Clebsch--Gordan normalization \eqref{eq:N_qk} gives the leading squared
endpoint OPE coefficient
\begin{align}\label{eq:general-q-lowest-defect-leading-OPE}
    \left.
    \left|
        \hatlambda^{qq}_{q,k,\widehat{\mathcal O}_{k,0}}
    \right|^2
    \right|_{\rm leading}
    =
    \binom{N+q+k-1}{q-k}\, .
\end{align}
We keep this coefficient unexpanded in $N$, since it is the purely group-theoretic normalization of the $R_k$ channel. 

For $k\ge 1$, the coefficients of $x^k$ and $x^k\log x$ are fixed by
\begin{align}\label{eq:general-q-F-q-qminusk-small-x}
    \begin{aligned}
    \mathscr F_{q,q-k}(1-x)
    &=
    \frac{
        12q(q+2)+8k(k-3)
    }{\pi^2}
    \log x \\
    &\quad+
    4k^2-8q(q-k)
    +\frac{8k(2q-5k)}{\pi^2}
    +O(x\log x)\, ,
    \end{aligned}
\end{align}
which follows from \eqref{eq:general-q-master-small-x-for-defect-data}, together with $2\gamma_q\log r(x)$.  With $\gamma_q$ given by
\eqref{eq:general-q-endpoint-dimension-bootstrap-main}, 
\begin{align}\label{eq:general-q-lowest-defect-small-x-result}
    &\
        \mathcal G^{qq}_{q,k}(x)
    =
    \binom{N+q+k-1}{q-k} 
    x^k
    \Bigg\{
        1
        +\frac1N
        \Bigg[
            \frac{8k(k-3)}{\pi^2}\log x
            \nonumber\\[-1mm]
    &\hspace{20mm}
            +4k^2-8q(q-k)
            +\frac{8k(2q-5k)}{\pi^2}
        \Bigg]
        +O(x\log x,N^{-2})
    \Bigg\}\, .
\end{align}
The absence of the correction linear in $q$ is just due to the charge conjugation symmetry of the bulk theory. Comparison with a single conformal block, using
\eqref{eq:general-q-small-x-block-main}, directly yields
\begin{align}\label{eq:general-q-lowest-defect-dimension-result}
    \widehat\Delta_{k,0}
    =
    k
    +\frac{8k(k-3)}{\pi^2N}
    +O(N^{-2})
    \, ,
    \quad
    k=1,\cdots,q\, .
\end{align}
The order-$1/N$ anomalous dimension depends only on the
flavor representation $R_k$ and not on the Wilson charge $q$.  The
non-logarithmic term gives the endpoint OPE coefficient,
\begin{align}\label{eq:general-q-lowest-defect-OPE-result}
    \begin{aligned}
    \left|
        \hatlambda^{qq}_{q,k,\widehat{\mathcal O}_{k,0}}
    \right|^2
    =
    \binom{N+q+k-1}{q-k}
    \Bigg[
        1+\frac1N
        \left(
            4k^2-8q(q-k)
            +\frac{8k(2q-5k)}{\pi^2}
        \right)
    \Bigg]
    +O\!\left(N^{q-k-2}\right).
    \end{aligned}
\end{align}
Equations \eqref{eq:general-q-lowest-defect-dimension-result} and
\eqref{eq:general-q-lowest-defect-OPE-result} are obtained directly in
position space.  For $q=1$ and $k=1$, the alpha-space analysis~\cite{Hogervorst:2017sfd} in appendix~\ref{sec:q-one-alpha-space-inversion} provides
an independent reconstruction of the anomalous dimension and the leading-order OPE weight from the pole positions and residues of the spectral density.

\subsection{Bulk--defect spectral relation at large $N$}
\label{sec:q-bulk-Wilson-relations}

In sections~\ref{subsec:bulk-CFT-data-bootstrap} and~\ref{subsec:general-q-bootstrap-extraction-main}, we extracted the lowest, derivative-free operators in each representation $R_k$ from the bulk and defect channels, respectively. Comparing equations~\eqref{eq:general-q-lowest-bulk-data} and~\eqref{eq:general-q-lowest-defect-dimension-result}, we find that their scaling dimensions agree through order $1/N$. This agreement is not an accidental property of the lowest operators. In fact, as we see below, the complete bulk and defect channel functions are related in a much stronger way. 

To this end, using the explicit expressions for the diagram integrals $\mathscr{A}(z),\mathscr{B}(z),\mathscr{C}(z)$, and $\mathscr{D}(z)$ in equations~\eqref{eq:general-q-A-master-function-main} and~\eqref{eq:general-q-BCD-master-functions-main}, we first notice
\begin{align}
    \begin{aligned}
        \mathscr{A}(1-x) &= \mathscr{B}(x) + \frac{12}{\pi^2}\log r(x) + 8\,, \\
        \mathscr{B}(1-x) &= \mathscr{A}(x) + \frac{12}{\pi^2}\log r(x) - 8\,, \\
        \mathscr{C}(1-x) &= \mathscr{C}(x) + \frac{24}{\pi^2}\log r(x)\,, \\
        \mathscr{D}(1-x) &= -\mathscr{D}(x)\,.
    \end{aligned}
\end{align}
Together with the explicit value of $\gamma_q$ in~\eqref{eq:general-q-endpoint-dimension-bootstrap-main}, these identities imply
\begin{align}
    \mathscr{F}_{q,q-k}(1-x) + 2\gamma_q\log r(x)
    =
    \mathscr{F}_{q,k}(x) -8q^2 + 16kq\,.
\end{align}
Substituting this identity into the defect-channel function~\eqref{eq:general-q-Wilson-channel-NLO} and using the bulk-channel result~\eqref{eq:general-q-channel-NLO}, together with the $g$-function in~\eqref{eq:general-q-g-function-bootstrap-main}, we arrive at the following spectral relation,\footnote{Here, the reason why the remainder starts from $N^{q-k-2}$ is due to the group factor appearing in \eqref{eq:general-q-lowest-defect-OPE-result}.} 
\begin{align}\label{eq:bulk-Wilson_relation_general_q}
    \mathcal{G}_{q,k}^{qq}(x)
    =
    \frac{1}{g_q}\left(1+\frac{16kq}{N}\right) \mathcal{G}_{q,k}^{00}(x)
    + O(N^{q-k-2})\,,\quad k=0,1,\cdots,q\, .
\end{align}
In the non-singlet case $k\ne 0$, this relation shows that the two channel functions have identical $x$-dependence through order $1/N$, differing only by an $x$-independent normalization factor. Consequently, their conformal-block decompositions are related level by level: the spectral data resolved by the two correlators coincide, while the corresponding OPE weights are related by the same level-independent multiplicative factor\footnote{This fixed-level relation between the OPE weights does not contradict the Tauberian result of Section~\ref{subsec:general-q-g-spectral-interpretation}, which concerns the large-dimension limit at finite $N$. The fixed-level $1/N$ expansion used here need not be uniform in this limit, where infinitely many operators contribute to reproduce the defect-channel identity. Thus, the subleading large-$N$ relation cannot in general be extrapolated directly to the Tauberian limit.}
\begin{align}
    \frac{1}{g_q}\left(1+\frac{16kq}{N}\right)\,.
\end{align}
This is also consistent with the lowest-operator OPE coefficients obtained in equations~\eqref{eq:general-q-lowest-squared-OPE} and~\eqref{eq:general-q-lowest-defect-OPE-result}: the relative $O(1/N)$ corrections differ by $8q(2k-q)$, as implied by the above multiplicative factor.

In the singlet case $k=0$, the same proportionality relation holds, although its spectral interpretation is slightly different. Since the OPE weights of non-identity singlet operators start at order $1/N$, the relation at the present order only probes their leading scaling dimensions and leading OPE weights, which agree between the two channels. To determine their $O(N^{-1})$ anomalous dimensions, the channel functions would need to be computed to $O(N^{-2})$.

\section{Higher-spin reconstruction from the fundamental endpoint}
\label{sec:higher-spin-reconstruction}

In the previous section, we studied the lowest-dimension operators on the Wilson line for general charge $q$. We now use the same endpoint four-point function to study operators of higher dimension. For simplicity, we set $q=1$ and focus on the adjoint representation of $\SU(N)$. We will show how bulk currents of different spins and their descendants contribute to the same one-dimensional defect conformal block, and determine their relative OPE weights.

The main issue is the degeneracy of the spectrum. Several distinct bulk conformal multiplets, including bulk descendants, can give rise to line primaries with the same dimension $n+1$ at $N=\infty$. Since these primaries multiply the same one-dimensional conformal block, the order-$1/N$ logarithmic correction measures an \emph{OPE-weighted average} of their anomalous dimensions. To understand this average structure, we must identify which bulk multiplets the endpoint pair couples to and determine their relative OPE weights.

For the fundamental endpoint, the spectral relation derived in section~\ref{sec:q-bulk-Wilson-relations} reads
\begin{align}\label{eq:q-one-adjoint-spectral-relation-main}
     \mathcal G^{11}_{1,1}(x) = g_1\,\mathcal G^{00}_{1,1}(x) +O(N^{-2})\, .
\end{align}
Thus the bulk and defect channels have the same OPE-weighted anomalous dimensions through this order, although their individual spectra need not coincide.

In this section, we first decompose the free bulk higher-spin multiplets into line primaries and determine their endpoint OPE weights. We then extract the averaged anomalous dimensions directly from the diagrammatic correlator computed in section~\ref{sec:analytical-bootstrap}. We also confirm that combining the branching weights with the known bulk higher-spin anomalous dimensions reproduces these averages at every level.  This reconstruction identifies the bulk contributions hidden within the collinear block expansion and provides an independent all-level check of the spectral constraints on the Wilson line.

\subsection{Bulk higher-spin multiplets restricted to the line}
\label{subsec:q-one-bulk-multiplets-on-line}
Because of~\eqref{eq:q-one-adjoint-spectral-relation-main}, it is enough to analyze the bulk channel. 
At $N=\infty$, the operators exchanged in $\mathcal G^{00}_{1,1}$ are the adjoint bilinears of $\Phi$ and $\Phi^\dagger$. 
We first organize them into conformal multiplets of the bulk conformal group $\mathrm{SO}(4,1)$. 
We then restrict these conformal multiplets to the line and determine how the endpoint pair couples to them.

We write a spacetime index as $\mu=(\tau,\alpha)$, where $\tau$ is the direction of the line and $\alpha=1,2$ label the transverse directions. 
A bilinear with derivatives is specified by a polynomial $f$,
\begin{align}\label{eq:collinear-polynomial-variables}
    \mathcal O^a[f] \equiv \Phi^\dagger T^a f(\xi,\eta)\,\Phi\, ,
    \quad \xi_\mu\equiv\overleftarrow{\partial_\mu}\, ,
    \quad \eta_\mu\equiv\overrightarrow{\partial_\mu}\, ,
\end{align}
where $\xi_\mu$ and $\eta_\mu$ are left and right spacetime derivatives acting on $\Phi^\dagger$ and $\Phi$, respectively.  

At $N=\infty$, a homogeneous polynomial of degree $n$ represents a bilinear of dimension~$n+1$. The free equations of motion and the Leibniz rule give
\begin{align}
    \xi^\mu\xi_\mu &=\eta^\mu\eta_\mu=0\, , \label{eq:bilinear-on-shell}\\
    \partial_\mu\mathcal O^a[f]
    &=
    \mathcal O^a[(\xi_\mu+\eta_\mu)f]\, . \label{eq:bilinear-total-derivative}
\end{align}
We use the same symbol for an operator and its polynomial, omitting the flavor index $a$ when referring to the latter. For example, the polynomial $\mathcal X_1(\xi,\eta)=\eta_\tau-\xi_\tau$
represents the bilinear
\begin{align*}
    \mathcal X_1^a
    =
    \mathcal O^a[\mathcal X_1]
    =
    \Phi^\dagger T^a\partial_\tau\Phi
    -
    (\partial_\tau\Phi^\dagger)T^a\Phi\, .
\end{align*}
Primary operators can be identified by acting directly on these polynomials with the special conformal generators. For a scalar primary of dimension $\Delta_\Phi$, their action
on polynomials in the derivative variables $\zeta_\mu$ is~\cite{Karateev:2017jgd}
\begin{align}\label{eq:special-conformal-generator}
    \mathcal K_\mu^{(\zeta)}
    = 2\left(\Delta_\Phi+\zeta_\nu\frac{\partial}{\partial \zeta_\nu}\right)\frac{\partial}{\partial \zeta^\mu}
    -\zeta_\mu\frac{\partial}{\partial \zeta^\nu}\frac{\partial}{\partial \zeta_\nu}\, .
\end{align}
For our bilinears, $\Delta_\Phi=1/2$ and $\mathcal K_\mu=\mathcal K_\mu^{(\xi)}
+\mathcal K_\mu^{(\eta)}$. The conformal symmetry along the line is $\mathrm{SL}(2,\mathbb R)$, while transverse rotations and reflections form $\mathrm O(2)_\perp$. The bulk and line primary conditions are, respectively,
\begin{align}
    \text{bulk primary } &:\ \mathcal K_\mu f=0
    \quad(\mu=\tau,1,2) \, , \label{eq:bulk-primary-condition}\\
    \text{line primary } &: \ \mathcal K_\tau f=0 \, . \label{eq:line-primary-condition}
\end{align}
Thus every bulk primary is a line primary, but the weaker line primary condition can also be satisfied by bulk descendants.

The solutions of the bulk primary condition \eqref{eq:bulk-primary-condition} are the standard free-scalar higher-spin currents. In the flavor-adjoint bilinear sector, there is one primary $J_s^a$ for each integer spin $s\geq0$~\cite{Flato:1978qz,Giombi:2016hkj}. Introducing a complex null polarization
$\varepsilon^\mu\varepsilon_\mu=0$ and writing $\xi_\varepsilon\equiv\varepsilon^\mu\xi_\mu$ and $\eta_\varepsilon\equiv \varepsilon^\mu\eta_\mu$, their derivative polynomials take the form~\cite{Zhou:2022pah}, 
\begin{align}\label{eq:ambient-Chebyshev-polynomial}
    \varepsilon^{\mu_1}\varepsilon^{\mu_2}\cdots\varepsilon^{\mu_s}
    (J_s)_{\mu_1\mu_2\cdots\mu_s}
    =
    (-1)^s(\xi_\varepsilon+\eta_\varepsilon)^s
    T_s\!\left(
        \frac{\xi_\varepsilon-\eta_\varepsilon}
             {\xi_\varepsilon+\eta_\varepsilon}
    \right)\, ,
\end{align}
where $T_s$ is the Chebyshev polynomial of the first kind. The cases $s=0$ and $s=1$ give the adjoint scalar $J_0^a=\Phi^\dagger T^a\Phi$ and the conserved flavor current, respectively.

At $N=\infty$, the bilinears decompose into the irreducible conformal multiplets generated by $J_s^a$~\cite{Flato:1978qz},
\begin{align}\label{eq:Flato-Fronsdal}
    V^{\mathrm{3d}}_{1/2,0}
    \otimes
    V^{\mathrm{3d}}_{1/2,0}
    =
    \bigoplus_{s=0}^{\infty}
    V^{\mathrm{3d}}_{s+1,s}\, .
\end{align}
Here $V^{\mathrm{3d}}_{\Delta,s}$ denotes the irreducible conformal module of a bulk primary of dimension $\Delta$ and spin $s$, with null states removed. At order $1/N$, we write the dimensions of these bulk primaries as
\begin{align}\label{eq:ambient-higher-spin-dimension}
    \Delta_{J_s}
    =
    s+1+\frac{\gamma_s^{\mathrm{bulk}}}{N}
    +O(N^{-2})\, .
\end{align}
The flavor current remains conserved, so $\gamma_1^{\mathrm{bulk}}=0$. A non-null descendant of $J_s^a$ with $\ell$ derivatives has dimension $\Delta_{J_s}+\ell$ and therefore inherits
the same anomalous dimension $\gamma_s^{\mathrm{bulk}}/N$.

Our interest here is in $\mathrm O(2)_\perp$-invariant line primaries, because the endpoints are scalars under transverse rotations and reflections, and their OPE can therefore contain only operators invariant under these transformations. Transverse indices can be contracted between two current indices, between a derivative and a current index, or between two derivatives. The first two possibilities are reduced by tracelessness and conservation at $N=\infty$:
\begin{align*}
    (J_s^a)^{\alpha}_{\ \alpha\nu_3\cdots\nu_s}
    =
    -(J_s^a)^{\tau}_{\ \tau\nu_3\cdots\nu_s}\, ,\quad 
    \partial^\alpha
    (J_s^a)_{\alpha\nu_2\cdots\nu_s}
    =
    -\partial^\tau
    (J_s^a)_{\tau\nu_2\cdots\nu_s}\, .
\end{align*}
The first identity implies that the transverse trace is not an independent operator:
tracelessness identifies it with a component containing two additional $\tau$ indices. It therefore does not give an additional line primary. The second produces a derivative along the line, which is a line descendant rather than a new primary. Thus the line primaries take the form
\begin{align}\label{eq:line-primaries-in-bulk-multiplet}
    \mathcal J_{s,k}^a
    =
    (\partial^\alpha\partial_\alpha)^k
    (J_s^a)_{\tau\cdots\tau}
    +\partial_\tau(\cdots)\, ,
    \qquad k=0,1,2,\cdots\, ,
\end{align}
where the total $\tau$-derivative terms are fixed by the line primary condition $\mathcal K_\tau\mathcal J_{s,k}=0$.

Each $\mathcal J_{s,k}^a$ has dimension $s+1+2k$ at $N=\infty$. The $\mathrm O(2)_\perp$-invariant part of the bulk multiplet therefore decomposes as
\begin{align}\label{eq:ambient-to-line-dimensions}
    \left.V^{\mathrm{3d}}_{s+1,s}
    \right|_{\mathrm O(2)_\perp\text{-invariant}}
    =
    \bigoplus_{k=0}^{\infty}
    V^{\mathrm{1d}}_{s+1+2k}\, ,
\end{align}
where $V^{\mathrm{1d}}_h$ denotes the conformal module generated by a line primary of dimension $h$. At a fixed dimension $n+1$, the condition $s+2k=n$ allows the bulk spins $s=n,n-2,\ldots$. The corresponding line primaries are
\begin{align}\label{eq:ambient-line-branching}
    \mathcal J_{n-2k,k}^a\, ,
    \quad
    k=0,1,\cdots,\left\lfloor\frac n2\right\rfloor\, .
\end{align}
Their multiplicity is therefore $1+\lfloor n/2\rfloor$. They belong to different bulk multiplets but contribute to the same one-dimensional block $g_{n+1}(x)$ at $N=\infty$. 

We next determine their relative weights in the endpoint OPE. At $N=\infty$, the adjoint part of the endpoint pair is represented by
\begin{align}\label{eq:endpoint-pair}
    \Phi^\dagger(\tau)T^a\Phi(0)
    =
    \mathcal O^a[e^{\tau\xi_\tau}]\, .
\end{align}
Its expansion contains only longitudinal derivatives. At degree $n$, the line-primary condition selects the polynomial
\begin{align}\label{eq:endpoint-primary}
    \mathcal X_n
    =
    (-1)^n(\xi_\tau+\eta_\tau)^n
    P_n\!\left(
        \frac{\xi_\tau-\eta_\tau}
             {\xi_\tau+\eta_\tau}
    \right)\, ,
\end{align}
where $P_n$ is the Legendre polynomial. The corresponding operator $\mathcal X_n^a=\mathcal O^a[\mathcal X_n]$ is the particular linear combination of \eqref{eq:ambient-line-branching} created by the endpoint pair. The free correlator 
\begin{align}\label{eq:free-adjoint-block-decomposition}
    \frac{x}{1-x}
    = 
    \sum_{n=0}^{\infty}
    \frac{(n!)^2}{(2n)!}g_{n+1}(x) 
\end{align}
fixes the sum of the squared endpoint
OPE coefficients at each level:
\begin{align}\label{eq:endpoint-OPE-weight}
    \left.
    \sum_{k=0}^{\lfloor n/2\rfloor}
    \left|
        \lambda^{00}_{1,1,\mathcal J_{n-2k,k}}
    \right|^2
    \right|_{N=\infty}
    =
    \frac{(n!)^2}{(2n)!}\, .
\end{align}
Resolving this sum into contributions from individual
bulk multiplets gives
\begin{align}\label{eq:spin-resolved-OPE-weight}
    \left.
    \left|
        \lambda^{00}_{1,1,\mathcal J_{s,k}}
    \right|^2
    \right|_{N=\infty}
    =
    \frac{(n!)^2}{(2n)!}\,\omega_{n,s}\, ,
    \quad n=s+2k\, .
\end{align}
Here $\omega_{n,s}$ is the fraction of the total squared OPE coefficient carried by the bulk spin-$s$ multiplet with $\sum_{s=n,n-2,\ldots}\omega_{n,s}=1$.
The operators labeling the OPE coefficients are understood to be unit normalized. The projection onto the bulk multiplets gives
\begin{align}\label{eq:branching-weight}
    \omega_{n,n-2k}
    =
    \left(2-\delta_{n,2k}\right)
    \frac{
        \binom{2k}{k}\binom{2n-2k}{n-k}
    }{4^n}\, ,
    \qquad
    0\leq k\leq\left\lfloor\frac n2\right\rfloor\, .
\end{align}
These weights also appear in the expansion of the
Legendre polynomial in Chebyshev polynomials:
\begin{align}\label{eq:Legendre-Chebyshev}
    P_n(\chi)
    =
    \sum_{k=0}^{\lfloor n/2\rfloor}
    \omega_{n,n-2k}\,T_{n-2k}(\chi)\, .
\end{align}
The projection calculation and the derivation of this identity are given in appendix~\ref{app:branching-weights}.

At free dimension $n+1$, the primaries $\mathcal J_{s,k}^a$ with $s+2k=n$ inherit the
anomalous-dimension coefficients $\gamma_s^{\mathrm{bulk}}$. We define their OPE-weighted average by
\begin{align}\label{eq:weighted-gamma-definition}
    \overline\gamma_n
    \equiv
    \frac{
        \displaystyle
        \sum_{k=0}^{\lfloor n/2\rfloor}
        \left.
        \left|
            \lambda^{00}_{1,1,\mathcal J_{n-2k,k}}
        \right|^2
        \right|_{N=\infty}
        \gamma_{n-2k}^{\mathrm{bulk}}
    }{
        \displaystyle
        \sum_{k=0}^{\lfloor n/2\rfloor}
        \left.
        \left|
            \lambda^{00}_{1,1,\mathcal J_{n-2k,k}}
        \right|^2
        \right|_{N=\infty}
    }\, .
\end{align}
Using \eqref{eq:spin-resolved-OPE-weight}, this becomes
\begin{align}\label{eq:weighted-gamma-bulk}
    \overline\gamma_n
    =
    \sum_{k=0}^{\lfloor n/2\rfloor}
    \omega_{n,n-2k}\,
    \gamma_{n-2k}^{\mathrm{bulk}}\, .
\end{align}
To see why this average appears, expand a block around its free dimension:
\begin{align*}
    g_{n+1+\gamma/N}(x)
    =
    g_{n+1}(x)
    +\frac{\gamma}{N}
    \left[
        g_{n+1}(x)\log x
        +(\text{terms analytic at }x=0)
    \right]
    +O(N^{-2})\, .
\end{align*}
All operators at level $n$ therefore contribute to the same function $g_{n+1}(x)\log x$, with coefficients proportional to their anomalous dimensions and leading squared OPE coefficients.\footnote{Note that corrections to the OPE coefficients themselves produce no logarithm at this order.} Summing these contributions, the coefficient of $\log x$
in $\mathcal G^{00}_{1,1}(x)$ is
\begin{align}\label{eq:log-coefficient}
    \frac1N
    \sum_{n=0}^{\infty}
    \frac{(n!)^2}{(2n)!}\,
    \overline\gamma_n\,g_{n+1}(x)\, .
\end{align}
Thus the logarithmic part of the correlator determines $\overline\gamma_n$, rather than the individual anomalous dimensions at a fixed level. We next extract these averages from the diagrammatic result and compare them with~\eqref{eq:weighted-gamma-bulk}.

\subsection{The adjoint $\mathrm{SL}(2,\mathbb R)$ spectrum}
\label{subsec:adjoint-SL2-spectrum}
Let us extract $\overline\gamma_n$ from the endpoint correlator and compare it with \eqref{eq:weighted-gamma-bulk}.
Setting $q=k=1$ in \eqref{eq:general-q-channel-NLO} and using \eqref{eq:general-q-four-master-functions-main}, the bulk-channel function is
\begin{align}\label{eq:adjoint-bulk-channel-function}
    \mathcal G^{00}_{1,1}(x)
    &= r(x)\left\{1+\frac{1}{N}\left[\mathscr B(x)+\mathscr D(x)\right]\right\}
    +O(N^{-2}) \notag \\ 
    &= r(x) +\frac{1}{N} \left[\mathcal L(x)\log x +(\text{terms without }\log x)\right]
    +O(N^{-2})\, ,
\end{align}
where \eqref{eq:general-q-BCD-master-functions-main} gives
\begin{align}\label{eq:adjoint-log-function}
    \mathcal L(x)
    = \frac{4}{\pi^2} \frac{x}{1-x}
    \left[\frac{3+x}{\sqrt{x}} \operatorname{arctanh}\sqrt{x} -7\right]\, .
\end{align}
Comparison with \eqref{eq:log-coefficient} gives
\begin{align}\label{eq:log-block-decomposition}
    \mathcal L(x) = \sum_{n=0}^{\infty} \frac{(n!)^2}{(2n)!}\,\overline\gamma_n\, g_{n+1}(x)\, .
\end{align}
Since $g_{n+1}(x)$ starts at $x^{n+1}$, the coefficients are determined recursively by the Taylor expansion around $x=0$.
We obtain
\begin{align}\label{eq:weighted-gamma-result}
    \overline\gamma_n
    = \frac{16}{\pi^2} \left[2H_n -H_{\lfloor n/2\rfloor} -\frac{7}{4}
    +\frac{3(-1)^n}{4(2n+1)}\right]\, ,
\end{align}
where $H_n\equiv \sum_{j=1}^{n}j^{-1}$ and $H_0\equiv 0$.
The first few values are
\begin{align}\label{eq:weighted-gamma-first-values}
    \overline\gamma_0 =-\frac{16}{\pi^2}\, ,
    \quad \overline\gamma_1 =0\, ,
    \quad \overline\gamma_2 =\frac{32}{5\pi^2}\, ,
    \quad \overline\gamma_3 =\frac{272}{21\pi^2}\, .
\end{align}

On the other hand, the anomalous dimensions of the adjoint currents in critical bosonic QED$_3$ are~\cite{Zhou:2022pah}
\begin{align}\label{eq:bulk-higher-spin-gamma}
    \gamma_s^{\mathrm{bulk}}
    = \frac{16}{\pi^2} \left[2H_{2s}-H_s -\frac{7s^2-1}{4s^2-1}\right]\, .
\end{align}
This also holds at $s=0$, and it gives $\gamma_1^{\mathrm{bulk}}=0$ for the conserved flavor current. 
Thus $\overline\gamma_n=\gamma_n^{\mathrm{bulk}}$ for $n=0, 1$, and $\overline\gamma_n\neq\gamma_n^{\mathrm{bulk}}$ for $n\geq2$.
The weighted sum \eqref{eq:weighted-gamma-bulk} accounts for this difference.
Using
\begin{align}\label{eq:weighted-sum-identities}
    \sum_s \omega_{n,s} \left(2H_{2s}-H_s\right) = 2H_n-H_{\lfloor n/2\rfloor}\, ,
    \quad \sum_s \frac{\omega_{n,s}}{4s^2-1} = \frac{(-1)^{n+1}}{2n+1}\, ,
\end{align}
where the sums run over $s=n,n-2,\cdots$, the right-hand side of \eqref{eq:weighted-gamma-bulk} becomes
\begin{align}\label{eq:weighted-gamma-reconstructed}
    \sum_{k=0}^{\lfloor n/2\rfloor} \omega_{n,n-2k}\, \gamma_{n-2k}^{\mathrm{bulk}}
    = \frac{16}{\pi^2} \left[2H_n-H_{\lfloor n/2\rfloor} -\frac74
    +\frac{3(-1)^n}{4(2n+1)}\right]\, .
\end{align}
This agrees with \eqref{eq:weighted-gamma-result} at every level.

Finally, the spectral relation \eqref{eq:q-one-adjoint-spectral-relation-main}
transfers these averaged data to the defect channel. The overall factor $g_1=1+8/N+O(N^{-2})$ rescales the summed squared OPE coefficients at each level,
while the OPE-weighted anomalous dimensions remain $\overline\gamma_n$ through order $1/N$.

This does not identify the individual Wilson-line dimensions at degenerate levels. Bulk spin is not preserved by the Wilson line, so operators originating from different bulk spins can mix. The bulk contributions reconstructed above therefore explain the shared averages, rather than the individual Wilson-line eigenstates.

At the non-degenerate levels $n=0,1$, however, the averages give individual dimensions:
\begin{align}\label{eq:Wilson-nondegenerate-dimensions}
    \hDelta_{n=0}
    =
    1-\frac{16}{\pi^2N}+O(N^{-2})\, , \quad 
    \hDelta_{n=1}
    =
    2+O(N^{-2})\, .
\end{align}
The first agrees with \eqref{eq:general-q-lowest-defect-dimension-result} at $q=k=1$.
We leave the extension of this all-level higher-spin reconstruction to $q\geq2$ as an open problem.

\section{Endpoint bootstrap for mixed correlators}
\label{sec:mixed-endpoint-correlators-general-q}

So far we have applied the endpoint bootstrap to correlators made
entirely from defect-changing operators.  We now apply the same idea
to a different correlator, with two endpoint operators and two
ordinary bulk operators inserted on the line.  This gives access to
new CFT data: the spectrum and OPE coefficients of defect-changing
operators created by fusing an endpoint with a bulk local operator.
We choose the lowest bulk singlet scalar, namely the Hubbard--Stratonovich
field, and carry out the calculation for an arbitrary fixed Wilson
charge $q$ through order $1/N$.

The organization parallels that of the four-endpoint correlator.  We
first give a brief review of bulk CFT data regarding the Hubbard--Stratonovich field. Next we describe the OPE structures and derive the crossing equation for the mixed correlator.  We then display all contributing diagrams and list the results. (The detailed evaluation of diagrams is relegated to appendix \ref{app:mixed_correlator}.) Finally, we solve the defect channel equation and extract the corresponding defect-changing CFT data.

\subsection{Known bulk CFT data regarding the Hubbard--Stratonovich field}

The scaling dimension of $\sigma$ is known from the ordinary bulk large-$N$
expansion~\cite{Benvenuti:2019ujm},
\begin{align}\label{eq:general-q-mixed-sigma-dimension}
    \Delta_\sigma
    =
    2+\frac{\gamma_\sigma}{N}+O(N^{-2})\, ,
    \quad
    \gamma_\sigma
    =
    -\frac{48}{\pi^2}\, .
\end{align}
Apart from the identity, the only bulk primaries of leading dimension $\Delta^{(0)}<4$ that can appear in $\sigma\times\sigma$ are $\sigma$ and the stress tensor.\footnote{Operators of odd spin are excluded by the exchange symmetry of the two identical scalars.} We normalize the bulk stress tensor by
\begin{align}\label{eq:general-q-mixed-CT-normalization}
\begin{aligned}
    \left\langle
        T_{\mu\nu}(x)T_{\rho\sigma}(0)
    \right\rangle
    &=
    \frac{C_T}{|x|^6}
    \mathcal I_{\mu\nu,\rho\sigma}(x)\, ,
    \\
    \mathcal I_{\mu\nu,\rho\sigma}(x)
    &=
    \frac12
    \left(
        I_{\mu\rho}(x)I_{\nu\sigma}(x)
        +I_{\mu\sigma}(x)I_{\nu\rho}(x)
    \right)
    -\frac13\delta_{\mu\nu}\delta_{\rho\sigma}\, ,
    \\
    I_{\mu\nu}(x)
    &=
    \delta_{\mu\nu}-2\frac{x_\mu x_\nu}{x^2}\, .
\end{aligned}
\end{align}
For a unit-normalized Hermitian bulk scalar primary $\mathcal O$ of dimension $\Delta_{\mathcal O}$, the conformal Ward--Takahashi identity fixes the stress-tensor contribution to its OPE~\cite{Osborn:1993cr}:
\begin{align}\label{eq:bulk-scalar-stress-Ward-identity}
    \mathcal O(x)\mathcal O(0)
    \supset
    \lambda_{\mathcal O\mathcal O T}
    \frac{x^\mu x^\nu}
         {|x|^{2\Delta_{\mathcal O}-1}}
    \frac{T_{\mu\nu}(0)}{\sqrt{C_T}}\, , \quad 
    \lambda_{\mathcal O\mathcal O T}
    =
    -\frac{3\Delta_{\mathcal O}}{8\pi\sqrt{C_T}}\, .
\end{align}
Here the factor $1/\sqrt{C_T}$ makes the exchanged bulk tensor unit-normalized. A free complex scalar contributes $3/(16\pi^2)$ to $C_T$. Thus, in critical bosonic QED$_3$,
\begin{align}\label{eq:general-q-mixed-bulk-stress-data}
    C_T
    =
    \frac{3N}{16\pi^2}+O(1)\, , \quad 
    \left|\lambda_{\sigma\sigma T}\right|^2
    =
    \frac{9\Delta_\sigma^2}{64\pi^2C_T}
    =
    \frac{3}{N}+O(N^{-2})\, .
\end{align}
Also, in three dimensions, the leading $O(N^{-1/2})$ contribution to $\langle\sigma\sigma\sigma\rangle$
vanishes at distinct insertion points, so that $\lambda_{\sigma\sigma\sigma}=O(N^{-3/2})$.

\subsection{Mixed crossing equation}
\label{subsec:general-q-mixed-crossing}

\begin{figure}[t]
\centering
\begin{tikzpicture}[
    wilson/.style={
        double,
        double distance=1.1pt,
        line width=0.45pt
    },
    insertion/.style={circle,fill=black,inner sep=2.1pt},
    every node/.style={font=\small}
]
    \draw[wilson] (-3.8,4.0)--(-1.5,4.0);
    \node[insertion] at (-3.8,4.0) {};
    \node[insertion] at (-1.5,4.0) {};
    \node[insertion] at (1.5,4.0) {};
    \node[insertion] at (3.8,4.0) {};

    \node[below=6pt] at (-3.8,4.0)
        {$\bar E_{(q)}^{I}(\tau_1)$};
    \node[below=6pt] at (-1.5,4.0)
        {$E^{(q)}_J(\tau_2)$};
    \node[below=6pt] at (1.5,4.0) {$\sigma(\tau_3)$};
    \node[below=6pt] at (3.8,4.0) {$\sigma(\tau_4)$};

    \draw[->,thick] (-0.15,2.90)--(-2.20,0.85)
        node[midway,above,sloped] {$(12)(34)$};
    \draw[->,thick] (0.15,2.90)--(2.20,0.85)
        node[midway,above,sloped] {$(14)(32)$};

    \node at (-2.55,-1.55)
        {bulk channel: $\mathcal H^{00}_{\one}$};
    \node at (-4.05,0) {$\displaystyle\sum_{\mathcal O}$};
    \node[insertion] at (-3.25,0) {};
    \node[insertion] at (-1.85,0) {};
    \node[below=5pt] at (-3.25,0) {$\mathcal O$};
    \node[below=5pt] at (-1.85,0) {$\mathcal O$};

    \node at (2.55,-1.55)
        {defect channel: $\mathcal H^{q0}_{S_q}$};
    \node at (0.55,0) {$\displaystyle\sum_{E^{(q)},\mathcal X}$};
    \draw[wilson] (1.85,0)--(3.25,0);
    \node[insertion] at (1.85,0) {};
    \node[insertion] at (3.25,0) {};
    \node[below=5pt] at (1.85,0) {$\bar{\mathcal X}$};
    \node[below=5pt] at (3.25,0) {$\mathcal X$};
\end{tikzpicture}
\caption{The configuration and the two OPE channels of the ordered
mixed correlator~\eqref{eq:general-q-mixed-correlator-definition},
with $\tau$ increasing from left to right. In the bulk channel,
the pairs $(12)$ and $(34)$ are expanded in flavor-singlet
operators in $\mathcal H^{00}_{\one}$, and the Wilson segment
$C_{12}$ shrinks. In the defect channel, the pairs $(14)$ and
$(32)$ are expanded in conjugate defect-changing operators.
The endpoint operator $\bar{\mathcal X}$ lies at the left end of the Wilson segment
and $\mathcal X$ at its right end. Here
$\mathcal X\in\mathcal H^{q0}_{S_q}$, and the sum also includes
the original endpoint $E^{(q)}$. The trivial defect is not drawn.}
\label{fig:general-q-mixed-OPE-channels}
\end{figure}
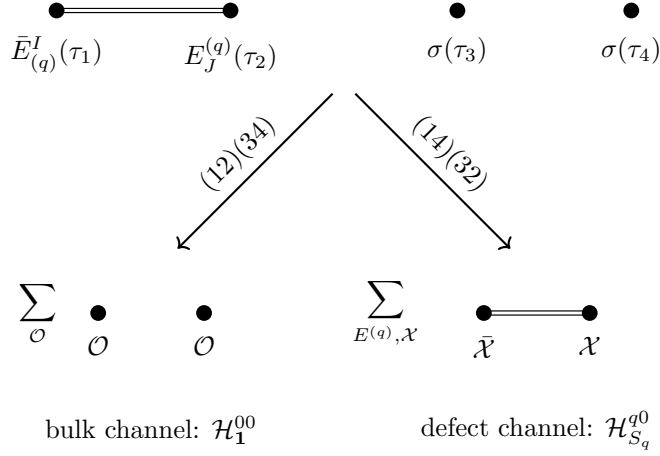

We place the four operators on the same straight line and choose the
ordering $\tau_1<\tau_2<\tau_3<\tau_4$.  The reduced mixed correlator
is defined by
\begin{align}\label{eq:general-q-mixed-correlator-definition}
    \left\langle
        \bar E_{(q)}^{I}(\tau_1)
        E^{(q)}_{J}(\tau_2)
        \sigma(\tau_3)
        \sigma(\tau_4)
    \right\rangle
    =
    \frac{\delta^I{}_J}
    {|\tau_{12}|^{2\Delta_{E_q}}|\tau_{34}|^{2\Delta_\sigma}}
    \cM_q(z)\, .
\end{align}
The corresponding ordered defect configuration is shown in
figure~\ref{fig:general-q-mixed-OPE-channels}.  The only
finite Wilson segment is $C_{12}$ and carries charge $q$.

In the bulk OPE channel, only the flavor-singlet component of
$\bar E_{(q)}\times E^{(q)}$ can also occur in
$\sigma\times\sigma$.  Hence
\begin{align}\label{eq:general-q-mixed-direct-channel}
    \cM_q(z)
    =
    \sum_{\cO\in\cH^{00}_{\one}}
    \lambda_{E_qE_q\cO}
    \lambda_{\sigma\sigma\cO}
    g_{\Delta_\cO}(z)\, .
\end{align}
On the other hand, the defect-channel OPE joins an endpoint to the bulk singlet,
\begin{align}\label{eq:general-q-mixed-crossed-OPE}
    E^{(q)}\times\sigma
    \subset
    \cH^{q0}_{S_q}\, , 
\end{align}
and gives rise to new endpoint operators. The endpoint $E^{(q)}$ itself is allowed to reappear in this OPE.  We denote the remaining primaries probed by the mixed correlator by
$\mathcal X_{q,n,\nu}$, where $n=0,1,\cdots$ labels the levels of the
large-$N$ tower derived below and $\nu$ distinguishes operators that
are degenerate at leading order.  Suppressing descendant
contributions, we define their OPE coefficients by
\begin{align}\label{eq:general-q-mixed-crossed-OPE-coefficients}
    \begin{aligned}
    E^{(q)}_I(0)\sigma(\tau)
    \sim{}&
    \lambda_{E_q\sigma E_q}
    |\tau|^{-\Delta_\sigma}E^{(q)}_I(0) \\
    &+
    \sum_{n,\nu}
    \lambda_{E_q\sigma\mathcal X_{q,n,\nu}}
    |\tau|^{
        \Delta_{q,n,\nu}-\Delta_{E_q}-\Delta_\sigma
    }
    \mathcal X_{q,n,\nu;I}(0)\, ,
    \quad \tau>0\, .
    \end{aligned}
\end{align}
Here $\Delta_{q,n,\nu}$ is the scaling dimension of
$\mathcal X_{q,n,\nu}$, and the flavor index $I$ makes explicit that
these operators transform in the same $S_q$ representation as the
endpoint. The corresponding one-dimensional block and spectral function are
\begin{align}\label{eq:general-q-mixed-block}
    k_h^{E_q\sigma}(x)
    &\equiv
    x^h\,{}_2F_1\!\left(
        h+\Delta_\sigma-\Delta_{E_q},
        h+\Delta_\sigma-\Delta_{E_q};
        2h;x
    \right),
    \\ \label{eq:general-q-mixed-crossed-spectral-function}
    \cM_q^{q0}(x)
    &\equiv
    \left|\lambda_{E_q\sigma E_q}\right|^2
    k_{\Delta_{E_q}}^{E_q\sigma}(x)
    +
    \sum_{n,\nu}
    \left|\lambda_{E_q\sigma\mathcal X_{q,n,\nu}}\right|^2
    k_{\Delta_{q,n,\nu}}^{E_q\sigma}(x)\, .
\end{align}
Matching the two OPEs yields the crossing relation for the mixed correlator, 
\begin{align}\label{eq:general-q-mixed-crossing-equation}
    \cM_q(z)
    =
    \frac{z^{2\Delta_\sigma}}
    {(1-z)^{\Delta_{E_q}+\Delta_\sigma}}
    \cM_q^{q0}(1-z)
    \, .
\end{align}
Note that, because $\sigma$ is a singlet, no flavor crossing matrix is needed.

\subsection{Large-$N$ diagrammatics}
\label{subsec:general-q-mixed-diagrammatics}

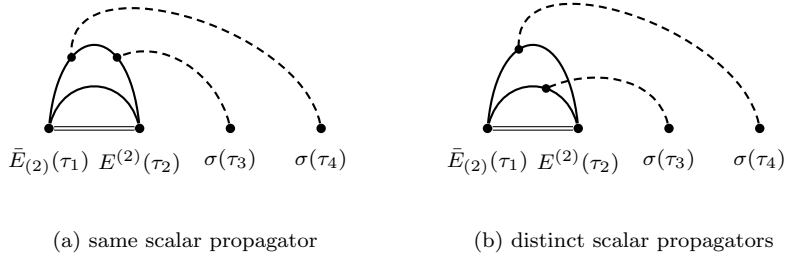
\begin{figure}[t]
\centering
\begin{tikzpicture}[
    scalar/.style={draw=black,thick},
    hs/.style={draw=black,thick,densely dashed},
    wilson/.style={double,double distance=0.9pt,line width=0.4pt},
    insertion/.style={circle,fill,inner sep=1.25pt},
    vertex/.style={circle,fill,inner sep=1.15pt},
    every node/.style={font=\scriptsize}
]
\begin{scope}[shift={(-4.9,0)}]
    \node[insertion] (a1) at (0,0) {};
    \node[insertion] (a2) at (1.2,0) {};
    \node[insertion] (a3) at (2.4,0) {};
    \node[insertion] (a4) at (3.6,0) {};

    \draw[wilson] (a2)--(a1);
    \draw[scalar]
        (a2) .. controls (1.05,1.45) and (0.15,1.45) ..
        node[pos=0.30,coordinate] (aR) {}
        node[pos=0.70,coordinate] (aL) {}
        (a1);
    \draw[scalar]
        (a2) .. controls (1.00,0.72) and (0.20,0.72) .. (a1);

    \node[vertex] at (aL) {};
    \node[vertex] at (aR) {};
    \draw[hs] (a4) to[out=102,in=95] (aL);
    \draw[hs] (a3) to[out=104,in=25] (aR);

    \node[below=4pt] at (a1) {$\bar E_{(2)}(\tau_1)$};
    \node[below=4pt] at (a2) {$E^{(2)}(\tau_2)$};
    \node[below=4pt] at (a3) {$\sigma(\tau_3)$};
    \node[below=4pt] at (a4) {$\sigma(\tau_4)$};
    \node at (1.8,-1.5) {(a) same scalar propagator};
\end{scope}
\begin{scope}[shift={(0.9,0)}]
    \node[insertion] (b1) at (0,0) {};
    \node[insertion] (b2) at (1.2,0) {};
    \node[insertion] (b3) at (2.4,0) {};
    \node[insertion] (b4) at (3.6,0) {};

    \draw[wilson] (b2)--(b1);
    \draw[scalar]
        (b2) .. controls (1.05,1.45) and (0.15,1.45) ..
        node[pos=0.62,coordinate] (bU) {}
        (b1);
    \draw[scalar]
        (b2) .. controls (1.00,0.72) and (0.20,0.72) ..
        node[pos=0.38,coordinate] (bD) {}
        (b1);

    \node[vertex] at (bU) {};
    \node[vertex] at (bD) {};
    \draw[hs] (b4) to[out=102,in=95] (bU);
    \draw[hs] (b3) to[out=104,in=25] (bD);

    \node[below=4pt] at (b1) {$\bar E_{(2)}(\tau_1)$};
    \node[below=4pt] at (b2) {$E^{(2)}(\tau_2)$};
    \node[below=4pt] at (b3) {$\sigma(\tau_3)$};
    \node[below=4pt] at (b4) {$\sigma(\tau_4)$};
    \node at (1.8,-1.5) {(b) distinct scalar propagators};
\end{scope}
\end{tikzpicture}
\caption{The two nonlocal diagram classes contributing to the mixed
correlator at order $1/N$, shown for $q=2$ with
$\tau_1<\tau_2<\tau_3<\tau_4$ from left to right. The endpoint
pair is joined by the charge-two Wilson segment and by two scalar
propagators. In panel (a), both external $\sigma$ fields attach to
the same scalar propagator while the other scalar propagator has no
$\sigma$ insertion. In panel (b), they attach to different scalar
propagators. Solid curves denote scalar propagators, dashed curves
denote external $\sigma$ propagators, and the double line denotes
the Wilson segment. For general charge $q$, the respective
multiplicities are $q$ and $q(q-1)$.}
\label{fig:general-q-mixed-diagrams}
\end{figure}

We expand
\begin{align}\label{eq:general-q-mixed-large-N-expansion}
    \cM_q(z)
    =
    \cM_q^{(0)}(z)
    +\frac1N\cM_q^{(1)}(z)
    +O(N^{-2})\, .
\end{align}
At $N=\infty$, the endpoint and $\sigma$ two-point functions
factorize, so that
\begin{align}\label{eq:general-q-mixed-leading-correlator}
    \cM_q^{(0)}(z)=1\, .
\end{align}
At order $1/N$, the two external $\sigma$ fields must attach to the
$q$ scalar propagators joining the endpoints.  There are two possibilities: both attach to the same scalar propagator, or
they attach to two distinct propagators.  These possibilities are
shown for $q=2$ in figure~\ref{fig:general-q-mixed-diagrams}.  For
general $q$, the remaining scalar propagators are spectators, and the
two classes have multiplicities $q$ and $q(q-1)$, respectively.\footnote{All self-energy, photon, and scalar--Wilson diagrams confined to the
endpoint pair cancel against the normalization by the charge-$q$
endpoint two-point function.  The analogous corrections confined to
the $\sigma$ pair cancel against its two-point function.  Moreover,
the leading separated-point $\sigma^3$ vertex vanishes in three
dimensions, while an effective $A_\mu\sigma\sigma$ vertex is forbidden
by charge conjugation.  Figure~\ref{fig:general-q-mixed-diagrams}
therefore exhausts all contributions at order $1/N$.}

Let $\cM_1^{(1)}$ denote the result with both insertions on a single scalar
propagator.  The two classes in figure~\ref{fig:general-q-mixed-diagrams} evaluate to
\begin{align}\label{eq:general-q-mixed-two-diagram-results}
    \left.\cM_q^{(1)}(z)\right|_{\rm same}
    &=
    q\cM_1^{(1)}(z)\, ,
    \\
    \left.\cM_q^{(1)}(z)\right|_{\rm distinct}
    &=
    \frac{2q(q-1)}{\pi^2}
    \frac{z^4}{(1-z)^2}\, ,
\end{align}
where
\begin{align}\label{eq:general-q-mixed-one-line-result}
    \cM_1^{(1)}(z)
    ={}&
    -\frac{z^2}{2(1-z)}\Lambda(z)
    +\frac{2z(2-z)}{\pi^2(1-z)}\log(1-z)
    +
    \frac{2z^2(z^2-2z+2)}
    {\pi^2(1-z)^2}\, . 
\end{align}
Here the function $\Lambda(z)$ is given in \eqref{eq:general-q-Lambda-main}. The detailed evaluation is relegated to appendix~\ref{app:mixed_correlator}. The complete result is therefore
\begin{align}\label{eq:general-q-mixed-correlator-final}
    \begin{aligned}
    \cM_q^{(1)}(z)
    =
    -\frac{qz^2}{2(1-z)}\Lambda(z)
    +\frac{2qz(2-z)}{\pi^2(1-z)}\log(1-z)
    +
    \frac{2qz^2(qz^2-2z+2)}
    {\pi^2(1-z)^2 }\, . 
    \end{aligned}
\end{align}
The two OPE limits provide useful checks.  In the bulk channel OPE limit $z\to0$, 
\begin{align}\label{eq:general-q-mixed-direct-limit-check}
    \cM_q^{(1)}(z)
    =
    \frac q2 z^3+O(z^4)\, .
\end{align}
The missing $z^2$ term agrees with $\lambda_{\sigma\sigma\sigma}=O(N^{-3/2})$.  The coefficient of $z^3$ is the
stress-tensor contribution.  With the bulk normalization~\eqref{eq:general-q-mixed-CT-normalization}, its
unit-normalized longitudinal component on the line is
\begin{align}
    \mathcal T(\tau)
    \equiv
    \sqrt{\frac{3}{2C_T}}\,T_{\tau\tau}(\tau)\, ,
    \quad
    \left\langle\mathcal T(\tau)\mathcal T(0)\right\rangle
    =
    \frac{1}{|\tau|^6}\, .
\end{align}
The bulk Ward--Takahashi identity \eqref{eq:bulk-scalar-stress-Ward-identity} and
\eqref{eq:general-q-mixed-bulk-stress-data} then give
\begin{align}
    \lambda_{\sigma\sigma\mathcal T}
    =
    \sqrt{\frac23}\lambda_{\sigma\sigma T}
    =
    -\sqrt{\frac2N}+O(N^{-3/2})\, .
\end{align}
At leading order, the Wilson line is trivial and the endpoint reduces to a normalized product of $q$
free scalars. Its stress-tensor coupling is therefore $q$ times that of a scalar of dimension $1/2$:
\begin{align}
    \lambda_{E_qE_q\mathcal T}
    =
    -\frac{q}{2\sqrt{2N}}+O(N^{-3/2})\, .
\end{align}
Consequently,
\begin{align}\label{eq:general-q-mixed-stress-check}
    \lambda_{E_qE_q\mathcal T}
    \lambda_{\sigma\sigma\mathcal T}
    =
    \frac{q}{2N}+O(N^{-2})\, .
\end{align}
Since $g_3(z)=z^3+O(z^4)$, this exchange contributes
$qz^3/(2N)$ to the full mixed correlator, reproducing \eqref{eq:general-q-mixed-direct-limit-check}.

On the other hand, in the defect OPE limit $x \to 0$,
\begin{align}\label{eq:general-q-mixed-crossed-limit-check}
    \cM_q^{(1)}(1-x)
    =
    \frac{2q^2}{\pi^2x^2}
    +O(1/x)\, .
\end{align}
This singularity is produced by the original charge-$q$ endpoint in
the $E^{(q)}\times\sigma$ OPE.  An independent three-point calculation gives
\begin{align}\label{eq:general-q-mixed-endpoint-three-point}
    \lambda_{E_qE_q\sigma}
    =
    \lambda_{E_q\sigma E_q}
    =
    \frac{q\sqrt2}{\pi\sqrt N}
    +O(N^{-3/2})\, ,
\end{align}
in agreement with \eqref{eq:general-q-mixed-crossed-limit-check}.\footnote{This result is also consistent with \eqref{eq:general-q-bulk-singlet}.}

\subsection{Defect-changing CFT data from mixed crossing}
\label{subsec:general-q-mixed-CFT-data}

We now use the explicit correlator~\eqref{eq:general-q-mixed-correlator-final} as the input to the crossing relation.
At $N=\infty$, the constant bulk-channel correlator decomposes as
\begin{align}\label{eq:general-q-mixed-GFF-decomposition}
    \frac{x^{q/2+2}}{(1-x)^4}
    =
    \sum_{n=0}^{\infty}
    \beta_{q,n}^{(0)}K_{q,n}(x)\, , \quad
    \beta_{q,n}^{(0)}
    =
    \frac{(q)_n(4)_n}
    {n!(q+n+3)_n}\, , 
\end{align}
where
\begin{align}\label{eq:general-q-mixed-leading-blocks}
    K_{q,n}(x)
    \equiv
    x^{q/2+2+n}
    {}_2F_1\!\left(
        n+4,n+4;q+2n+4;x
    \right)\, .
\end{align}
Thus the leading dimensions in $\cH^{q0}_{S_q}$ seen by this
correlator are
\begin{align}\label{eq:general-q-mixed-leading-dimensions}
    \Delta_{q,n}^{(0)}
    =
    \frac q2+2+n\, ,
    \quad n=0,1,2,\cdots\, .
\end{align}
At finite $N$, several operators can in principle contribute at the
same leading one-dimensional dimension.  In \eqref{eq:general-q-mixed-crossed-OPE-coefficients}, we write such CFT data as
\begin{align}\label{eq:general-q-mixed-data-expansion}
    \begin{aligned}
    \Delta_{q,n,\nu}
    &=
    \frac q2+2+n
    +\frac{\gamma_{q,n,\nu}}{N}
    +O(N^{-2})\, ,
    \\
    \sum_\nu
    \left|\lambda_{E_q\sigma\mathcal X_{q,n,\nu}}\right|^2
    &=
    \beta_{q,n}^{(0)}
    +\frac{\beta_{q,n}^{(1)}}{N}
    +O(N^{-2})\, .
    \end{aligned}
\end{align}
For each exchanged primary, we write
\begin{align}
    \left|\lambda_{E_q\sigma\mathcal X_{q,n,\nu}}\right|^2
    =
    \beta_{q,n,\nu}^{(0)}
    +\frac{\beta_{q,n,\nu}^{(1)}}{N}
    +O(N^{-2})\, .
\end{align}
The correlator does not resolve the degeneracy label $\nu$ and hence
determines only the total OPE weights
\begin{align}\label{eq:general-q-mixed-total-OPE-weights}
    \beta_{q,n}^{(r)}
    \equiv
    \sum_\nu\beta_{q,n,\nu}^{(r)}\, ,
    \quad r=0,1\, .
\end{align}
The mixed correlator determines the OPE-weighted anomalous dimension
\begin{align}\label{eq:general-q-mixed-weighted-gamma}
\overline\gamma_{q,n}
\equiv
\frac{1}{\beta_{q,n}^{(0)}}
\sum_\nu
\beta_{q,n,\nu}^{(0)}\gamma_{q,n,\nu}\, .
\end{align}
At $N=\infty$, factorization gives the level-$n$ operators the leading
dimension
$\Delta_{E_q}^{(0)}+\Delta_\sigma^{(0)}+n=q/2+2+n$, as found in
\eqref{eq:general-q-mixed-leading-dimensions}.  When the two external
dimensions receive their order-$1/N$ corrections, this sum is shifted
by the already known amount $\gamma_q+\gamma_\sigma$.  We define the
additional shift of the exchanged operators by
\begin{align}\label{eq:general-q-mixed-eta-definition}
    \overline\gamma_{q,n}
    =
    \gamma_q+\gamma_\sigma+\eta_{q,n}\, .
\end{align}
It is $\eta_{q,n}$ that will be determined by mixed crossing.

We now explain how this information is isolated from the calculated
correlator.  Rewriting~\eqref{eq:general-q-mixed-crossing-equation} for the defect channel
function gives
\begin{align}\label{eq:general-q-mixed-crossed-reconstruction}
    \cM_q^{q0}(x)
    =
    \frac{x^{\Delta_{E_q}+\Delta_\sigma}}
    {(1-x)^{2\Delta_\sigma}}
    \cM_q(1-x)\, .
\end{align}
Write
\begin{align}\label{eq:general-q-mixed-crossed-large-N-expansion}
    \cM_q^{q0}(x)
    =
    \cM_q^{q0,(0)}(x)
    +\frac1N\cM_q^{q0,(1)}(x)
    +O(N^{-2})\, , \quad 
    \cM_q^{q0,(0)}(x)
    =
    \frac{x^{q/2+2}}{(1-x)^4}\, .
\end{align}
Expanding \eqref{eq:general-q-mixed-crossed-reconstruction} to first
order gives
\begin{align}\label{eq:general-q-mixed-crossed-order-one}
    \cM_q^{q0,(1)}(x)
    =
    \cM_q^{q0,(0)}(x)
    \left[
        \cM_q^{(1)}(1-x)
        +
        (\gamma_q+\gamma_\sigma)\log x
        -2\gamma_\sigma\log(1-x)
    \right].
\end{align}
The block for $\mathcal X_{q,n,\nu}$ contains the factor $x^{\Delta_{q,n,\nu}}$. Expanding its dimension as in \eqref{eq:general-q-mixed-data-expansion}, we find
\begin{align}
    x^{\Delta_{q,n,\nu}}
    =
    x^{q/2+2+n}
    \left[
        1+\frac{\gamma_{q,n,\nu}}{N}\log x
        +O(N^{-2})
    \right].
\end{align}
The hypergeometric function multiplying this factor has a power-series expansion around $x=0$.
Changing its parameters changes the coefficients of this series but introduces no $\log x$.
Likewise, the order-$1/N$ correction to the squared OPE coefficient multiplies the leading block and
produces no logarithm.

Thus, the $\log x$ term from each primary is proportional to $\beta_{q,n,\nu}^{(0)}\gamma_{q,n,\nu}$.
Summing over the primaries and using \eqref{eq:general-q-mixed-weighted-gamma}, we obtain
\begin{align}\label{eq:general-q-mixed-log-function-definition}
    \cM_q^{q0,(1)}(x)
    =
    \left[
        \sum_{n=0}^{\infty}
        \beta_{q,n}^{(0)}
        \overline\gamma_{q,n}K_{q,n}(x)
    \right]\log x
    + (\text{terms without }\log x )\, .
\end{align}
Using
\eqref{eq:general-q-mixed-GFF-decomposition} and
\eqref{eq:general-q-mixed-eta-definition}, we define
$\mathcal L_q(x)$ by
\begin{align}\label{eq:general-q-mixed-log-function-subtraction}
    \cM_q^{q0,(1)}(x)
    =
    \left[
        (\gamma_q+\gamma_\sigma)
        \cM_q^{q0,(0)}(x)
        +
        \mathcal L_q(x)
    \right]\log x
    +( \text{terms without }\log x )\, .
\end{align}
The endpoint exchange
\eqref{eq:general-q-mixed-endpoint-three-point} contributes no
$\log x$ at this order and is included among the terms without
$\log x$.  Substituting the calculated correlator
\eqref{eq:general-q-mixed-correlator-final} into
\eqref{eq:general-q-mixed-crossed-order-one} gives
\begin{align}\label{eq:general-q-mixed-log-function}
    \mathcal L_q(x)
    =
    \frac{2q}{\pi^2}
    \left[
        \frac{x^{q/2+1}(1+x)}{(1-x)^3}
        -
        \frac{x^{(q+1)/2}}{1-x}
        \operatorname{arctanh}\sqrt x
    \right].
\end{align}
Comparing \eqref{eq:general-q-mixed-log-function-definition} with \eqref{eq:general-q-mixed-log-function-subtraction} gives
\begin{align}\label{eq:general-q-mixed-log-block-equation}
    \mathcal L_q(x)
    =
    \sum_{n=0}^{\infty}
    \beta_{q,n}^{(0)}\eta_{q,n}K_{q,n}(x)
    \, .
\end{align}
To solve this equation for arbitrary $n$, expand
$\mathcal L_q(x)$ around $x=0$:
\begin{align}\label{eq:general-q-mixed-log-series}
    \mathcal L_q(x)
    =
    \sum_{m=0}^{\infty}
    \ell_{q,m}\,x^{q/2+2+m}\, , \quad
    \ell_{q,m}
    =
    \frac{2q}{\pi^2}
    \left[
        (m+2)^2-H_{2m+3}
        +\frac12H_{m+1}
    \right].
\end{align}
The block expansion following from
\eqref{eq:general-q-mixed-leading-blocks} is
\begin{align}\label{eq:general-q-mixed-block-series}
    K_{q,n}(x)
    =
    x^{q/2+2+n}
    \sum_{j=0}^{\infty}
    \frac{(n+4)_j^2}
    {j!(q+2n+4)_j}\,x^j\, .
\end{align}
Matching the coefficient of $x^{q/2+2+r}$ in
\eqref{eq:general-q-mixed-log-block-equation} gives
\begin{align}\label{eq:general-q-mixed-eta-coefficient-equation}
    \ell_{q,r}
    =
    \sum_{n=0}^{r}
    \frac{(q)_n(4)_n}
    {n!(q+n+3)_n}
    \eta_{q,n}
    \frac{
        (n+4)_{r-n}^{\,2}
    }{
        (r-n)!(q+2n+4)_{r-n}
    }\, .
\end{align}
Solving these equations gives the following finite sum for every
$n\geq0$:
\begin{align}\label{eq:general-q-mixed-eta-general-result}
    \begin{aligned}
    \eta_{q,n}
    ={}&
    \frac{
        2q\,n!(q+n+3)_n
    }{
        \pi^2(q)_n(4)_n
    }
    \sum_{m=0}^{n}
    (-1)^{n-m}
    \frac{
        (m+4)_{n-m}^{\,2}
    }{
        (n-m)!(q+n+m+3)_{n-m}
    }
    \left[
        (m+2)^2-H_{2m+3}
        +\frac12H_{m+1}
    \right].
    \end{aligned}
\end{align}
In particular, the lowest defect-changing dimension is
\begin{align}\label{eq:general-q-mixed-lowest-dimension}
    \overline\Delta_{q,0}
    =
    \frac q2+2
    -\frac{2(9q^2+10q+72)}{3\pi^2N}
    +O(N^{-2})
    \, .
\end{align}
Here and below a bar denotes the OPE-weighted value if the leading level is degenerate.

The terms without $\log x$ determine the OPE coefficients.  We first
separate the variation implied by the external dimensions:
\begin{align}\label{eq:general-q-mixed-beta-decomposition}
    \beta_{q,n}^{(1)}
    =
    \beta_{q,n}^{(0)}
    \left(
        \gamma_q\kappa_{q,n}^{(E)}
        +\gamma_\sigma\kappa_{q,n}^{(\sigma)}
    \right)
    +\delta\beta_{q,n}\, .
\end{align}
The coefficients $\kappa_{q,n}^{(E)}$ and $\kappa_{q,n}^{(\sigma)}$ are the logarithmic derivatives
of the factorized OPE weights with respect to the external dimensions $\Delta_{E_q}$ and $\Delta_\sigma$,
respectively, evaluated at $\Delta_{E_q}=q/2$ and $\Delta_\sigma=2$.
Explicitly,
\begin{align}\label{eq:general-q-mixed-kinematic-kappa}
    \kappa_{q,n}^{(E)}
    &=
    2\left(
        H_{q+n-1}-H_{q-1}
        -H_{q+2n+2}+H_{q+n+2}
    \right),
    \\
    \kappa_{q,n}^{(\sigma)}
    &=
    2\left(
        H_{n+3}-H_3
        -H_{q+2n+2}+H_{q+n+2}
    \right).
\end{align}
To determine $\delta\beta_{q,n}$, we first remove the external-dimension contribution in
\eqref{eq:general-q-mixed-crossed-order-one} and subtract the endpoint exchange:
\begin{align}
    \cM_q^{q0,(0)}(x)\cM_q^{(1)}(1-x)
    -
    \frac{2q^2}{\pi^2}
    x^{q/2}{}_2F_1(2,2;q;x)
    =
    \mathcal L_q(x)\log x+\mathcal R_q(x)\, .
\end{align}
The part without $\log x$ is
\begin{align}\label{eq:general-q-mixed-regular-function}
    \mathcal R_q(x)
    =
    \frac{2q}{\pi^2}x^{q/2}
    \Bigg[
        \frac{\sqrt x}{1-x}
        \left(
            \Li_2(\sqrt x)-\Li_2(-\sqrt x)
        \right)
        +\frac{1+x^2}{(1-x)^2}
        -q\,{}_2F_1(2,2;q;x)+q-1
    \Bigg].
\end{align}
Expanding this around $x=0$ gives
\begin{align}
    \mathcal R_q(x)
    =
    x^{q/2+2}\sum_{p=0}^{\infty}R_{q,p}x^p\, , \quad 
    R_{q,p}
    =
    \frac{2q}{\pi^2}
    \left[
        2H_{2p+3}^{(2)}
        -\frac12H_{p+1}^{(2)}
        +2p+4
        -q\frac{(2)_{p+2}^2}
        {(p+2)!(q)_{p+2}}
    \right].
\end{align}
The coefficients $R_{q,p}$ are now known. To extract the OPE coefficients, we must subtract the
contributions of the lower-level blocks, including their already determined dimension corrections.
Matching successive powers of $x$ gives
\begin{align}\label{eq:general-q-mixed-delta-beta-recursion}
    \delta\beta_{q,p}
    =
    R_{q,p}
    -\sum_{n=0}^{p-1}
    c_{q;n,p-n}
    \left[
        \delta\beta_{q,n}
        +\beta_{q,n}^{(0)}\eta_{q,n}
        d_{q;n,p-n}
    \right]
    \, ,
\end{align}
where the coefficients $c_{q;n,m}$ and $d_{q;n,m}$ are defined by
\begin{align}\label{eq:general-q-mixed-block-series-data}
    c_{q;n,m}
    &\equiv
    \frac{(n+4)_m^2}
    {m!(q+2n+4)_m}\, ,
    \\
    d_{q;n,m}
    &\equiv
    2\left[
        H_{n+m+3}-H_{n+3}
        -H_{q+2n+m+3}+H_{q+2n+3}
    \right]\, .
\end{align}
This determines the OPE correction at every level.  At the lowest level, this yields the sum rule of the squared OPE coefficients, 
\begin{align}\label{eq:general-q-mixed-lowest-OPE}
    \sum_\nu
    \left|
        \lambda_{E_q\sigma\mathcal X_{q,0,\nu}}
    \right|^2
    =
    1+
    \frac{4q(28q-53)}
    {9(q+1)\pi^2N}
    +O(N^{-2})
    \; .
\end{align}

The three correlators $\langle \bar E_{(q)} E^{(q)} \bar E_{(q)} E^{(q)}\rangle$, $\langle\bar E_{(q)}E^{(q)}\sigma\sigma\rangle$ and $\langle\sigma\sigma\sigma\sigma\rangle$ together give a system of crossing equations amenable to semidefinite programming, and we are pursuing its numerical implementation~\cite{Maeda:2026numerical}.

\subsection{Simplification at $q=1$}
\label{subsec:q-one-mixed-simplification}

For general $q$ and $n$, the mixed bootstrap determines the CFT data
through the sum rules
\eqref{eq:general-q-mixed-total-OPE-weights} and
\eqref{eq:general-q-mixed-weighted-gamma}.  Operators with the same
leading dimension contribute to the same
$\mathrm{SL}(2,\mathbb R)$ block, so this correlator does not resolve
their individual anomalous dimensions and OPE coefficients.  If the
relevant OPE sector has multiplicity one, however, each sum contains
only one term.  The bootstrap equations then determine the individual
defect-changing CFT data, rather than an OPE-weighted average.

For $q=1$, the lowest level is non-degenerate.  Indeed, the
$E^{(1)}\times\sigma$ OPE selects the parity-even,
transverse-spin-zero fundamental sector, and at free dimension
$5/2$ this sector contains a single $\mathrm{SL}(2,\mathbb R)$
primary. An explicit count of the independent operators in this
sector is given in appendix~\ref{app:q-one-mixed-lowest-multiplicity}. It may be represented at $N=\infty$ as\footnote{This expression gives the leading representative
of the interacting primary. At finite $N$, contributions from the lower-dimensional endpoint and its descendants are subtracted in defining the renormalized primary.}
\begin{align}\label{eq:q-one-mixed-lowest-primary}
    \mathcal X_{1,0}^{(0)}
    =
    :\!\sigma E^{(1)}\!:\, ,
    \quad
    \Delta_{\mathcal X_{1,0}}^{(0)}
    =
    \frac52\, .
\end{align}
Consequently, the sums over $\nu$ in
\eqref{eq:general-q-mixed-data-expansion} collapse to one term for
$(q,n)=(1,0)$, and the results~\eqref{eq:general-q-mixed-lowest-dimension} and \eqref{eq:general-q-mixed-lowest-OPE} give the individual CFT data
\begin{align}\label{eq:q-one-mixed-lowest-individual-data}
    \begin{aligned}
    \Delta_{\mathcal X_{1,0}}
    &=
    \frac52-
    \frac{182}{3\pi^2N}
    +O(N^{-2})\, ,
    \\
    \left|\lambda_{E_1\sigma\mathcal X_{1,0}}\right|^2
    &=
    1-
    \frac{50}{9\pi^2N}
    +O(N^{-2})\, .
    \end{aligned}
\end{align}
As an independent consistency check, in appendix~\ref{app:q-one-mixed-lowest-two-point-check} we compute the two-point function of the renormalized primary represented by $:\!\sigma E^{(1)}\!:$, which reproduces the anomalous dimension given in \eqref{eq:q-one-mixed-lowest-individual-data}.

For higher charge, or at higher levels, several primaries may share the same leading dimension, and the present correlator again determines only their OPE-weighted combinations.  We leave the
systematic extraction of the individual defect-changing CFT data at general $q$ from an enlarged endpoint-bootstrap system as an interesting direction for future work.

\section{Conclusions and future prospects}

In this work, we have shown how to extract conformal data associated with Wilson lines in critical bosonic QED$_3$ by combining endpoint crossing relations with diagrammatic calculations of four-point functions involving their endpoints. We carried out this analysis through first subleading order
in the large-$N$ expansion for arbitrary positive integer charge $q$ held fixed as $N\to\infty$.

There are many directions in which this viewpoint can be developed. We conclude by highlighting some broader questions for which endpoint bootstrap may provide a useful probe of defect physics.

\begin{itemize}
    \item \emph{Screening, string breaking, and phases of Wilson lines.} 
    Endpoint bootstrap gives direct access to the CFT data that
    diagnose whether a conformal Wilson line is stable against
    screening.  The spectrum in $\mathcal H^{qq}$ identifies relevant
    line operators that can destabilize $\mathcal D_q$, while
    correlators of junctions between different charge sectors determine
    the operators and OPE coefficients through which
    $\mathcal D_q$ can flow to a partially or completely screened line.
    Comparing their $g$-functions then distinguishes the ultraviolet
    and infrared defect fixed points.  Applying this bootstrap
    simultaneously to several charges, and eventually to charges that
    grow with $N$, could therefore locate the onset of screening,
    identify the associated fixed-point mergers, and reconstruct the
    phase diagram of conformal Wilson lines
    \cite{Aharony:2023amq}. 
    \item \emph{Dualities of line defects and their endpoints.} 
    Three-dimensional dualities exchange electric, magnetic, and
    vortex degrees of freedom, and should consequently act on Wilson
    lines, vortex lines, and the operators at which they end.  Applying
    endpoint bootstrap to fermionic QED$_3$ and Chern--Simons matter
    theories would make it possible to follow these extended objects
    through particle--vortex and bosonization dualities
    \cite{Karch:2016sxi,Seiberg:2016gmd}.  Matching endpoint
    dimensions, defect $g$-functions, and junction OPE coefficients on
    the two sides would test the duality beyond the spectrum of local
    bulk operators.
    \item \emph{Gauge-invariant spectroscopy of fractionalized excitations.}
    At a deconfined quantum critical point, the elementary spinons are
    gauge charged and are therefore invisible to correlators of local
    gauge-invariant operators
    \cite{Senthil:2003eed,Kaul:2008xw,Senthil:2023vqd}.  An open Wilson
    line terminated by matter provides a gauge-invariant insertion of
    such an excitation.  Endpoint correlators could thus
    determine spinon scaling dimensions, multi-spinon fusion data, and
    possible bound states, while mixed correlators would describe how
    spinons couple to the ordinary order-parameter sector.  Endpoint
    bootstrap may therefore provide a direct spectroscopy of the
    fractionalized sector of deconfined critical theories.
    \item \emph{Critical endpoints at anyon-condensation transitions.}
    In Chern--Simons matter theories, Wilson lines represent the
    worldlines of anyons.  When an anyon condenses, line operators may
    end, split, or become identified, while other anyons become
    confined~\cite{Bais:2008ni,Burnell:2017otf}.  It would be
    interesting to study such continuous transitions using endpoint
    bootstrap, by considering correlators of endpoints and junctions
    that terminate or change an anyon line.  Fusion rules restrict the allowed OPE channels, while
    crossing constrains the spectra and OPE coefficients in the corresponding line and junction sectors.  These data can diagnose which anyons condense
    or become identified and determine the critical exponents for
    their absorption and transmutation.
\end{itemize}
In each direction, the relevant bulk physics is encoded by line
operators.  Endpoint bootstrap may translate processes such as
screening, spinon creation, and anyon absorption into the dimensions
and OPE coefficients of endpoint and junction operators.  It thereby
provides a unified way to extract quantitative data about these
processes in interacting gauge theories.

\acknowledgments
We thank Gabriel Cuomo for useful discussion. 
 JM is supported by JST SPRING, Grant
Number JPMJSP2110.
 TO is supported by JST SPRING, Grant
Number JPMJSP2110.
 SS is supported by Grant-in-Aid for JSPS Fellows No.~26KJ0182.  
We used OpenAI Codex (versions 6 and 6.1) to assist with checking the equations and with writing and proofreading of the manuscript.

\appendix

\section{Representation theory for $\overline{S_q}\otimes S_q$}
\label{app:representation_theory}

\subsection{Clebsch--Gordan tensors for $\overline{S_q}\otimes S_q$}
\label{app:tensor_index_CG}

In this appendix, we give an explicit realization of the
Clebsch--Gordan tensors used in the decomposition \eqref{eq:decomposition}.
The representation $S_q=\mathrm{Sym}^q\mathbf N$ is realized by a
completely symmetric tensor with $q$ lower fundamental indices,
\begin{align}
v_{j_1\cdots j_q}
=
v_{(j_1\cdots j_q)}\, ,
\end{align}
while its conjugate $\overline{S_q}$ is represented by a completely
symmetric tensor with $q$ upper fundamental indices,
\begin{align}
\bar v^{i_1\cdots i_q}
=
\bar v^{(i_1\cdots i_q)} \, .
\end{align}
Accordingly, an element of
$\overline{S_q}\otimes S_q$ may be represented as
\begin{align}
X^{i_1\cdots i_q}{}_{j_1\cdots j_q}\, ,
\end{align}
which is separately symmetric in the $q$ upper and the $q$ lower
indices.
The irreducible representation $R_k=[k,0,\cdots,0,k]$ can be realized by a tensor
\begin{align}
\left(
t^{(k)A}
\right)^{a_1\cdots a_k}{}_{b_1\cdots b_k} \, ,
\end{align}
which is separately symmetric in its upper and lower indices,
\begin{align}
\left(
t^{(k)A}
\right)^{a_{\sigma(1)}\cdots a_{\sigma(k)}}{}_{
b_{\rho(1)}\cdots b_{\rho(k)}
}
=
\left(
t^{(k)A}
\right)^{a_1\cdots a_k}{}_{b_1\cdots b_k} \, ,
\quad
\sigma,\rho\in \mathfrak{S}_k \, ,
\end{align}
and traceless under contraction of any upper index with any lower
index,
\begin{align}
\left(
t^{(k)A}
\right)^{c\,  a_2\cdots a_k}{}_{c\, b_2\cdots b_k}
=
0\, .
\end{align}
We choose an orthonormal basis of $R_k$ with respect to the natural Hilbert--Schmidt inner product,
\begin{align}
\left(
t_A^{(k)\dagger}
\right)^{b_1\cdots b_k}{}_{a_1\cdots a_k}
\left(
t^{(k)B}
\right)^{a_1\cdots a_k}{}_{b_1\cdots b_k}
=
\delta_A{}^B\, .
\end{align}
For notational convenience, we introduce the upper and lower
multi-indices
\begin{align}
\bm{i}
\equiv
(i_1,\cdots,i_q)\, ,
\quad
\bm{j}
\equiv
(j_1,\cdots,j_q)\, .
\end{align}
We denote by $\mathfrak S_q$ the permutation group of $q$ objects,
in order to distinguish it from the symmetric $\SU(N)$ representation
$S_q=\mathrm{Sym}^q\mathbf N$. Independent unit-weight
symmetrization over the upper and lower indices is defined by
\begin{align}
\operatorname{Sym}_{\bm{i}}
\operatorname{Sym}_{\bm{j}}
X^{i_1\cdots i_q}{}_{j_1\cdots j_q}
\equiv
\frac{1}{(q!)^2}
\sum_{\sigma,\rho\in\mathfrak S_q}
X^{
i_{\sigma(1)}
\cdots
i_{\sigma(q)}
}{}_{
j_{\rho(1)}
\cdots
j_{\rho(q)}
}\, .
\end{align}
A convenient tensor realization of the Clebsch--Gordan coefficients for
symmetric representations of unitary groups is well known; see for
example~\cite{Alisauskas:1972sto,Alex:2010wi}. Adapting these tensor
methods to our normalization convention,
\begin{align}\label{eq:explicit-Sq-CG-tensor}
    \mathcal C_q^{
        (k)A\,i_1\cdots i_q
    }{}_{
        j_1\cdots j_q
    }
    &=
    \mathcal N_{q,k}\,
    \operatorname{Sym}_{\bm{i}}\,
    \operatorname{Sym}_{\bm{j}}\,
    \Bigg[
        \mathcal{T}_q^{(k)A\,i_1\cdots i_q}{}_{j_1\cdots j_q}
    \Bigg]\, , 
\end{align}
where
\begin{align}\label{eq:T_q^k}
    \mathcal{T}_q^{(k)A\,i_1\cdots i_q}{}_{j_1\cdots j_q}
    =
    \left(
        t^{(k)A}
    \right)^{
        i_1\cdots i_k
    }{}_{
        j_1\cdots j_k
    }
    \delta^{i_{k+1}}{}_{j_{k+1}}
    \cdots
    \delta^{i_q}{}_{j_q}
\end{align}
and
\begin{align}\label{eq:N_qk}
\mathcal N_{q,k}
=
\binom{q}{k}
\left[
\binom{N+q+k-1}{q-k}
\right]^{-1/2} \, .
\end{align}
With this normalization, the Clebsch--Gordan coefficients satisfy
\begin{align}
\sum_{I,J}
\left(
\mathcal C_q^{(k)A,I}{}_{J}
\right)^*
\mathcal C_q^{(\ell)B,I}{}_{J}
=
\delta_{k\ell}\delta^{AB}\, .
\end{align}
For $k=0$, this gives the normalized singlet intertwiner
\begin{align}
\mathcal C_q^{(0),I}{}_{J}
=
\frac{1}{\sqrt{n_q}}
\delta^{I}{}_{J}\, ,
\quad
n_q
=
\binom{N+q-1}{q}\, .
\end{align}
For $k=1$, choosing
\begin{align}
\left(t^{(1)a}\right)^i{}_{j}
=
\sqrt{2}
(T^a)^i{}_{j}
\end{align}
gives the adjoint intertwiner. In particular, for $q=1$ one recovers
\begin{align}
\mathcal C_1^{(1)a,i}{}_{j}
=
\sqrt{2}
(T^a)^i{}_{j} \, ,
\end{align}
in agreement with the flavor normalization used in the main text.


\subsection{Computation of $\Pi_{km}^{(\mathrm{s}/\mathrm{t})}$}
\label{app:Pi_km}

We determine how each flavor monomial
\begin{align}\label{eq:Xm_app}
    \mathbb X_m
    =
    (\bar u_1\cdot u_2)^{q-m}
    (\bar u_3\cdot u_4)^{q-m}
    (\bar u_1\cdot u_4)^m
    (\bar u_3\cdot u_2)^m
\end{align}
contributes to the irreducible flavor channels. Its expansion in the orthonormal
bulk-channel basis is
\begin{align}
    \mathbb X_m
    =
    \sum_{k=0}^{q}
    \sqrt{d_k}\,
    \Pi_{km}^{(\mathrm s)}
    \mathbb T_k^{(\mathrm s)}\, , \quad 
    \Pi_{km}^{(\mathrm s)}
    =
    \frac{1}{\sqrt{d_k}}
    \bigl(\mathbb T_k^{(\mathrm s)},\mathbb X_m\bigr)\, .
\end{align}
In this appendix, we evaluate $\Pi_{km}^{(\mathrm s)}$. The defect-channel coefficient $\Pi_{km}^{(\mathrm t)}$ can be deduced via the crossing relation~\eqref{eq:general-q-t-from-s-relabeling}.

First consider the factor $(\bar u_1\cdot u_2)^{q-m}$ in \eqref{eq:Xm_app}. It contracts $q-m$ upper and lower indices within each Clebsch--Gordan tensor. The resulting partial trace is
\begin{align}
    \mathcal S_{q,m}^{(k)A\,i_1\cdots i_m}
    {}_{j_1\cdots j_m}
    \equiv
    \sum_{a_{m+1},\ldots,a_q}
    \mathcal C_q^{(k)A\,i_1\cdots i_m
    a_{m+1}\cdots a_q}
    {}_{j_1\cdots j_m a_{m+1}\cdots a_q}\, .
\end{align}
Similarly, for the factor $(\bar u_3\cdot u_4)^{q-m}$, the contraction gives rise to the complex conjugate of $\mathcal S_{q,m}^{(k)A\,i_1\cdots i_m}{}_{j_1\cdots j_m}$.

The remaining factors
$(\bar u_1\cdot u_4)^m(\bar u_3\cdot u_2)^m$
then connect the untraced indices of the two tensors:
the $(14)$ contractions identify the indices at endpoints
$1$ and $4$, while the $(32)$ contractions identify those at
endpoints $3$ and $2$. Together, they give the Hilbert--Schmidt norm of the partial trace. Consequently, using \eqref{eq:general-q-two-flavor-bases}, we obtain
\begin{align}
    \begin{aligned}
    \Pi_{km}^{(\mathrm s)}
    &=
    \frac{1}{d_k}
    \sum_{A=1}^{d_k}
    \sum_{\substack{i_1,\ldots,i_m\\j_1,\ldots,j_m}}
    \left|
        \mathcal S_{q,m}^{(k)A\,i_1\cdots i_m}
        {}_{j_1\cdots j_m}
    \right|^2 \\
    &=
    \frac{1}{d_k}
    \sum_{A=1}^{d_k}
    \bigl(
        \mathcal S_{q,m}^{(k)A},
        \mathcal S_{q,m}^{(k)A}
    \bigr)\, .
    \end{aligned}
\end{align}

It remains to determine the partial trace. Because contraction with a Kronecker delta commutes with $\SU(N)$ transformations, $\mathcal S_{q,m}^{(k)A}$ transforms
in $R_k$, if it is nonzero. On the other hand, its remaining
indices belong to
\begin{align}
    \overline{S_m}\otimes S_m
    =
    \bigoplus_{\ell=0}^{m}R_\ell\, .
\end{align}
Hence $\mathcal S_{q,m}^{(k)A}=0$ for $k>m$. On the other hand, for $k\leq m$, the representation $R_k$ occurs only once, so the partial trace must be proportional to the normalized Clebsch--Gordan tensor with $m$ upper and lower indices:
\begin{align}\label{eq:S_propotional_to_C}
    \mathcal S_{q,m}^{(k)A}
    =
    s_{q,m,k}\, \mathcal C_m^{(k)A}\, .
\end{align}
Their unit normalization therefore gives
\begin{align}
    \Pi_{km}^{(\mathrm s)}
    =
    \begin{cases}
        0\, , & k>m\, ,\\
        |s_{q,m,k}|^2\, , & k\leq m\, .
    \end{cases}
\end{align}
The remaining task is to compute $s_{q,m,k}$ for $k\leq m$. Note that
\begin{align}\label{eq:cal N}
    \Big(\mathcal{T}_m^{(k)A},\mathcal{S}_{q,m}^{(k)A}\Big)
    =
    \Big(\mathcal{T}_q^{(k)A},\mathcal{C}_q^{(k)A}\Big)
    =
    \frac{1}{\mathcal{N}_{q,k}}\,.
\end{align}
Combining this with equation~\eqref{eq:S_propotional_to_C}, we obtain
\begin{align}
    s_{q,m, k} 
    = \frac{1}{\mathcal{N}_{q,k}}
    \frac{1}{(\mathcal{T}_m^{(k)A},\mathcal{C}_m^{(k)A})} 
    = \frac{\mathcal{N}_{m,k}}{\mathcal{N}_{q,k}}\,.
\end{align}
The contribution with the highest power of $N$ corresponds to $m=k$, for which we find
\begin{align}
    \binom{q}{k}^2\Pi_{kk}^{(\mathrm{s})}
    = \binom{N+q+k-1}{q-k}\,.
\end{align}
The next contribution comes from $m=k+1$, yielding
\begin{align}\label{eq:NLO_Pi_km}
    \binom{q}{k+1}^2\Pi_{k,k+1}^{(\mathrm{s})} 
    = \binom{q}{k}^2\Pi_{kk}^{(\mathrm{s})}
    \cdot \frac{(q-k)^2}{N+2k}\,.
\end{align}
All other terms are subleading at higher orders in $1/N$.

\section{General-charge diagrammatics at first subleading order}
\label{app:general-q-diagrammatics_single}

\subsection{Evaluation of single correlator $\langle \bar E E \bar E E \rangle$}
This appendix evaluates the four functions $\mathscr A, \mathscr B, \mathscr C$ and $\mathscr D$ appearing in \eqref{eq:general-q-four-master-functions-main}. Summing the order-$1/N$ diagrams in figure~\ref{fig:general-q-diagram-classes} and applying the two-point normalization in \eqref{eq:general-q-microscopic-to-reduced-correlator} gives
\begin{align}
    \mathscr F_{q,m} &=(q-m)^2\mathscr A +m^2\mathscr B +m(q-m)\mathscr C +m\mathscr D\, .
\end{align}
After Wick contractions, the four functions have the following integral expressions:
\begin{align}\label{eq:general-q-four-master-definitions-app}
\begin{aligned}
    \mathscr A
    &=\mathcal I_\sigma^{12|34}-\mathcal I_A^{12|34}
    -\mathcal I_{AW}^{12;34}
    -\mathcal I_{AW}^{34;12}-\mathcal I_{WW}^{12|34}\, ,
    \\
    \mathscr B
    &=\mathcal I_\sigma^{14|32}-\mathcal I_A^{14|32}
    \\
    &\quad+\frac12\Bigl[
        \mathcal I_\sigma^{14|14}+\mathcal I_\sigma^{32|32}
        -\mathcal I_\sigma^{12|12}-\mathcal I_\sigma^{34|34}
        -\mathcal I_A^{14|14}-\mathcal I_A^{32|32}
        +\mathcal I_A^{12|12}+\mathcal I_A^{34|34}
    \Bigr]
    \\
    &\quad+\mathcal I_{AW}^{12;12}+\mathcal I_{AW}^{34;34}
    -\sum_{rs\in\{14,32\}}
    \left(\mathcal I_{AW}^{rs;12}+\mathcal I_{AW}^{rs;34}\right)
    -\mathcal I_{WW}^{12|34}\, ,
    \\
    \mathscr C
    &=\sum_{rs\in\{12,34\}}\sum_{uv\in\{14,32\}}
    \left(\mathcal I_\sigma^{rs|uv}-\mathcal I_A^{rs|uv}\right)
    \\
    &\quad-\mathcal I_\sigma^{12|12}-\mathcal I_\sigma^{34|34}
    +\mathcal I_A^{12|12}+\mathcal I_A^{34|34}
    +\mathcal I_{AW}^{12;12}+\mathcal I_{AW}^{34;34}
    -\mathcal I_{AW}^{12;34}-\mathcal I_{AW}^{34;12}
    \\
    &\quad-\sum_{rs\in\{14,32\}}
    \left(\mathcal I_{AW}^{rs;12}+\mathcal I_{AW}^{rs;34}\right)
    -2\mathcal I_{WW}^{12|34}\, ,
    \\
    \mathscr D
    &=\Sigma_{14}+\Sigma_{32}-\Sigma_{12}-\Sigma_{34}
    \\
    &\quad-\frac12\Bigl[
        \mathcal I_\sigma^{14|14}+\mathcal I_\sigma^{32|32}
        -\mathcal I_\sigma^{12|12}-\mathcal I_\sigma^{34|34}
        -\mathcal I_A^{14|14}-\mathcal I_A^{32|32}
        +\mathcal I_A^{12|12}+\mathcal I_A^{34|34}
    \Bigr]\, . 
    \end{aligned}
\end{align}
The diagram integrals $\mathcal I$ are defined by
\begin{align}\label{eq:general-q-exchange-integrals-app}
    \begin{aligned}
    \mathcal I_\sigma^{rs|uv}
    &=g_{\sigma\Phi\Phi}^2 \int_{y,y'}p_{rs}(y) D_\sigma(y-y')p_{uv}(y')\, ,
    \\
    \mathcal I_A^{rs|uv}
    &=\int_{y,y'}j_{rs}^{\mu}(y)
    \widehat D^A_{\mu\nu}(y-y')j_{uv}^{\nu}(y')\, ,
    \\
    \mathcal I_{AW}^{rs;uv}
    &=\int\d^3y\int_{C_{uv}}\d x'^{\nu}\,
    j_{rs}^{\mu}(y)\widehat D^A_{\mu\nu}(y-x')\, ,
    \\
    \mathcal I_{WW}^{rs|uv}
    &=\int_{C_{rs}}\d x^{\mu}\int_{C_{uv}}\d x'^{\nu}\,
    \widehat D^A_{\mu\nu}(x-x')\, ,
    \end{aligned}
\end{align}
where 
\begin{align}
    g_{\sigma\Phi\Phi}^2 = \frac{8}{\pi^2}\,.
\end{align}
The scalar self-energy corrections $\Sigma_{rs}$ are defined through the bare scalar propagator,
\begin{align}\label{eq:general-q-scalar-self-energy-definition-app}
    D_\Phi^B(x_r-x_s)
    =
    D_{rs}
    \left[
        1+\frac{\Sigma_{rs}}{N}+O(N^{-2})
    \right]\, .
\end{align}
Here $\int_{y,y'}\equiv\int\d^3y\,\d^3y'$ and $C_{uv}$ is oriented from $x_u$ to $x_v$. The four integrals correspond, in order, to panels~(c)--(f) of figure~\ref{fig:general-q-diagram-classes},
while $\Sigma_{rs}$ includes the scalar self-energy diagrams from $\sigma$ and $A_\mu$ in panels~(a) and~(b).
In particular, $\mathcal I_\sigma^{rs|rs}$ and $\mathcal I_A^{rs|rs}$ describe an exchange between two different scalar propagators with the same endpoints, not a self-energy
correction on a single propagator. 

Each exchange integral is divided by the free scalar propagators that it replaces. Accordingly, the scalar insertions in these definitions are
\begin{align}\label{eq:general-q-normalized-scalar-kernels}
    p_{rs}(y)
    \equiv
    \frac{D_\Phi(x_r-y)D_\Phi(y-x_s)}{D_{rs}}\, , \quad 
    j_{rs}^{\mu}(y)
    \equiv 
    -\frac{D_\Phi(x_r-y)\overleftrightarrow{\partial_{y^\mu}}
    D_\Phi(y-x_s)}{D_{rs}}\, ,
\end{align}
where $f\overleftrightarrow{\partial_\mu}g \equiv f\,\partial_\mu g-(\partial_\mu f)g$. The free scalar propagator and the effective propagators are
\begin{align}
    D_\Phi(x)=\frac{1}{4\pi|x|}\, ,\quad
    D_{rs}=D_\Phi(x_r-x_s)\, ,
\end{align}
\begin{align}\label{eq:general-q-rescaled-effective-propagators}
    \widetilde{D}_\sigma(p)&=-\pi^2|p|\, ,\quad
    \widetilde{\widehat D}^{A}_{\mu\nu}(p)
    =\frac{16}{|p|}
    \left(\delta_{\mu\nu}-\zeta\frac{p_\mu p_\nu}{p^2}\right)\, .
\end{align}
Here $D^A_{\mu\nu}=\widehat D^A_{\mu\nu}/N$, and $\zeta$ is the gauge parameter. Remark that the two-point subtraction terms are already included in~\eqref{eq:general-q-four-master-definitions-app}. In what follows, we give the detailed evaluation of each integral in \eqref{eq:general-q-exchange-integrals-app} and \eqref{eq:general-q-scalar-self-energy-definition-app}. 

\subsubsection{Scalar exchange integrals: $\mathcal I_\sigma$}
\label{app:general-q-kernel-evaluation}

We first derive a parameter representation for $\mathcal I_\sigma^{rs|uv}$ and then evaluate it for the endpoint pairs appearing in \eqref{eq:general-q-four-master-definitions-app}.
We choose the collinear conformal frame
\begin{align}\label{eq:collinear-frame}
    x_r=\tau_r n\, ,\quad
    (\tau_1,\tau_2,\tau_3,\tau_4)=(0,1,h,h+1)\, ,\quad
    h=z^{-1/2}>1\, ,
\end{align}
where $n$ is the unit vector along the line. We write $d_{rs}\equiv\tau_r-\tau_s$ and $L_{rs}\equiv |d_{rs}|$.

With the Fourier convention
$\widetilde f(k)=\int\d^3y\,e^{-\i k\cdot y}f(y)$,
Feynman parameterization of $p_{rs}$ gives
\begin{align}
    \widetilde p_{rs}(k)
    =
    \frac{L_{rs}}{2|k|}
    \int_0^1\frac{\d\alpha}{\sqrt{\alpha(1-\alpha)}}\,
    e^{-\i k\cdot[\alpha x_r+(1-\alpha)x_s]
       -|k|L_{rs}\sqrt{\alpha(1-\alpha)}}\, .
\end{align}
Setting $\alpha=(1-\cos\theta)/2$, we define
\begin{align}\label{eq:Feynman-parameter-angular-variable}
    x_{rs}(\theta)
    \equiv 
    \frac{\tau_r+\tau_s-d_{rs}\cos\theta}{2}\, , \quad  y_{rs}(\theta)\equiv \frac{L_{rs}}{2}\sin\theta\, .
\end{align}
The measure becomes $\d\alpha/\sqrt{\alpha(1-\alpha)}=\d\theta$. Writing $\int_k=\int\d^3k/(2\pi)^3$, the momentum integral is
\begin{align}
    \int_k\frac{e^{-\i(k\cdot n)X-|k|Y}}{|k|}
    =\frac{1}{2\pi^2(X^2+Y^2)}\, ,
    \qquad Y>0\, .
\end{align}
Together with $\widetilde{D}_\sigma(k)=-\pi^2 |k|$, this gives
\begin{align}\label{eq:lambda-exchange-Feynman-parameter}
    \mathcal I_\sigma^{rs|uv}
    =
    -\frac{L_{rs}L_{uv}}{\pi^2}
    \int_0^\pi\d\theta\int_0^\pi\d\phi\,\frac{1}{D}\, ,
\end{align}
where
\begin{align}
    \begin{gathered}
        X_1=x_{rs}(\theta)\, ,\quad
        Y_1=y_{rs}(\theta)\, ,\quad
        X_2=x_{uv}(\phi)\, ,\quad
        Y_2=y_{uv}(\phi)\, , \quad 
        D=(X_1-X_2)^2+(Y_1+Y_2)^2\, .
    \end{gathered}
\end{align}
We regulate the parameter endpoints by imposing
\begin{align}
    y_{rs}(\theta)\ge a_{\rm UV}
    \quad\Longleftrightarrow\quad
    \delta_{rs}\le\theta\le\pi-\delta_{rs}\, ,
    \quad
    \delta_{rs}\equiv \arcsin\frac{2a_{\rm UV}}{L_{rs}}\, .
\end{align}
The same length cutoff $a_{\rm UV}$ is used for every propagator. Unless the limits $0,\pi$ are written explicitly, angular integrals below carry these regulated limits. Remark that divergent integrals are combined before taking $a_{\rm UV}\to0$.

\paragraph{Four distinct endpoints.}
For $(r,s,u,v)=(1,2,3,4)$ in $\mathcal I_\sigma^{rs|uv}$, the integral \eqref{eq:lambda-exchange-Feynman-parameter} is finite and the cutoff can be removed. Equation~\eqref{eq:lambda-exchange-Feynman-parameter} becomes
\begin{align}
    \begin{aligned}
    \mathcal I_\sigma^{12|34}
    &=
    -\frac{1}{\pi^2}
    \int_0^\pi\d\theta\int_0^\pi\d\phi\,
    \frac{1}{
        \left(h+\frac{-\cos\theta+\cos\phi}{2}\right)^2
        +\left(\frac{\sin\theta+\sin\phi}{2}\right)^2
    }
    \\
    &=
    -\frac{2}{\pi^2h}
    \int_0^\pi\frac{\d\theta}{h-\cos\theta}
    \left[
        \frac{\pi}{2}
        -\arctan\frac{\sin\theta}{2h(h-\cos\theta)}
    \right]
    \\
    &=
    -\frac{2}{\pi^2h\sqrt{h^2-1}}
    \int_{-1}^1\frac{\d u}{\sqrt{1-u^2}}
    \left[
        \frac{\pi}{2}
        -\arctan\frac{\sqrt{1-u^2}}{2h\sqrt{h^2-1}}
    \right]\, .
    \end{aligned}
\end{align}
The second line follows by integrating over $\phi$.
In the last line we used
\begin{align}
    u=\frac{-1+h\cos\theta}{h-\cos\theta}\, ,
    \qquad
    \frac{\sin\theta}{h-\cos\theta}
    =\frac{\sqrt{1-u^2}}{\sqrt{h^2-1}}\, .
\end{align}
To evaluate the remaining integral, we set $t=\sqrt{1-z}=\sqrt{h^2-1}/h$ and define
\begin{align}
    f(t)
    =
    \int_{-1}^1\frac{\d u}{\sqrt{1-u^2}}
    \left[
        \frac{\pi}{2}
        -\arctan\left(
            \frac{1-t^2}{2t}\sqrt{1-u^2}
        \right)
    \right]\, .
\end{align}
Differentiating under the integral gives
\begin{align}
    f'(t)
    &=
    \frac{1+t^2}{2t^2}
    \int_{-1}^1
    \frac{\d u}{
        1+\dfrac{(1-t^2)^2}{4t^2}(1-u^2)
    }
    =
    -\frac{4\log t}{1-t^2}\, .
\end{align}
Since $f(0)=0$, we obtain
\begin{align}
    f(t)
    =-4\int_0^t\frac{\log v}{1-v^2}\,\d v
    =2\bigl[\Li_2(t)-\Li_2(-t)\bigr]
      -4\log t\,\operatorname{arctanh}t\, .
\end{align}
Consequently,
\begin{align}\label{eq:direct-paring-lambda-exchange}
    \mathcal I_\sigma^{12|34}
    =
    -\frac{2}{\pi^2h\sqrt{h^2-1}}\,
    f\!\left(\sqrt{1-z}\right)
    =
    \Lambda(z)\, ,
\end{align}
with $\Lambda$ defined in
\eqref{eq:general-q-Lambda-main}.
Interchanging $2$ and $4$ sends $z$ to $1-z$, giving the other
integral with four distinct endpoints,
\begin{align}\label{eq:general-q-crossed-lambda-exchange-app}
    \mathcal I_\sigma^{14|32}=\Lambda(1-z)\, .
\end{align}

\paragraph{Two propagators with the same endpoints.}
For $r=u$ and $s=v$, the denominator simplifies to
\begin{align}
    D=L_{rs}^2\sin^2\frac{\theta+\phi}{2}\, .
\end{align}
The regulated integral is therefore
\begin{align}\label{eq:diagonal-scalar-integral}
    \begin{aligned}
    \mathcal I_\sigma^{rs|rs}
    &=
    -\frac{1}{\pi^2}
    \int_{\delta_{rs}}^{\pi-\delta_{rs}}\d\theta
    \int_{\delta_{rs}}^{\pi-\delta_{rs}}\d\phi\,
    \frac{1}{\sin^2\frac{\theta+\phi}{2}}
    \\
    &=
    \frac{8}{\pi^2}\log\sin\delta_{rs}
    =
    -\frac{8}{\pi^2}\log\frac{L_{rs}}{2a_{\rm UV}}\, .
    \end{aligned}
\end{align}
In particular, the cutoff cancels in the difference
\begin{align}\label{eq:general-q-diagonal-lambda-difference-app}
    \mathcal I_\sigma^{14|14}
     +\mathcal I_\sigma^{32|32}
     -\mathcal I_\sigma^{12|12}
     -\mathcal I_\sigma^{34|34}
    =-\frac{8}{\pi^2}
      \log\frac{L_{14}L_{32}}{L_{12}L_{34}}
    =\frac{8}{\pi^2}\log r(z)\, .
\end{align}

\paragraph{Two propagators ending at the same point.}
\begin{figure}[t]
\centering
\begin{tikzpicture}[
    scalar/.style={thick},
    hs/.style={thick,densely dashed},
    wilson/.style={
        double,double distance=1.1pt,line width=0.45pt
    },
    endpoint/.style={circle,fill=black,inner sep=1.8pt},
    vertex/.style={circle,fill=black,inner sep=1.4pt},
    every node/.style={font=\small}
]
\foreach \panelshift/\vertexpos/\othervertexpos in {
    0/0.50/0.64,
    6.2/0.85/0.92
}{
    \begin{scope}[xshift=\panelshift cm]
        \coordinate (x1) at (0,0);
        \coordinate (x2) at (1.2,0);
        \coordinate (x3) at (2.7,0);
        \coordinate (x4) at (3.9,0);

        \draw[wilson] (x2)--(x1);
        \draw[wilson] (x4)--(x3);

        \draw[scalar]
            (x2) .. controls (1.1,1.0) and (0.1,1.0) ..
            node[pos=\vertexpos,coordinate] (w) {}
            (x1);

        \draw[scalar]
            (x4) .. controls (3.6,3.0) and (0.3,3.0) ..
            node[pos=\othervertexpos,coordinate] (wp) {}
            (x1);

        \draw[hs] (w)--(wp);
        \node[vertex] at (w) {};
        \node[vertex] at (wp) {};
        \node[right=1pt] at (w) {$w$};
        \node[left=2pt] at (wp) {$w'$};

        \foreach \i in {1,2,3,4}{
            \node[endpoint] at (x\i) {};
            \node[below=5pt] at (x\i) {$x_{\i}$};
        }
    \end{scope}
}

\draw[->,thick] (4.35,1.15)--(5.70,1.15)
    node[midway,above=6pt] {$w,w'\to x_1$};

\draw[gray,densely dotted]
    (6.2,0) circle[radius=0.80];
\end{tikzpicture}
\caption{The integration region responsible for the logarithmic
divergence of $\mathcal I_\sigma^{12|14}$.
The dashed $\sigma$ propagator joins an interaction vertex $w$
on the scalar propagator $(12)$ to a vertex $w'$ on $(14)$.
On the right, both vertices approach $x_1$, while all external
endpoints remain fixed.  Double lines denote the physical Wilson
segments $C_{12}$ and $C_{34}$; the other scalar propagators are
omitted.  The curves indicate propagator connectivity, not particle
trajectories.}
\label{fig:general-q-lambda-endpoint-limit}
\end{figure}
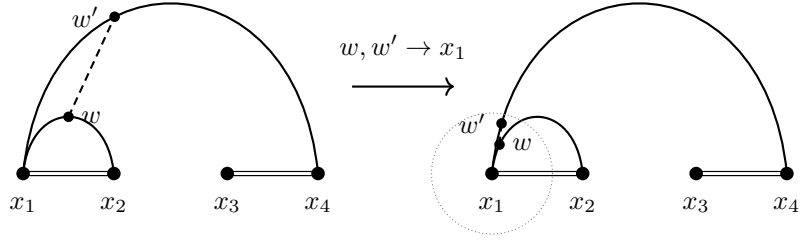
Finally, we consider $\mathcal I_\sigma^{12|14}$. Both interaction vertices can approach $x_1$, producing the logarithmic divergence illustrated in figure~\ref{fig:general-q-lambda-endpoint-limit}. Since $p_{rs}=p_{sr}$, we may write $\mathcal I_\sigma^{12|14}=\mathcal I_\sigma^{21|41}$ and choose both angular parameters to start at $x_1$. Introducing
\begin{align}
    \xi=(h+1)\cot\frac{\theta}{2}\, ,
    \quad
    \eta=\cot\frac{\phi}{2}\, ,
\end{align}
reduces the integral to
\begin{align}
    \mathcal I_\sigma^{12|14}
    =
    -\frac{4}{\pi^2}
    \int_0^M\d\xi\int_0^M\d\eta\,
    \frac{1}{(\xi+\eta)^2+h^2}
    +o(1)\, ,
    \quad M=\frac{h+1}{a_{\rm UV}}\, .
\end{align}
Setting $v=\xi+\eta$ gives
\begin{align}
    \begin{aligned}
    \int_0^M\d\xi\int_0^M\d\eta\,
    \frac{1}{(\xi+\eta)^2+h^2}
    &=
    \int_0^M\frac{v\,\d v}{v^2+h^2}
    +\int_M^{2M}\frac{(2M-v)\,\d v}{v^2+h^2}
    \\
    &=\log\frac{M}{2h}+1+o(1)\, .
    \end{aligned}
\end{align}
The same calculation for the remaining endpoint pairs yields
\begin{align}\label{eq:scalar-integral-one-sahred-endpoint}
    \begin{aligned}
    \mathcal I_\sigma^{12|14}
    =\mathcal I_\sigma^{34|14}
    &=
    -\frac{4}{\pi^2}
    \left[
        \log\frac{h+1}{2ha_{\rm UV}}+1
    \right]+o(1)\, ,
    \\
    \mathcal I_\sigma^{12|32}
    =\mathcal I_\sigma^{34|32}
    &=
    -\frac{4}{\pi^2}
    \left[
        \log\frac{h-1}{2ha_{\rm UV}}+1
    \right]+o(1)\, .
    \end{aligned}
\end{align}
Together with~\eqref{eq:diagonal-scalar-integral}, these expressions give
\begin{align}\label{eq:general-q-lambda-mixed-combination-app}
    &\lim_{a_{\rm UV}\to0}
    \left[
        \sum_{rs\in\{12,34\}}\sum_{uv\in\{14,32\}}
        \mathcal I_\sigma^{rs|uv}
        -\mathcal I_\sigma^{12|12}
        -\mathcal I_\sigma^{34|34}
    \right]
    =-\frac{16}{\pi^2}
     -\frac{8}{\pi^2}\log(1-z)\, .
\end{align}
This completes the scalar exchange integrals required in~\eqref{eq:general-q-four-master-definitions-app}.

\subsubsection{Photon exchange integrals: $\mathcal I_A, \mathcal I_{AW}, \mathcal I_{WW}$}
We use the gauge $\zeta=0$, in which the photon propagator is
\begin{align}
    \widehat D^A_{\mu\nu}(k)
    =\frac{16}{|k|}\delta_{\mu\nu}\, ,
    \quad
    \widehat D^A_{\mu\nu}(x)
    =\frac{8}{\pi^2|x|^2}\delta_{\mu\nu}\, .
\end{align}
We keep the angular variables and the cutoff prescription of subsection~\ref{app:general-q-kernel-evaluation}. As there, $X_1,Y_1$ refer to the propagator $(rs)$ and $X_2,Y_2$ to $(uv)$. Primes denote differentiation with respect to their respective angular variables, $\theta$ and $\phi$.

\paragraph{Scalar--scalar and scalar--Wilson exchanges.}
The Fourier transform of the scalar current is
\begin{align}
    \widetilde j_{rs}^{\mu}(k)
    =
    \int\d\theta\,
    e^{-\i\omega X_1-|k|Y_1}
    \left[
        X_1'n^\mu
        -\i Y_1'\frac{k^\mu}{|k|}
    \right]\, ,
    \quad \omega=k\cdot n\, .
\end{align}
In addition to the momentum integral used in the previous subsection, we need
\begin{align}
    \int_k\frac{e^{-\i\omega X-|k|Y}}{|k|^2}
    =\frac{g(X,Y)}{2\pi^2}\, ,
    \quad
    g(X,Y)=\frac{1}{X}\arctan\frac{X}{Y}\, ,
    \quad
    g(0,Y)=\frac{1}{Y}\, .
\end{align}
Here $Y>0$, and
$\partial_Yg(X,Y)=-1/(X^2+Y^2)$.
Contracting the two scalar currents gives
\begin{align}\label{eq:scalar-scalar-exchange}
    \begin{aligned}
    \mathcal I_A^{rs|uv}
    &=
    \frac{8}{\pi^2}\int\d\theta\,\d\phi\,
    \Bigg[
        \frac{X_1'X_2'+Y_1'Y_2'}{D}
        +(Y_1'X_2'-X_1'Y_2')
        \partial_Xg(X_1-X_2,Y_1+Y_2)
    \Bigg]
    \\
    &=
    \frac{8}{\pi^2}\Bigg[
        \int\d\theta\,\d\phi\,
        \frac{X_1'X_2'-Y_1'Y_2'}{D}
        -\int\d\theta\,Y_1'
        \Big[g(X_1-X_2,Y_1+Y_2)\Big]
        _{\phi=\delta_{uv}}^{\phi=\pi-\delta_{uv}}
        \\
    &\qquad\qquad 
        -\int\d\phi\,Y_2'
        \Big[g(X_1-X_2,Y_1+Y_2)\Big]
        _{\theta=\delta_{rs}}^{\theta=\pi-\delta_{rs}}
    \Bigg]\, .
    \end{aligned}
\end{align}
The second equality follows by integrating by parts, using
\begin{align}
    &(Y_1'X_2'-X_1'Y_2')\partial_Xg
    =-Y_1'\partial_\phi g-Y_2'\partial_\theta g
      -\frac{2Y_1'Y_2'}{D}\, .
\end{align}
The boundary terms in \eqref{eq:scalar-scalar-exchange} must be kept at finite cutoff.

For an attachment to the oriented straight path $C_{uv}$,
\begin{align}
    \int_{C_{uv}}\d x^\nu\,e^{-\i k\cdot x}
    =n^\nu\int_{\tau_u}^{\tau_v}\d t\,e^{-\i\omega t}\, .
\end{align}
Contracting this with the scalar current gives
\begin{align}\label{eq:scalar-Wilson-exchange}
    \begin{aligned}
    \mathcal I_{AW}^{rs;uv}
    =
    \frac{8}{\pi^2}\int\d\theta\,
    \Bigg[
        &X_1'\int_{\tau_u}^{\tau_v}
        \frac{\d t}{(X_1-t)^2+Y_1^2}
    +Y_1'
        \bigl\{
            g(X_1-\tau_u,Y_1)
            -g(X_1-\tau_v,Y_1)
        \bigr\}
    \Bigg]\, .
    \end{aligned}
\end{align}
\paragraph{Combining the boundary terms.}
The two scalar--Wilson attachments contain the endpoint terms
needed to combine with
\eqref{eq:scalar-scalar-exchange}. It is therefore convenient
to calculate the sum
\begin{align}\label{eq:general-q-photon-P-definition-app}
    \mathcal P^{rs|uv}
    \equiv
    \mathcal I_A^{rs|uv}
    +\mathcal I_{AW}^{rs;uv}
    +\mathcal I_{AW}^{uv;rs}\, .
\end{align}
The paths in this auxiliary expression are straight paths
between the indicated endpoints; they do not introduce additional
physical Wilson segments.
The straight-line identity
\begin{align}\label{eq:integral-path-identity}
    \int_{C_{14}}+\int_{C_{32}}
    =
    \int_{C_{12}}+\int_{C_{34}}
\end{align}
allows the scalar--Wilson terms in
\eqref{eq:general-q-four-master-definitions-app} to be
rewritten in terms of these sums.

We define $F\equiv \arctan[(X_1-X_2)/(Y_1+Y_2)]$.
Substitution of \eqref{eq:Feynman-parameter-angular-variable} gives
\begin{align}
    \begin{aligned}
    \frac{X_1'X_2'-Y_1'Y_2'}{D}
    =
    \frac{d_{uv}}{L_{uv}}\partial_\theta F
    -\frac{d_{rs}}{L_{rs}}\partial_\phi F
    -\frac{d_{rs}d_{uv}}{2L_{rs}L_{uv}}
    -\frac{d_{rs}d_{uv}}{8L_{rs}L_{uv}}\,
    \frac{
        L_{rs}^2+L_{uv}^2
        -(\tau_r+\tau_s-\tau_u-\tau_v)^2
    }{D}\, .
    \end{aligned}
\end{align}
The first total derivative gives
\begin{align}
    \frac{d_{uv}}{L_{uv}}
    \int\d\phi\int\d\theta\,\partial_\theta F
    =
    -\int\d\phi\,X_2'
    \int_{\tau_r}^{\tau_s}
    \frac{\d t}{(X_2-t)^2+Y_2^2}
    +o(1)\, .
\end{align}
This cancels the first term of $\mathcal I_{AW}^{uv;rs}$ in
\eqref{eq:scalar-Wilson-exchange}. The $\partial_\phi F$ term cancels the corresponding term of
$\mathcal I_{AW}^{rs;uv}$.

The $g$-terms require more care. At finite cutoff, the scalar boundaries lie at $Y=a_{\rm UV}$, whereas the Wilson endpoints lie at $Y=0$. Their difference is

\begin{align}
    \begin{aligned}
        &\frac{8}{\pi^2}\Bigg[\int\d\theta\,Y_1'\bigg\{g(X_1-\tau_u,Y_1)-g(X_1-\tau_v,Y_1) \\
        &\hspace{3cm} -\Big[g(X_1-X_2,Y_1+Y_2)\Big]_{\phi=\delta_{uv}}^{\phi=\pi-\delta_{uv}}\bigg\} + (rs \leftrightarrow uv)\Bigg] \\
        &\qquad = c_{\rm end}N_{rs|uv} + o(1)\,.
    \end{aligned}
\end{align}
\normalsize
In the term denoted by $(rs\leftrightarrow uv)$, the variables $(X_1,Y_1,\theta)$ and $(X_2,Y_2,\phi)$ are also interchanged. Here $N_{rs|uv}$ counts the shared endpoints, and $c_{\rm end}$ is independent of the interval lengths and includes both boundary contributions at each shared endpoint. The value of $c_{\rm end}$ is unnecessary, since these constants cancel in the combinations below. Combining these results and using~\eqref{eq:lambda-exchange-Feynman-parameter} to express the remaining double integral in terms of $\mathcal{I}_{\sigma}^{rs|uv}$, we obtain
\begin{align}\label{eq:photon-lambda-relation}
    \mathcal{P}^{rs|uv} = \frac{L_{rs}^2+L_{uv}^2-(\tau_r+\tau_s-\tau_u-\tau_v)^2}{d_{rs}d_{uv}}\mathcal{I}_{\sigma}^{rs|uv} - 4\frac{d_{rs}d_{uv}}{L_{rs}L_{uv}} + c_{\rm end}N_{rs|uv} + o(1)\,.
\end{align}
In what follows, we will evaluate this quantity for each $(r,s,u,v)$. 

\begin{itemize}
    \item $(r,s,u,v)=(1,2,3,4).$ \\
    Here $N_{12|34}=0$, $d_{12}d_{34}=L_{12}L_{34}=1$, and $(\tau_1+\tau_2-\tau_3-\tau_4)^2=4h^2$. Thus~\eqref{eq:photon-lambda-relation} gives
\begin{align}\label{eq:general-q-direct-photon-sum-app}
    \mathcal P^{12|34}
    =
    -\frac{2(2-z)}{z}\Lambda(z)-4\, .
\end{align}
For the same endpoint pairs, the Wilson--Wilson integral is
\begin{align}\label{eq:Wilson-Wilson-integral}
    \begin{aligned}
    \mathcal I_{WW}^{12|34}
    &=
    \frac{8}{\pi^2}
    \int_{\tau_1}^{\tau_2}\d t
    \int_{\tau_3}^{\tau_4}\d t'\,
    \frac{1}{(t-t')^2}
    \\
    &=
    \frac{8}{\pi^2}
    \int_0^1\d t\int_h^{h+1}\d t'\,
    \frac{1}{(t-t')^2}
    \\
    &=
    \frac{8}{\pi^2}\log\frac{h^2}{h^2-1}
    =
    -\frac{8}{\pi^2}\log(1-z)\, .
    \end{aligned}
\end{align}
Exchanges within a single physical Wilson segment cancel against the two-point normalization.

\item $(r,s,u,v)=(1,4,3,2).$ \\
Again the four endpoints are distinct, so $N_{14|32}=0$. Now $d_{14}d_{32}=-(h^2-1)$ and
$\tau_1+\tau_4-\tau_3-\tau_2=0$. Consequently,
\begin{align}\label{eq:general-q-crossed-photon-sum-app}
    \mathcal P^{14|32}
    =
    4-\frac{2(1+z)}{1-z}\Lambda(1-z)\, .
\end{align}
\item $(u,v)=(r,s)$. \\
This case includes $12|12$, $34|34$, $14|14$, and $32|32$. For these values, $\mathcal P^{rs|rs}$ becomes
\begin{align}\label{eq:general-q-photon-P-diagonal-app}
    \begin{aligned}
    \mathcal P^{rs|rs}
    &=
    2\mathcal I_\sigma^{rs|rs}-4+2c_{\rm end}+o(1)
    \\
    &=
    -\frac{16}{\pi^2}
    \log\frac{L_{rs}}{2a_{\rm UV}}
    -4+2c_{\rm end}+o(1)\, ,
    \end{aligned}
\end{align}
where $L_{12}=L_{34}=1$, $L_{14}=h+1$, and $L_{32}=h-1$.

The separate scalar--Wilson integral can also be evaluated
directly. Translation and reflection allow us to take
$(\tau_s,\tau_r)=(0,L_{rs})$. Then $X_1'=Y_1$ and
\begin{align}
    \int_0^{L_{rs}}
    \frac{\d t}{(X_1-t)^2+Y_1^2}
    =\frac{\pi}{2Y_1}\, .
\end{align}
Substitution into
\eqref{eq:scalar-Wilson-exchange} gives
\begin{align}
    \begin{aligned}
    \mathcal I_{AW}^{rs;rs}
    &=
    -\frac{2}{\pi^2}
    \int_{\delta_{rs}}^{\pi-\delta_{rs}}\d\theta\,
    \left[
        (\pi-\theta)\sec^2\frac{\theta}{2}
        +\theta\csc^2\frac{\theta}{2}
    \right]
    \\
    &=
    -\frac{4}{\pi^2}
    \left[
        -2\theta\cot\frac{\theta}{2}
        +4\log\sin\frac{\theta}{2}
    \right]_{\delta_{rs}}^{\pi-\delta_{rs}}
    \\
    &=
    -\frac{16}{\pi^2}
    \left[\log\frac{L_{rs}}{a_{\rm UV}}+1\right]
    +o(1)\, .
    \end{aligned}
\end{align}
Using $\mathcal I_A^{rs|rs} =\mathcal P^{rs|rs}-2\mathcal I_{AW}^{rs;rs}$, we therefore find
\begin{align}
    \mathcal I_A^{rs|rs}
    =
    \frac{16}{\pi^2}\log\frac{L_{rs}}{a_{\rm UV}}
    +\text{constant}+o(1)\, .
\end{align}
The constant includes $c_{\rm end}$ but is independent of the endpoint pair. In particular,
\begin{align}
    \mathcal I_\sigma^{rs|rs}-\mathcal I_A^{rs|rs}
    =
    -\frac{24}{\pi^2}\log\frac{L_{rs}}{a_{\rm UV}}
    +\text{constant}+o(1)\, .
\end{align}

\item $(r,s,u,v)=(1,2,1,4)$ or $(3,4,1,4)$. \\
Each pair has one coincident endpoint. Using the integral \eqref{eq:scalar-integral-one-sahred-endpoint} gives
\begin{align}
    \mathcal P^{12|14}
    =\mathcal P^{34|14}
    =
    -\frac{8}{\pi^2}
    \left[
        \log\frac{h+1}{2ha_{\rm UV}}+1
    \right]
    -4+c_{\rm end}+o(1)\, .
\end{align}

\item $(r,s,u,v)=(1,2,3,2)$ or $(3,4,3,2)$. \\
Each pair again has one coincident endpoint, and 
\begin{align}
    \mathcal P^{12|32}
    =\mathcal P^{34|32}
    =
    -\frac{8}{\pi^2}
    \left[
        \log\frac{h-1}{2ha_{\rm UV}}+1
    \right]
    +4+c_{\rm end}+o(1)\, .
\end{align}

\end{itemize}
Combining these results, the cutoff and $c_{\rm end}$ cancel in the combinations entering
\eqref{eq:general-q-four-master-definitions-app}:
\begin{align}
    &\lim_{a_{\rm UV}\to0}
    \bigl[
        \mathcal P^{14|14}+\mathcal P^{32|32}
        -\mathcal P^{12|12}-\mathcal P^{34|34}
    \bigr]
    =\frac{16}{\pi^2}\log r(z)\, ,
    \label{eq:general-q-photon-diagonal-difference-app}
    \\
    &\lim_{a_{\rm UV}\to0}
    \Bigg[
        -\sum_{rs\in\{12,34\}}\sum_{uv\in\{14,32\}}
        \mathcal P^{rs|uv}
        +\mathcal P^{12|12}+\mathcal P^{34|34}
    \Bigg]
    =-8+\frac{32}{\pi^2}
      +\frac{16}{\pi^2}\log(1-z)\, .
    \label{eq:general-q-photon-mixed-combination-app}
\end{align}
Similarly, we also have
\begin{align}\label{eq:general-q-diagonal-exchange-difference-app}
    \begin{aligned}
    &\lim_{a_{\rm UV}\to0}
    \Bigl[
        \mathcal I_\sigma^{14|14}+\mathcal I_\sigma^{32|32}
        -\mathcal I_\sigma^{12|12}-\mathcal I_\sigma^{34|34}
        \\[-1mm]
    &\hspace{24mm}
        -\mathcal I_A^{14|14}-\mathcal I_A^{32|32}
        +\mathcal I_A^{12|12}+\mathcal I_A^{34|34}
    \Bigr]
    =\frac{24}{\pi^2}\log r(z)\, .
    \end{aligned}
\end{align}

\subsubsection{Scalar self-energy: $\Sigma$}
It remains to evaluate the self-energy diagrams in panels~(a) and~(b) of
figure~\ref{fig:general-q-diagram-classes}. We write $\Sigma_{rs}=\Sigma^\sigma_{rs}+\Sigma^A_{rs}$ and continue to use the gauge $\zeta=0$.\footnote{Momentum-independent terms, including the seagull tadpole, are removed by tuning the scalar mass to criticality.} The two diagrams give
\begin{align}
    \begin{aligned}
    D_{rs}\Sigma^\sigma_{rs}
    &=
    -8\int_p
    \frac{e^{\i p\cdot(x_r-x_s)}}{(p^2)^2}
    \int_k
    \left[
        \frac{|k|}{(p-k)^2}-\frac{1}{|k|}
    \right]\, ,
    \\
    D_{rs}\Sigma^A_{rs}
    &=
    16\int_p
    \frac{e^{\i p\cdot(x_r-x_s)}}{(p^2)^2}
    \int_k
    \left[
        \frac{(2p-k)^2}{|k|(p-k)^2}
        -\frac{1}{|k|}
    \right]\, ,
    \end{aligned}
\end{align}
where $\int_p\equiv \int\d^3p/(2\pi)^3$. Only the ultraviolet logarithms are needed, since all
pair-independent constants cancel in $\Sigma_{14}+\Sigma_{32}-\Sigma_{12}-\Sigma_{34}$.
For $|k|\gg|p|$, angular integration gives
\begin{align}
    \frac{1}{4\pi}\int\d\Omega_k\,
    \frac{|k|}{(p-k)^2}
    &=
    \frac{1}{|k|}
    +\frac{p^2}{3|k|^3}
    +O\!\left(\frac{|p|^4}{|k|^5}\right)\, ,
    \\
    \frac{1}{4\pi}\int\d\Omega_k\,
    \frac{(2p-k)^2}{|k|(p-k)^2}
    &=
    \frac{1}{|k|}
    +\frac{5p^2}{3|k|^3}
    +O\!\left(\frac{|p|^4}{|k|^5}\right)\, .
\end{align}
The terms proportional to $1/|k|^3$ produce
\begin{align}
    \int_{|p|<|k|<\Lambda_{\rm UV}}
    \frac{\d^3k}{(2\pi)^3|k|^3}
    =
    \frac{1}{2\pi^2}
    \log\frac{\Lambda_{\rm UV}}{|p|}\, .
\end{align}
Fourier transformation then uses
\begin{align}
    \int_p
    \frac{e^{\i p\cdot x}}{p^2}
    \log\frac{\Lambda_{\rm UV}}{|p|}
    =
    D_\Phi(x)
    \bigl[
        \log(\Lambda_{\rm UV}|x|)
        +\text{constant}
    \bigr]\, ,
    \quad x\neq0\, .
\end{align}
Identifying $\Lambda_{\rm UV}$ with $a_{\rm UV}^{-1}$ up to a finite rescaling, which changes only the constants, we obtain
\begin{align}
    \Sigma^\sigma_{rs}
    &=
    -\frac{4}{3\pi^2}
    \log\frac{L_{rs}}{a_{\rm UV}}
    +\text{constant}+o(1)\, ,
    \\
    \Sigma^A_{rs}
    &=
    \frac{40}{3\pi^2}
    \log\frac{L_{rs}}{a_{\rm UV}}
    +\text{constant}+o(1)\, .
\end{align}
Each constant is independent of the endpoint pair. Their sum therefore gives
\begin{align}\label{eq:general-q-self-energy-length-app}
    \Sigma_{rs}
    =
    \frac{12}{\pi^2}
    \log\frac{L_{rs}}{a_{\rm UV}}
    +\text{constant}+o(1)\, .
\end{align}
Using $L_{12}=L_{34}=1$, $L_{14}=h+1$, and $L_{32}=h-1$, we find
\begin{align}\label{eq:general-q-self-energy-difference-app}
    \Sigma_{14}+\Sigma_{32}-\Sigma_{12}-\Sigma_{34}
    =
    \frac{12}{\pi^2}
    \log\frac{L_{14}L_{32}}{L_{12}L_{34}}
    =
    -\frac{12}{\pi^2}\log r(z)\, .
\end{align}
Both the cutoff and the additive constant have canceled as expected.

All combinations of integrals entering \eqref{eq:general-q-four-master-definitions-app}
have now been evaluated. Using \eqref{eq:general-q-photon-P-definition-app} and \eqref{eq:integral-path-identity} to reorganize the photon terms,
we substitute the scalar exchange results
\eqref{eq:direct-paring-lambda-exchange},
\eqref{eq:general-q-crossed-lambda-exchange-app},
\eqref{eq:general-q-diagonal-lambda-difference-app}, and
\eqref{eq:general-q-lambda-mixed-combination-app},
the photon exchange results
\eqref{eq:general-q-direct-photon-sum-app},
\eqref{eq:general-q-crossed-photon-sum-app},
\eqref{eq:general-q-photon-diagonal-difference-app},
\eqref{eq:general-q-photon-mixed-combination-app},
\eqref{eq:general-q-diagonal-exchange-difference-app}, and
\eqref{eq:Wilson-Wilson-integral},
together with the self-energy difference
\eqref{eq:general-q-self-energy-difference-app},
into \eqref{eq:general-q-four-master-definitions-app}.
This reproduces the four functions given in \eqref{eq:general-q-A-master-function-main} and
\eqref{eq:general-q-BCD-master-functions-main}.

\subsection{Evaluation of mixed correlator $\langle \bar E E \sigma \sigma \rangle$}
\label{app:mixed_correlator}

We next evaluate the two diagrams in figure~\ref{fig:general-q-mixed-diagrams}. The propagator of the Hubbard--Stratonovich field is given in equation~\eqref{eq:general-q-mixed-sigma-definition}, but it should be understood as a distribution rather than an ordinary function. We define this distribution through the differential representation
\begin{align}
    D_{\sigma}(x) = \frac{1}{2}\partial^2\frac{1}{|x|^2}\,.
\end{align}
Equivalently, for a smooth function $f(X)$,
\begin{align}\label{eq:regularization_finite_part}
    \operatorname{FP}\int\d^3x\frac{f(x)}{|x|^4} = \lim_{\epsilon\to0}\left[\int_{|x|>\epsilon}\d^3x\frac{f(x)}{|x|^4} - \frac{4\pi}{\epsilon}f(0)\right]\,.
\end{align}
In this prescription, the analytically continued star-triangle identity is given by
\begin{align}\label{eq:star-triangle-identity}
    \operatorname{FP}\int\frac{\d^3y}{|y-x_a||y-x_b||y-u|^4} = -2\pi\frac{L_{ab}}{|x_a-u|^2|x_b-u|^2}\,.
\end{align}

\paragraph{Distinct scalar propagators.}
We first consider the case in which the $\sigma$ propagators attach two distinct scalar propagators in panel~(b) of figure~\ref{fig:general-q-mixed-diagrams}. Using~\eqref{eq:star-triangle-identity}, the corresponding $\sigma$ insertions are given by
\begin{align}
    \frac{1}{D_{\Phi}(x_1-x_2)}\operatorname{FP}\int\d^3y\,D_{\Phi}(x_1-y)D_{\Phi}(x_2-y)D_{\sigma}(x_i-y) = -\frac{L_{12}^2}{2L_{1i}^2L_{2i}^2}, \quad i=3,4\,.
\end{align}
There are $q(q-1)$ ordered choices of distinct propagators. Dividing by $D_{\sigma}(x_3-x_4)$, we obtain
\begin{align}
    \left.\cM_q^{(1)}(z)\right|_{\rm distinct} = \frac{2q(q-1)}{\pi^2}\frac{L_{12}^4L_{34}^4}{L_{13}^2L_{23}^2L_{14}^2L_{24}^2} = \frac{2q(q-1)}{\pi^2}\frac{z^4}{(1-z)^2}\,.
\end{align}

\paragraph{The same scalar propagator.}
We next evaluate the diagram in panel~(a) of figure~\ref{fig:general-q-mixed-diagrams}. Since both $\sigma$ propagators attach to the same scalar line, there are $q$ possible choices. Fixing one such scalar line and denoting the position of the vertices by $y,u$, we obtain
\begin{align}
    \begin{aligned}
        \cM_1^{(1)} &= \frac{g_{\sigma\Phi\Phi}^2}{D_{\Phi}(x_1-x_2)D_{\sigma}(x_3-x_4)}\operatorname{FP}\int\d^3y\,\d^3u \\
        &\qquad\qquad \left[D_{\Phi}(x_1-y)D_{\Phi}(y-u)D_{\Phi}(u-x_2)D_{\sigma}(y-x_3)D_{\sigma}(u-x_4) + (x_1 \leftrightarrow x_2)\right]\,.
    \end{aligned}
\end{align}
Performing the integration over $y$ using equation~\eqref{eq:star-triangle-identity},
\begin{align}
    \cM_1^{(1)} = -\frac{L_{12}L_{34}^4}{\pi^3}\operatorname{FP}\int\frac{\d^3u}{|u-x_1||u-x_2||u-x_3|^2|u-x_4|^4}\left(\frac{|u-x_1|^2}{L_{13}^2} + \frac{|u-x_2|^2}{L_{23}^2}\right)\,.
\end{align}
We again evaluate this integral in the collinear conformal frame~\eqref{eq:collinear-frame}. Since for $i=1,2$,
\begin{align}
    |u-x_i|^2 = \frac{L_{i4}}{L_{34}}|u-x_3|^2 - \frac{L_{i3}}{L_{34}}|u-x_4|^2 + L_{i3}L_{i4}\,,
\end{align}
we obtain
\begin{align}
    \begin{aligned}
        &\frac{|u-x_1|^2}{L_{13}^2} + \frac{|u-x_2|^2}{L_{23}^2} \\
        &\qquad = \left(\frac{h+1}{h^2}+\frac{h}{(h-1)^2}\right)|u-x_3|^2
        - \left(\frac{1}{h}+\frac{1}{h-1}\right)|u-x_4|^2
        + \frac{h+1}{h} + \frac{h}{h-1}\,.
    \end{aligned}
\end{align}
Therefore, $\cM_1^{(1)}$ can be written as
\begin{align}
    \cM_1^{(1)} 
    = -\frac{1}{\pi^3}\left(\frac{h+1}{h^2}+\frac{h}{(h-1)^2}\right)J_1
    +\frac{1}{\pi^3}\left(\frac{1}{h}+\frac{1}{h-1}\right)J_2
    -\frac{1}{\pi^3}\left(\frac{h+1}{h} + \frac{h}{h-1}\right)J_3\,,
\end{align}
where
\begin{align}\label{eq:general-position-integral}
    \begin{aligned}
        J_1 &= \operatorname{FP}\int\frac{\d^3u}{|u-x_1||u-x_2||u-x_4|^4}\,, \\
        J_2 &= \int\frac{\d^3u}{|u-x_1||u-x_2||u-x_3|^2|u-x_4|^2}\,, \\
        J_3 &= \operatorname{FP}\int\frac{\d^3u}{|u-x_1||u-x_2||u-x_3|^2|u-x_4|^4}\,.
    \end{aligned}
\end{align}
The first integral can be evaluated by using~\eqref{eq:star-triangle-identity},
\begin{align}
    J_1 &= -2\pi\frac{L_{12}}{L_{14}^2L_{24}^2} = -\frac{2\pi}{h^2(h+1)^2}\,.
\end{align}
The second integral is related to the $\mathcal{I}_{\sigma}^{12|34}$. Evaluating $\mathcal{I}_{\sigma}^{12|34}$ in position space, we obtain
\begin{align}
    \begin{aligned}
        \mathcal{I}_{\sigma}^{12|34}
        &= \frac{L_{12}L_{34}}{2\pi^4}\int\frac{\d^3y\,\d^3u}{|x_1-u||x_2-u||x_3-y||x_4-y||y-u|^4} \\
        &= -\frac{L_{12}L_{34}^2}{\pi^3}\int\frac{\d^3u}{|x_1-u||x_2-u||x_3-u|^2|x_4-u|^2} \\
        &= -\frac{L_{12}L_{34}^2}{\pi^3}J_2\,.
    \end{aligned}
\end{align}
Thus
\begin{align}
    J_2 = -\frac{\pi^3}{L_{12}L_{34}^2}\Lambda(z)\,.
\end{align}

To reduce $J_3$, differentiate the general-position integral~\eqref{eq:general-position-integral} with respect to $\tau_4$, keeping $\tau_1,\tau_2,\tau_3$ fixed, before imposing the frame. Since
\begin{align}
    \partial_{\tau_4}\frac{1}{|u-x_4|^2} = \frac{|u-x_3|^2-|u-x_4|^2-L_{34}^2}{L_{34}|u-x_4|^4}\,,
\end{align}
we obtain
\begin{align}
    L_{34}\partial_{\tau_4}J_2 = J_1 - J_2 - L_{34}^2J_3\,.
\end{align}
Since $\partial_{\tau_4}z = z(\frac{1}{L_{34}}-\frac{1}{L_{24}})$, we obtain
\begin{align}
    L_{34}\partial_{\tau_4}J_2 = \frac{2\pi^3}{L_{12}L_{34}^2}\Lambda(z) - \frac{\pi^3}{L_{12}L_{34}}\left(\frac{1}{L_{34}}-\frac{1}{L_{24}}\right)z\Lambda'(z)\,.
\end{align}
Using equation~\eqref{eq:direct-paring-lambda-exchange},
\begin{align}
    \Lambda'(z) = \frac{2-z}{2z(1-z)}\Lambda(z)-\frac{2}{\pi^2(1-z)}\log(1-z)\,,
\end{align}
and therefore
\begin{align}
    J_3 = -\pi^3\frac{\sqrt{z}(2+\sqrt{z})}{2(1+\sqrt{z})}\Lambda(z) - \frac{2\pi z}{1+\sqrt{z}}\log(1-z) - \frac{2\pi z^2}{(1+\sqrt{z})^2}\,.
\end{align}
Consequently, we obtain
\begin{align}
    \cM_1^{(1)} 
    = -\frac{z^2}{2(1-z)}\Lambda(z)
    +\frac{2z(2-z)}{\pi^2(1-z)}\log(1-z)
    +
    \frac{2z^2(z^2-2z+2)}
    {\pi^2(1-z)^2}\, .
\end{align}

\section{Comparison with independent calculations}

\subsection{Defect $g$-function}
\label{app:defect-g-function-check}

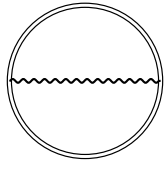
\begin{figure}[t]
    \centering
    \begin{tikzpicture}[
        wilson/.style={
            double,
            double distance=1.1pt,
            line width=0.45pt
        },
        photon/.style={thick,decorate,decoration={snake,amplitude=0.7pt,segment length=4pt}},
        insertion/.style={circle,fill=black,inner sep=2.1pt},
        every node/.style={font=\small}
    ]
        \draw[wilson] (0,0) circle (1cm);
        \draw[photon] (-1,0)--(1,0);
    \end{tikzpicture}
    \caption{The diagram contributing to the expectation value of the circular Wilson line at $O(N^{-1})$.}
    \label{fig:leading-diagram-of-g-function}
\end{figure}

In this appendix, we compute the $g$-function $g_q$ of the charge-$q$ Wilson line in the large-$N$ expansion. We evaluate the expectation value of the charge-$q$ Wilson line placed on a circle of radius $R$ in flat space and subtract the scheme-dependent part proportional to $R$. Expanding the exponential, the expectation value of the Wilson line takes the form
\begin{align}
    \langle W_q(C) \rangle
    = 1
    + \i q\int_C \d x^{\mu}\,\langle A_{\mu}(x) \rangle
    - \frac{q^2}{2} \int_C\d x^{\mu}\int_C\d y^{\nu}\,\langle A_{\mu}(x)A_{\nu}(y) \rangle + \cdots\,.
\end{align}
Since the one-point function of $A_{\mu}$ vanishes, the leading correction comes from the quadratic contribution, represented by the diagram in figure~\ref{fig:leading-diagram-of-g-function}. Contributions from cubic and higher-order terms are subleading and will be neglected.
In the gauge convention of~\cite{Iqbal:2025yop, Benvenuti:2019ujm, Kaul:2008xw}, the effective photon propagator is given by
\begin{align}
    \langle A_{\mu}(x)A_{\nu}(0) \rangle
    = \frac{8}{\pi^2N}\left[(1-\zeta)\frac{\delta_{\mu\nu}}{|x|^2} + 2\zeta\frac{x_{\mu}x_{\nu}}{|x|^4}\right]\,.
\end{align}
Then, the expectation value of the Wilson line is given by
\begin{align}
    \langle W_q(C) \rangle
    = 1
    - \frac{4q^2}{\pi^2N}\int_C\d x^{\mu}\int_C\d y^{\nu}\,
    \left[(1-\zeta)\frac{\delta_{\mu\nu}}{|x-y|^2} + 2\zeta\frac{(x-y)_{\mu}(x-y)_{\nu}}{|x-y|^4}\right]
    +O(N^{-2})\,.
\end{align}
Parameterizing $C$ as $R(\cos\theta,\sin\theta,0)$, the second term becomes
\begin{align}
    \begin{aligned}
        -\frac{4q^2}{\pi^2N}\int_0^{2\pi}\d\theta_1\int_0^{2\pi}\d\theta_2\,
        &\left[
        (1-\zeta)\frac{\cos(\theta_1-\theta_2)}{4\sin^2\frac{\theta_1-\theta_2}{2}}
        + 2\zeta\frac{\sin^2(\theta_1-\theta_2)}{16\sin^4\frac{\theta_1-\theta_2}{2}}
        \right] \\
        &=-\frac{8q^2}{\pi N}\int_0^{2\pi}\d\phi\,
        \left[
        (1-\zeta)\frac{\cos\phi}{4\sin^2\frac{\phi}{2}}
        + 2\zeta\frac{\sin^2\phi}{16\sin^4\frac{\phi}{2}}
        \right]\,,
    \end{aligned}
\end{align}
where $\phi=\theta_1-\theta_2$. Since this integral is divergent, we introduce a short-distance cutoff $a$, or equivalently an angle cutoff $\delta$. They are related by
\begin{align}
    \delta = \frac{a}{R} + O\left(\frac{a^3}{R^3}\right)\,.
\end{align}
With the angular cutoff $\delta$, the integration region is given by
\begin{align}
    \delta < \phi < 2\pi-\delta\,.
\end{align}
The integral can then be evaluated as
\begin{align}
    \begin{aligned}
        -\frac{8q^2}{\pi N}\int_{\delta}^{2\pi-\delta}\d\phi\,
        \left[
        (1-\zeta)\frac{\cos\phi}{4\sin^2\frac{\phi}{2}}
        + 2\zeta\frac{\sin^2\phi}{16\sin^4\frac{\phi}{2}}
        \right]
        &= -\frac{8q^2}{\pi N}\int_{\delta}^{2\pi-\delta}\d\phi\,
        \left[
        \frac{1+\zeta}{4\sin^2\frac{\phi}{2}}
        - \frac{1}{2}
        \right] \\
        &= -\frac{8q^2}{\pi N}
        \left[
        (1+\zeta)\cot\frac{\delta}{2}
        - \frac{1}{2}(2\pi-2\delta)
        \right] \\
        &= -\frac{16q^2(1+\zeta)}{\pi N\delta} + \frac{8q^2}{N} + O(\delta)\,.
    \end{aligned}
\end{align}
The first term is proportional to $R$, since $\delta\sim\frac{a}{R}$, and can therefore be removed by a defect-local cosmological counterterm. Thus, the universal part of the Wilson-line expectation value, which is identified with the $g$-function, is given by
\begin{align}
    g_q = 1 + \frac{8q^2}{N} + O(N^{-2}),
\end{align}
in agreement with~\eqref{eq:general-q-g-function-bootstrap-main}.

\subsection{Anomalous dimensions of endpoint operators}
\label{app:q-one-mixed-lowest-two-point-check}
In this appendix, we compute the anomalous dimensions of endpoint operators in the large-$N$ expansion. As calculated in section \ref{subsec:q-one-mixed-simplification}, the anomalous dimension of the operator $\mathcal X_{q,0}^{(0)} =:\!\sigma E^{(q)}\!:$, which appears in the OPE for $E^{(q)} \times \sigma$, is given by equation \eqref{eq:general-q-mixed-lowest-dimension}. We reproduce this anomalous dimension by a direct large-$N$ perturbative calculation.

At order $1/N$, the two-point function can be factorized as follows:
\begin{align}\label{eq:foctrized-x}
    \langle\overline{\mathcal{X}}_{q, 0}^{(0),I}(x_1) \mathcal{X}_{q, 0, J}^{(0)}(x_2) \rangle 
    =\langle \sigma\sigma\rangle\langle \overline{E}_{(q)}^I E^{(q)}_J\rangle 
    \left[ 
    1 + \frac{1}{N}\mathcal{V}(x_{12})
    \right]
    + O({N^{-2}})\, ,
\end{align}
Each $\sigma\Phi^{\dagger}\Phi$ vertex carries $N^{-1/2}$, while the $\sigma$ and photon propagators are $O(1)$ and $O(N^{-1})$, respectively. The correlator contains two external $\sigma$ fields, and they either contract with each other or each terminate on a $\sigma\Phi^{\dagger}\Phi$ vertex. 
In the first case the remaining correction is confined to the $\sigma$ pair or to the endpoint pair and is already accounted for by $\gamma_\sigma$ and $\gamma_q$ in \eqref{eq:general-q-mixed-sigma-dimension} and \eqref{eq:general-q-endpoint-dimension-bootstrap-main}. 
In the second case the two vertices saturate the order, so any additional photon or Hubbard--Stratonovich exchange would contribute only at higher order. Thus, $\mathcal{V}$ contains the corrections from the two $\sigma\Phi^{\dagger}\Phi$ vertices. 
Comparing with \eqref{eq:general-q-mixed-eta-definition}, the logarithmic term in $\mathcal{V}$ measures $\eta_{q,0}$ alone,
\begin{align}
    \mathcal{V}\Big|_{\mathrm{log}} = -2\eta_{q, 0} \log \frac{|x_{12}|}{a}, 
\end{align}
with a short-distance cutoff $a$. 

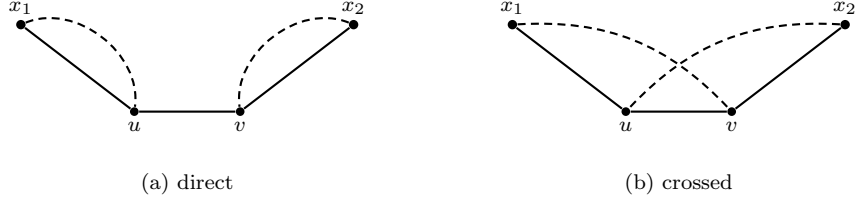
\begin{figure}[tb]
\centering
\begin{tikzpicture}[
    scalar/.style={draw=black,thick},
    hs/.style={draw=black,thick,densely dashed},
    insertion/.style={circle,fill,inner sep=1.25pt},
    vertex/.style={circle,fill,inner sep=1.15pt},
    every node/.style={font=\scriptsize}
]
\begin{scope}
    \node[insertion] (a) at (0,0) {};
    \node[insertion] (b) at (4.4,0) {};
    \node[vertex] (u) at (1.5,-1.15) {};
    \node[vertex] (v) at (2.9,-1.15) {};
    \draw[scalar] (a) -- (u) -- (v) -- (b);
    \draw[hs] (a) to[bend left=60] (u);
    \draw[hs] (b) to[bend right=60] (v);
    \node[above] at (a) {$x_1$};
    \node[above] at (b) {$x_2$};
    \node[below] at (u) {$u$};
    \node[below] at (v) {$v$};
    \node at (2.2,-2.1) {(a) direct};
\end{scope}
\begin{scope}[shift={(6.5,0)}]
    \node[insertion] (a) at (0,0) {};
    \node[insertion] (b) at (4.4,0) {};
    \node[vertex] (u) at (1.5,-1.15) {};
    \node[vertex] (v) at (2.9,-1.15) {};
    \draw[scalar] (a) -- (u) -- (v) -- (b);
    \draw[hs] (a) to[bend left=25] (v);
    \draw[hs] (b) to[bend right=25] (u);
    \node[above] at (a) {$x_1$};
    \node[above] at (b) {$x_2$};
    \node[below] at (u) {$u$};
    \node[below] at (v) {$v$};
    \node at (2.2,-2.1) {(b) crossed};
\end{scope}
\end{tikzpicture}
\caption{The two order-$1/N$ topologies contributing to $\mathcal{V}$.  Solid
lines are scalar propagators and dashed lines are $\sigma$
propagators.  The Wilson segment does not enter at this order.}
\label{fig:app-two-point-diagrams}
\end{figure}

In diagrams associated with $\mathcal{V}$, two topologies contribute, depending on whether each external $\sigma$ ends at the vertex adjacent to its own insertion point or at the far one (see figure~\ref{fig:app-two-point-diagrams}). 
Since $\sigma$ is unit-normalized in \eqref{eq:general-q-mixed-sigma-definition}, each external $\sigma$ line is simply $|x_i -w|^{-4}$. The three scalar propagators $1/(4\pi|x|)$ along the chain supply $(4\pi)^{-3}$. The integrals for these two diagrams are given by
\begin{align}
        I_{\mathrm{dir}} &= \frac{1}{(4\pi)^3} \int \d^3u\, \d^3v\, \frac{1}{|x_1-u|^5|u-v||x_2-v|^5}, \\
        I_{\mathrm{cr}} &= \frac{1}{(4\pi)^3} \int \d^3u\, \d^3v\, \frac{1}{|x_1-u|^4|x_2-u||u-v||x_2-v|^4|x_1-v|}.
\end{align}
Using these integrals, $\mathcal{V}$ is given by 
\begin{align}
    \mathcal{V} = 4\pi g_{\sigma\Phi\Phi}^2|x_{12}|^5(I_{\mathrm{dir}}+ I_{\mathrm{cr}})\, . 
\end{align}

For $q=1$, because $\Delta_\sigma + 2\Delta_\Phi = d$, the inner vertex of the crossed diagram is evaluated using the star-triangle relation. The inner integral of the direct diagram is simply a chain, regulated by setting $\Delta_\sigma = 2 - \delta$. Thus
\begin{align}
    \int \frac{\d^3u}{|x_1 - u|^4 |x_2 - u| |u - v|} = -\frac{2\pi |x_2 - v|}{|x_{12}|^2 |x_1 - v|^2} ,
    \quad \int \frac{\d^3u}{|x_1 - u|^5 |u - v|} = -\frac{2\pi}{3} \frac{1}{|x_1 - v|^3}, 
\end{align}
and therefore 
\begin{align}\label{eq:integran-v-log}
    \begin{aligned}
        I_{\mathrm{cr}} &= -\frac{2\pi}{(4\pi)^3 |x_{12}|^2} \int \frac{\d^3v}{|x_1 - v|^3 |x_2 - v|^3} ,\\ I_{\mathrm{dir}} &= -\frac{2\pi}{3(4\pi)^3} \int \frac{\d^3v}{|x_1 - v|^3 |x_2 - v|^5}.
    \end{aligned}
\end{align}
The remaining integrals diverge logarithmically at $v \to x_1$ and $v \to x_2$. 
From $I_{\mathrm{cr}}$, we find that the log-divergent part is given by
\begin{align}
    I_{\mathrm{cr}}\Big|_{\log} = -\frac{2\pi}{(4\pi)^3 |x_{12}|^2} \times \left( 8\pi \frac{1}{|x_{12}|^3} \log \frac{|x_{12}|}{a} \right) = -\frac{16\pi^2}{(4\pi)^3} \frac{1}{|x_{12}|^5} \log \frac{|x_{12}|}{a},
\end{align}
and from $I_{\mathrm{dir}}$, the log-divergent part is
\begin{align}
    I_{\mathrm{dir}}\Big|_{\log} = -\frac{2\pi}{3(4\pi)^3} \times \left( 8\pi|x_{12}|^{-5} \log \frac{|x_{12}|}{a} \right) = -\frac{16\pi^2}{3(4\pi)^3} \frac{1}{|x_{12}|^5} \log \frac{|x_{12}|}{a} .
\end{align}
Collecting the two topologies and using $g^2_{\sigma\Phi\Phi} = 8/\pi^2$, 
\begin{align}
    \mathcal{V} \Big|_{\mathrm{log}} = -\frac{4 g_{\sigma\Phi\Phi}^2}{3} \log\frac{|x_{12}|}{a} = -\frac{32}{3\pi^2}\log\frac{|x_{12}|}{a}, \quad \eta_{1,0} = \frac{16}{3\pi^2}.
\end{align}

For $q\geq 2$, the two external $\sigma$ fields may attach to the same scalar line or to two distinct ones. Each of the $q$ same-line terms reproduces the computation above. Each distinct-line term factorizes into two single-vertex insertions, whose logarithmic contribution is proportional to $\partial^2_u |u - x_2|^{-1} = 0$ and hence vanishes. This gives $\eta_{q,0} = q \eta_{1,0}$, which is the linear $q$ dependence of \eqref{eq:general-q-mixed-eta-general-result}.  Adding $\gamma_q$ from \eqref{eq:general-q-endpoint-dimension-bootstrap-main} and $\gamma_\sigma$ from \eqref{eq:general-q-mixed-sigma-dimension},
\begin{align}
    \begin{aligned}
        \overline\Delta_{q,0} &= \frac{q}{2} + 2 + \frac{1}{3\pi^2 N} \left(-18q^2 - 36q - 144 + 16q\right) + O(N^{-2}) \\
        &= \frac{q}{2} + 2 - \frac{2(9q^2 + 10q + 72)}{3\pi^2 N} + O(N^{-2}),
    \end{aligned}
\end{align}
in agreement with \eqref{eq:general-q-mixed-lowest-dimension}.

\section{Multiplicity of the lowest mixed-channel primary}
\label{app:q-one-mixed-lowest-multiplicity}

The operator $:\!\sigma E^{(1)}\!:$ has leading dimension $5/2$. To establish that this level is non-degenerate, we must check whether there is another independent primary with the same dimension and quantum numbers. Since $\sigma$ is gauge neutral and a flavor singlet, such an operator must carry the same charge and flavor representation as $E^{(1)}$. It must also be invariant under transverse rotations and reflections.

We therefore classify operators of leading dimension $5/2$ constructed from the bulk fields at the Wilson-line endpoint. In the expressions below, we display only the local fields and the Wilson line attached to the insertion is left implicit.

Before imposing the equations of motion, the candidate expressions in this sector are
\begin{align}\label{eq:q-one-multiplicity-candidates}
    \begin{aligned}
    &\sigma\Phi\, ,\quad
    D_\tau^2\Phi\, ,\quad
    D^\alpha D_\alpha\Phi\, ,
    \\
    &(\Phi^\dagger\Phi)D_\tau\Phi\, ,\quad
    (\Phi^\dagger D_\tau\Phi)\Phi\, ,\quad
    ((D_\tau\Phi)^\dagger\Phi)\Phi\, ,
    \\
    &(\Phi^\dagger\Phi)^2\Phi\, .
    \end{aligned}
\end{align}
The second line contains the flavor-fundamental combinations of three scalar fields and one derivative, while the third line contains the corresponding five-scalar expression.
Composites written using flavor-adjoint bilinears or currents are therefore already included in this list.\footnote{ Expressions involving the field strength also have leading dimension $5/2$, but are excluded by transverse symmetry. Indeed, $F_{12}\Phi$ is odd under a transverse reflection, while $F_{\tau\alpha}\Phi$ carries nonzero transverse spin.} We now impose the equations of motion and remove descendants of $E^{(1)}$ to determine whether any independent primary remains besides $:\!\sigma E^{(1)}\!:$.

The action~\eqref{eq:CP-action} gives the equations of motion
\begin{align}\label{eq:q-one-multiplicity-equations-of-motion}
    \Phi^\dagger\Phi
    &=\frac{N}{g_c}\, ,
    \\
    \Phi^\dagger D_\mu\Phi
    -(D_\mu\Phi)^\dagger\Phi
    &=0\, ,
    \\
    D^\mu D_\mu\Phi
    &=\lambda\Phi
    =\frac{2\sqrt2}{\pi\sqrt N}\sigma\Phi\, .
\end{align}
These equations are applied to regulated products before taking the coincident-point limit.\footnote{
In a correlator containing a Wilson line, variation of the gauge
field also acts on the Wilson line action and produces the source
$\int_C \d x^\mu\,\delta^{(3)}(y-x)$.
The gauge equation displayed above therefore holds away from
the line. Terms supported on the line or at the other insertions
must be included in the local subtractions defining the
renormalized endpoint products. The reductions below are
understood modulo these terms.}

First consider the two-derivative expressions in the first line of \eqref{eq:q-one-multiplicity-candidates}. The insertion $D_\tau^2\Phi$, with its Wilson line attachment,
gives the line descendant $\partial_\tau^2E^{(1)}$. The scalar equation of motion then gives
\begin{align}\label{eq:q-one-multiplicity-derivative-reduction}
    D^\alpha D_\alpha\Phi
    =
    \frac{2\sqrt2}{\pi\sqrt N}\sigma\Phi
    -D_\tau^2\Phi\, .
\end{align}
Thus, after removing the longitudinal descendant, the transverse
Laplacian does not provide an operator independent of
$\sigma\Phi$.
Note that this relation is used before taking $N\to\infty$.

Next consider the three-scalar expressions in the second line of \eqref{eq:q-one-multiplicity-candidates}. Differentiating the constraint gives
\begin{align}
    0
    =
    \partial_\mu(\Phi^\dagger\Phi)
    =
    (D_\mu\Phi)^\dagger\Phi
    +\Phi^\dagger D_\mu\Phi\, .
\end{align}
Together with the gauge equation, this implies
\begin{align}\label{eq:q-one-multiplicity-scalar-contractions}
    \Phi^\dagger D_\mu\Phi
    =
    (D_\mu\Phi)^\dagger\Phi
    =
    0\, .
\end{align}
Consequently, the second line of \eqref{eq:q-one-multiplicity-candidates} reduces to
\begin{align}\label{eq:q-one-multiplicity-three-scalar-reduction}
    (\Phi^\dagger\Phi)D_\tau\Phi
    =\frac{N}{g_c}D_\tau\Phi\, , \quad 
    (\Phi^\dagger D_\tau\Phi)\Phi
    =0\, , \quad 
    ((D_\tau\Phi)^\dagger\Phi)\Phi
    =0\, .
\end{align}
The only surviving expression is proportional to the longitudinal derivative of the original endpoint, and all of them do not give new endpoint operators.

Finally, the five-scalar expression obeys
\begin{align}\label{eq:q-one-multiplicity-five-scalar-reduction}
    (\Phi^\dagger\Phi)^2\Phi
    =
    \left(\frac{N}{g_c}\right)^2\Phi\, .
\end{align}
This is proportional to the original endpoint itself. The coefficients involving $1/g_c$ depend on the ultraviolet regulator and supply the missing mass dimensions.
These expressions therefore contribute to the subtraction of lower-dimensional operators, rather than defining new primaries of dimension $5/2$.

Thus the endpoint sector probed by $E^{(1)}\times\sigma$ contains a single primary at leading dimension $5/2$, represented by $:\!\sigma E^{(1)}\!:$. This level is therefore non-degenerate.

\section{Branching weights of the endpoint state}
\label{app:branching-weights}
In this appendix we prove \eqref{eq:branching-weight} and \eqref{eq:Legendre-Chebyshev}.
We write $\omega_{n,s}$ as a ratio of norms and evaluate it in the oscillator realization of the bilinears~\cite{Flato:1978qz}.

The two-point functions of the bilinears at $N=\infty$ define an inner product on the polynomials $f(\xi,\eta)$. 
It is fixed by $\langle 1 | 1 \rangle = 1$ and 
\begin{align}\label{eq:bilinear-inner-product}
    \langle\xi_\mu f|g\rangle=\langle f|\mathcal K^{(\xi)}_\mu g\rangle\, ,
    \quad
    \langle\eta_\mu f|g\rangle=\langle f|\mathcal K^{(\eta)}_\mu g\rangle\, .    
\end{align}
Multiplication by $\xi_\tau + \eta_\tau$ is the adjoint of $\mathcal K_\tau$, so line primaries are orthogonal to line descendants. 
The polynomials of degree $n$ in $\xi_\tau$ and $\eta_\tau$ are spanned by  $\mathcal X_n$ and the line descendants $(\xi_\tau+\eta_\tau)h$, where $h$ is a polynomial of degree $n-1$ in $\xi_\tau$ and $\eta_\tau$. 
The degree-$n$ part of the endpoint pair \eqref{eq:endpoint-pair} is then 
\begin{align}
    \frac{\tau^n\xi_\tau^n}{n!}
    =\frac{\langle\mathcal X_n|e^{\tau\xi_\tau}\rangle}{\langle\mathcal X_n|\mathcal X_n\rangle}\,\mathcal X_n
    +(\xi_\tau+\eta_\tau)h\, ,
\end{align}
with 
\begin{align}\label{eq:endpoint-primary-norm}
    \langle\mathcal X_n|\mathcal X_n\rangle=(2n)!\, ,
    \quad
    \langle\mathcal X_n|e^{\tau\xi_\tau}\rangle=(-\tau)^n\,n!\, .
\end{align}
For the unit-normalized $\mathcal J_{s,k}^a$ with $s+2k=n$ this gives 
\begin{align}
    \left.\left|\lambda^{00}_{1,1,\mathcal J_{s,k}}\right|^2\right|_{N=\infty}
    =\frac{\left|\langle\mathcal J_{s,k}|e^{\tau\xi_\tau}\rangle\right|^2}
    {\tau^{2n}\,\langle\mathcal J_{s,k}|\mathcal J_{s,k}\rangle}
    =\frac{(n!)^2}{(2n)!}\,\omega_{n,s}\, ,
\end{align}
where 
\begin{align}\label{eq:branching-weight-definition}
    \omega_{n,s}
    =\frac{\left|\langle\mathcal J_{s,k}|\mathcal X_n\rangle\right|^2}
    {\langle\mathcal J_{s,k}|\mathcal J_{s,k}\rangle\,\langle\mathcal X_n|\mathcal X_n\rangle}\, .
\end{align} 
This reproduces \eqref{eq:endpoint-OPE-weight} and \eqref{eq:spin-resolved-OPE-weight}. 
The bulk multiplets $V^{\mathrm{3d}}_{s+1,s}$ are mutually orthogonal and are mapped into themselves by $\mathcal K_\mu$ and by multiplication by $\xi_\mu+\eta_\mu$.
The orthogonal projection $\Pi_s$ onto $V^{\mathrm{3d}}_{s+1,s}$ therefore commutes with $\mathcal K_\tau$ and with $\mathrm O(2)_\perp$.
Then $\Pi_s\mathcal X_n$ is an $\mathrm O(2)_\perp$-invariant line primary of degree $n$ in $V^{\mathrm{3d}}_{s+1,s}$, which is proportional to $\mathcal J_{s,k}$.
Hence $\omega_{n,s}=\langle\Pi_s\mathcal X_n|\Pi_s\mathcal X_n\rangle/\langle\mathcal X_n|\mathcal X_n\rangle$ is the fraction of the norm of $\mathcal X_n$ carried by $V^{\mathrm{3d}}_{s+1,s}$.

To evaluate this fraction, let $y$ and $\tilde y$ be two-component column vectors of commuting variables, and let $\partial_y$ and $\partial_{\tilde y}$ be the column vectors of the corresponding derivatives.
The on-shell conditions \eqref{eq:bilinear-on-shell} are solved by
\begin{align}\label{eq:oscillator-variables}
    \xi_\mu=\frac12\,y^{\mathrm T}\Sigma_\mu\,y\, ,
    \quad
    \eta_\mu=\frac12\,\tilde y^{\mathrm T}\Sigma_\mu\,\tilde y\, ,
    \quad
    \Sigma_\tau=\mathbf 1\, ,
    \quad
    \Sigma_1=\i\sigma_3\, ,
    \quad
    \Sigma_2=\i\sigma_1\, ,
\end{align}
where $\sigma_1$ and $\sigma_3$ are Pauli matrices.
Polynomials in $\xi_\mu$ and $\eta_\mu$ modulo \eqref{eq:bilinear-on-shell} are in one-to-one correspondence with polynomials in $y$ and $\tilde y$ that are even in $y$ and even in $\tilde y$.
On these polynomials the generators \eqref{eq:special-conformal-generator} with $\Delta_\Phi=1/2$ become
\begin{align}\label{eq:oscillator-K}
    \mathcal K_\mu
    =\frac12\left(\partial_y^{\mathrm T}\Sigma_\mu^*\,\partial_y
    +\partial_{\tilde y}^{\mathrm T}\Sigma_\mu^*\,\partial_{\tilde y}\right)\, ,
\end{align}
and the inner product \eqref{eq:bilinear-inner-product} becomes
\begin{align}\label{eq:oscillator-inner-product}
    \langle F|G\rangle
    =\left.\overline F(\partial_y,\partial_{\tilde y})\,G(y,\tilde y)\right|_{y=\tilde y=0}\, ,
\end{align}
where $\overline F$ is $F$ with complex-conjugated coefficients.
With respect to \eqref{eq:oscillator-inner-product}, multiplication by $\xi_\mu$ is the adjoint of $\frac12\partial_y^{\mathrm T}\Sigma_\mu^*\,\partial_y$, which is $\mathcal K^{(\xi)}_\mu$ in \eqref{eq:oscillator-K}.
We use \eqref{eq:oscillator-K} and \eqref{eq:oscillator-inner-product} on all polynomials in $y$ and $\tilde y$.

The rotation
\begin{align}\label{eq:oscillator-rotation}
    R_\theta\, :
    \quad
    y\to y\cos\theta+\tilde y\sin\theta\, ,
    \quad
    \tilde y\to -y\sin\theta+\tilde y\cos\theta
\end{align}
leaves $\xi_\mu+\eta_\mu$ and $\mathcal K_\mu$ invariant.
It therefore commutes with multiplication by $\xi_\mu+\eta_\mu$ and with $\mathcal K_\mu$, which generate the conformal algebra.
It is an orthogonal transformation of $(y,\tilde y)\in\mathbb R^4$ and preserves \eqref{eq:oscillator-inner-product}.
On $w\equiv y+\i\tilde y$ and $\bar w\equiv y-\i\tilde y$ it acts as $w\to e^{-\i\theta}w$ and $\bar w\to e^{\i\theta}\bar w$.
For a null vector $\varepsilon^\mu$, the symmetric matrix $\varepsilon^\mu\Sigma_\mu$ has $\det(\varepsilon^\mu\Sigma_\mu)=\varepsilon^\mu\varepsilon_\mu=0$, so $\varepsilon^\mu\Sigma_\mu=\kappa\kappa^{\mathrm T}$ for a two-component vector $\kappa$.
Then $\xi_\varepsilon=\frac12(\kappa^{\mathrm T}y)^2$ and $\eta_\varepsilon=\frac12(\kappa^{\mathrm T}\tilde y)^2$, and \eqref{eq:ambient-Chebyshev-polynomial} becomes
\begin{align}\label{eq:oscillator-current}
    \varepsilon^{\mu_1}\cdots\varepsilon^{\mu_s}(J_s)_{\mu_1\cdots\mu_s}
    =\frac{(-1)^s}{2^{s+1}}\left[(\kappa^{\mathrm T}w)^{2s}+(\kappa^{\mathrm T}\bar w)^{2s}\right]\, .
\end{align}
The two terms are eigenvectors of $R_\theta$ with eigenvalues $e^{-2\i s\theta}$ and $e^{2\i s\theta}$.
The multiplet $V^{\mathrm{3d}}_{s+1,s}$ is obtained from the components of $J_s$ by multiplication by $\xi_\mu+\eta_\mu$.
Every $v\in V^{\mathrm{3d}}_{s+1,s}$ is a sum $v=v_++v_-$ with $R_\theta v_\pm=e^{\mp2\i s\theta}v_\pm$.
The reflection $\tilde y\to-\tilde y$ exchanges $w$ and $\bar w$, preserves the inner product, and leaves $v$ invariant.
It exchanges $v_+$ and $v_-$, so $\langle v_+|v_+\rangle=\langle v_-|v_-\rangle$.
For $s\geq1$ the two terms have different eigenvalues of the unitary operator $R_\theta$ and are orthogonal. For $s=0$, $R_\theta v=v$.
For $v\in V^{\mathrm{3d}}_{s+1,s}$ and $v'\in V^{\mathrm{3d}}_{s'+1,s'}$ with $s'\neq s$ we obtain
\begin{align}\label{eq:rotation-matrix-elements}
    \langle v|R_\theta|v\rangle=\cos(2s\theta)\,\langle v|v\rangle\, ,
    \quad
    \langle v'|R_\theta|v\rangle=0\, .
\end{align}
Writing $\mathcal X_n=\sum_s\Pi_s\mathcal X_n$, we obtain
\begin{align}\label{eq:weight-generating-function}
    \frac{\langle\mathcal X_n|R_\theta|\mathcal X_n\rangle}
    {\langle\mathcal X_n|\mathcal X_n\rangle}
    =\sum_{s=n,n-2,\cdots}\omega_{n,s}\cos(2s\theta)\, .
\end{align}

We now evaluate the left-hand side.
By \eqref{eq:oscillator-K}, $\mathcal K_\tau=\frac12\left(\partial_y^{\mathrm T}\partial_y+\partial_{\tilde y}^{\mathrm T}\partial_{\tilde y}\right)$ is one half of the Laplacian on $\mathbb R^4$.
Line primaries of degree $n$ are harmonic polynomials of degree $2n$.
Harmonic polynomials of degree $2n$ on $\mathbb R^4$ form the representation of $\mathrm{SO}(4)=[\mathrm{SU}(2)\times\mathrm{SU}(2)]/\mathbb Z_2$ with spins $(n,n)$.
Since $\xi_\tau=\frac12y^{\mathrm T}y$ and $\eta_\tau=\frac12\tilde y^{\mathrm T}\tilde y$, the polynomial $\mathcal X_n$ in \eqref{eq:endpoint-primary} is invariant under independent rotations of the $(y_1,y_2)$ plane and of the $(\tilde y_1,\tilde y_2)$ plane.
These rotations form a maximal torus of $\mathrm{SO}(4)$, and $\mathcal X_n$ is proportional to the state $|n,0\rangle\otimes|n,0\rangle$.
The rotation $R_\theta$ rotates the $(y_1,\tilde y_1)$ and $(y_2,\tilde y_2)$ planes by the same angle.
It belongs to one of the two $\mathrm{SU}(2)$ factors, in which it is a rotation by the angle $2\theta$ about an axis orthogonal to the quantization axis.
Its matrix element is the Wigner function
\begin{align}\label{eq:Wigner-function}
    \frac{\langle\mathcal X_n|R_\theta|\mathcal X_n\rangle}
    {\langle\mathcal X_n|\mathcal X_n\rangle}
    =d^{\,n}_{00}(2\theta)
    =P_n(\cos2\theta)\, .
\end{align}
Comparing with \eqref{eq:weight-generating-function} at $\chi=\cos2\theta$, where $\cos(2s\theta)=T_s(\chi)$, gives \eqref{eq:Legendre-Chebyshev}.
The generating function of the Legendre polynomials,
\begin{align}\label{eq:Legendre-generating-function}
    \sum_{n=0}^{\infty}t^nP_n(\cos2\theta)
    =\left(1-te^{2\i\theta}\right)^{-1/2}\left(1-te^{-2\i\theta}\right)^{-1/2}\, ,
    \quad
    (1-u)^{-1/2}=\sum_{k=0}^{\infty}\binom{2k}{k}\frac{u^k}{4^k}\, ,
\end{align}
gives
\begin{align}
    P_n(\cos2\theta)
    =\frac{1}{4^n}\sum_{k=0}^{n}\binom{2k}{k}\binom{2n-2k}{n-k}e^{2\i(n-2k)\theta}\, .
\end{align}
Combining the terms with $k$ and $n-k$ gives \eqref{eq:branching-weight}.

\section{Alpha-space inversion of the endpoint crossing equation}
\label{sec:q-one-alpha-space-inversion}

The adjoint spectrum in
subsection~\ref{subsec:adjoint-SL2-spectrum} was extracted by
expanding the logarithmic part of the correlator around the bulk OPE
limit and recursively matching one-dimensional conformal blocks.  In this appendix, we
reproduce the same data by alpha-space inversion \cite{Hogervorst:2017sfd}. This transform maps a conformal-block decomposition to a meromorphic function. 
Conformal dimensions become pole locations, leading OPE weights become
residues, and anomalous dimensions become double-pole coefficients.
We will recover the leading dimensions and OPE weights, together with
the complete set of OPE-weighted anomalous dimensions
$\overline\gamma_n$ using the alpha-space inversion. 

\subsection{Alpha-space transform and pole dictionary}
\label{subsec:q-one-alpha-transform}

The one-dimensional Casimir acting on the conformal blocks
\eqref{eq:general-q-SL2-block} is
\begin{align}\label{eq:q-one-alpha-Casimir-main}
    \mathcal D
    =
    z^2(1-z)\partial_z^2-z^2\partial_z\, ,
    \quad
    \mathcal Dg_\Delta(z)
    =
    \Delta(\Delta-1)g_\Delta(z)\, .
\end{align}
Writing the eigenvalue as
$\Delta(\Delta-1)=\alpha^2-1/4$, introduce the even Casimir
eigenfunction~\cite{Hogervorst:2017sfd}
\begin{align}\label{eq:q-one-alpha-basis-main}
    \Psi_\alpha(z)
    =
    {}_2F_1\!\left(
        \frac12+\alpha,
        \frac12-\alpha;
        1;
        \frac{z-1}{z}
    \right)\, .
\end{align}
For a sufficiently well-behaved function $f(z)$, its alpha transform
and inverse are\footnote{When the integral is divergent at an endpoint, the transform is
defined by analytic continuation from a region in which it converges.
Equivalently, one may subtract the endpoint asymptotic terms, evaluate
the resulting convergent integral, and restore the transforms of the
subtracted terms by analytic continuation.  Possible analytic
ambiguities associated with this prescription do not affect the pole
positions and residues extracted below.}
\begin{align}
    \widehat f(\alpha)
    &=
    \int_0^1\frac{\d z}{z^2}\,
    f(z)\Psi_\alpha(z)\, , \label{eq:q-one-alpha-transform-main}
    \\
    f(z)
    &=
    \frac{1}{2\pi\i}
    \int_{-\i\infty}^{\i\infty}
    \frac{\d\alpha}{\mathcal N(\alpha)}\,
    \widehat f(\alpha)\Psi_\alpha(z)\, , \label{eq:q-one-alpha-transform-main-inverse}
\end{align}
where
\begin{align}\label{eq:q-one-alpha-measure-main}
    \mathcal N(\alpha)
    =
    \frac{
        \Gamma(\alpha)\Gamma(-\alpha)
    }{
        2\pi
        \Gamma\!\left(\frac12+\alpha\right)
        \Gamma\!\left(\frac12-\alpha\right)
    }\, .
\end{align}
In this normalization, a single block transforms as
\begin{align}\label{eq:q-one-alpha-single-block-main}
    \widehat g_h(\alpha)
    =
    \frac{\mathcal C(h)}{
        \alpha^2-\left(h-\frac12\right)^2
    }\, ,
    \quad
    \mathcal C(h)
    =
    -\frac{\Gamma(2h)}{\Gamma(h)^2}\, .
\end{align}
Its pole in the right half-plane is at
$\alpha=h-1/2$, with residue
\begin{align}\label{eq:q-one-alpha-single-block-residue-main}
    \underset{\alpha=h-1/2}{\operatorname*{Res}}
    \widehat g_h(\alpha)
    =
    -\frac{\Gamma(2h-1)}{\Gamma(h)^2}\, .
\end{align}
The spectrum and OPE weights of a correlator are therefore encoded in
the locations and residues of the poles of its alpha-space density.

\subsection{Crossing relation in alpha space}
\label{subsec:q-one-alpha-crossing}

We now apply the alpha transform to the position-space crossing
relation and thereby obtain its alpha-space form.  The adjoint
component of the $q=1$ crossing equation is
\begin{align}\label{eq:q-one-alpha-adjoint-crossing-main}
    \mathcal G^{00}_{1,1}(z)
    =
    \frac{\rho_1(z)}{N}
    \left[
        \mathcal G^{11}_{1,0}(1-z)
        -\mathcal G^{11}_{1,1}(1-z)
    \right]\, .
\end{align}
Thus the bulk-channel function that we wish to transform is completely
determined by the two defect-channel functions on the right-hand side.
Recall that their diagrammatic expressions to the required order are
\begin{align}\label{eq:q-one-alpha-Wilson-input-main}
    \frac1N\mathcal G^{11}_{1,0}(x)
    &=
    \frac1{g_1}
    +\frac1N\left[\mathscr A(x)+r(x)\right]
    +O(N^{-2})\, ,
    \\
    \mathcal G^{11}_{1,1}(x)
    &=
    r(x)+O(N^{-1})\, .
\end{align}
The two terms proportional to $r(1-z)$ cancel after substitution in
\eqref{eq:q-one-alpha-adjoint-crossing-main}.  Defining
\begin{align}\label{eq:q-one-alpha-crossing-definitions-main}
    p
    \equiv
    2\Delta_{E_1}
    =
    1+\frac{2\gamma_1}{N}+O(N^{-2})\, , \quad 
    \rho_1(z)
    =
    \left(\frac{z}{1-z}\right)^p\, ,
\end{align}
the position-space crossing relation becomes
\begin{align}\label{eq:q-one-alpha-crossing-decomposition-main}
    \mathcal G^{00}_{1,1}(z)
    =
    \frac1{g_1}\rho_1(z)
    +\frac1N r(z)\mathscr A(1-z)
    +O(N^{-2})\, .
\end{align}
By linearity of the transform, its alpha-space form is therefore
\begin{align}\label{eq:q-one-alpha-crossing-relation-main}
    \widehat{\mathcal G}^{\,00}_{1,1}(\alpha)
    =
    \frac1{g_1}\widehat{\rho}_1(\alpha)
    +\frac1N
    \int_0^1\frac{\d z}{z^2}\,
    r(z)\mathscr A(1-z)\Psi_\alpha(z) 
    +O(N^{-2})\, .
\end{align}
The first term comes from the defect-channel identity, while the second
is the order-$1/N$ interaction correction.  We analyze their pole
structures separately below.  In particular, at strict $N=\infty$
the defect-channel identity block generates the complete bulk adjoint tower.

\subsection{Leading poles and OPE weights}
\label{subsec:q-one-alpha-leading-poles}

After a simple computation, the alpha transform of the identity contribution is
\begin{align}\label{eq:q-one-alpha-rho-transform-main}
    \widehat{\rho}_1(\alpha)
    =
    \frac{\Gamma(1-p)}{\Gamma(p)}
    \frac{
        \Gamma\!\left(p-\frac12+\alpha\right)
        \Gamma\!\left(p-\frac12-\alpha\right)
    }{
        \Gamma\!\left(\frac12+\alpha\right)
        \Gamma\!\left(\frac12-\alpha\right)
    }\, .
\end{align}
The integral representation converges on the principal series for
$1/2<\operatorname{Re}p<1$. Outside this strip,
equation \eqref{eq:q-one-alpha-rho-transform-main} defines its analytic
continuation.  In the right half-plane, its poles are distributed as
\begin{align}\label{eq:q-one-alpha-rho-poles-main}
    \alpha_n(p)
    =
    p+n-\frac12 \, ,
    \quad
    n\in \ZZ_{\geq 0}\, .
\end{align}
They correspond to blocks of dimensions
\begin{align}\label{eq:q-one-alpha-rho-dimensions-main}
    \Delta_n(p)
    =
    p+n\, .
\end{align}
Taking the ratio between the residue of
\eqref{eq:q-one-alpha-rho-transform-main} and the single-block residue
\eqref{eq:q-one-alpha-single-block-residue-main} gives the squared OPE
coefficient
\begin{align}\label{eq:q-one-alpha-rho-OPE-main}
    a_n(p)
    =
    \frac{(p)_n^2}{n!(2p+n-1)_n}\, .
\end{align}
Equivalently,
\begin{align}\label{eq:q-one-alpha-rho-block-decomposition-main}
    \rho_1(z)
    =
    \sum_{n=0}^{\infty}
    \frac{(p)_n^2}{n!(2p+n-1)_n}
    g_{p+n}(z)\, .
\end{align}
Taking the limit $p\to1$ therefore gives
\begin{align}\label{eq:q-one-alpha-leading-data-main}
    \Delta_n^{(0)}
    =
    n+1\, ,
    \qquad
    a_n^{(0)}
    =
    \frac{(n!)^2}{(2n)!}\, ,
\end{align}
reproducing the free block decomposition
\eqref{eq:free-adjoint-block-decomposition} directly from
the poles and residues of the defect-channel identity contribution.

\subsection{Double poles and weighted anomalous dimensions}
\label{subsec:q-one-alpha-double-poles}

We next extract the order-$1/N$ shift of these poles.  
A block of dimension $n + 1 + \gamma/N$ produces a pole at
\begin{align}
    \alpha
    =
    n+\frac12+\frac{\gamma}{N}+O(N^{-2})\, .
\end{align}
Expanding the shifted pole around
$\alpha_n^{(0)}=n+1/2$ generates a double pole.  Using the explicit value~\eqref{eq:q-one-alpha-leading-data-main}, the singular part of the full adjoint density takes the universal
form
\begin{align}\label{eq:q-one-alpha-double-pole-form-main}
    \widehat{\mathcal G}^{\,00}_{1,1}(\alpha)
    =
    -\frac1{\alpha-\alpha_n^{(0)}}
    -\frac1N
    \frac{\overline\gamma_n}{
        \left(\alpha-\alpha_n^{(0)}\right)^2
    }
    +O\!\left(
        \frac{N^{-1}}{\alpha-\alpha_n^{(0)}}
    \right)\, .
\end{align}
Consequently,
\begin{align}\label{eq:q-one-alpha-gamma-residue-rule-main}
    \overline\gamma_n
    =
    -N
    \underset{\alpha=\alpha_n^{(0)}}{\operatorname*{Res}}
    \left[
        \left(\alpha-\alpha_n^{(0)}\right)
        \widehat{\mathcal G}^{\,00}_{1,1}(\alpha)
    \right]
    +O(N^{-1})\, .
\end{align}

The two terms in
\eqref{eq:q-one-alpha-crossing-relation-main} can be treated
separately.  Consider first the term proportional to
$\widehat{\rho}_1(\alpha)$, which is the alpha transform of the
defect-channel identity contribution.  Since
$p=1+2\gamma_1/N+O(N^{-2})$, its $n$-th pole is located at
\begin{align}
    \alpha_n(p)
    =
    \alpha_n^{(0)}
    +\frac{2\gamma_1}{N}
    +O(N^{-2})\, .
\end{align}
Comparison with the shifted pole in
\eqref{eq:q-one-alpha-double-pole-form-main} shows that the
defect-channel identity contributes $2\gamma_1$ to
$\overline\gamma_n$ at every level. The factor $g_1^{-1}$ changes the simple-pole residue and hence the order-$1/N$ OPE coefficient, but it does not move the pole.

For the order-$1/N$ interaction term, only its explicit logarithm can
generate a double pole.  Separating this term gives
\begin{align}\label{eq:q-one-alpha-dynamical-log-main}
    \begin{aligned}
        r(z)\mathscr A(1-z)
        &=
        \mathcal L_{\mathrm{dyn}}(z)\log z
        +\mathcal R_{\mathrm{dyn}}(z)\, ,
        \\
        \mathcal L_{\mathrm{dyn}}(z)
        &=
        \frac4{\pi^2}\frac{z}{1-z}
        \left[
            \frac{3+z}{\sqrt z}
            \operatorname{arctanh}\sqrt z
            +2
        \right]\, .
    \end{aligned}
\end{align}
Here $\mathcal R_{\mathrm{dyn}}(z)$ contains no $\log z$ around the
bulk OPE limit and contributes at most simple poles.\footnote{The
defect-channel identity term supplies $2\gamma_1r(z)\log z$. Using
$\gamma_1=-18/\pi^2$, its sum with $\mathcal L_{\mathrm{dyn}}(z)\log z$ reproduces the total logarithmic function \eqref{eq:adjoint-log-function}.} To evaluate the dynamical double poles, expand
\begin{align}\label{eq:q-one-alpha-Ldyn-series-main}
    \mathcal L_{\mathrm{dyn}}(z)
    =
    \sum_{m=1}^{\infty}d_mz^m\, ,
    \quad
    d_m
    \equiv 
    \frac4{\pi^2}
    \left[
        2+4H_{2m-2}-2H_{m-1}
        +\frac3{2m-1}
    \right]\, .
\end{align}
Note that the alpha transform of a general power is
\begin{align}\label{eq:q-one-alpha-power-transform-main}
    \widehat{z^\beta}(\alpha)
    =
    \frac{
        \Gamma\!\left(\beta-\frac12+\alpha\right)
        \Gamma\!\left(\beta-\frac12-\alpha\right)
    }{
        \Gamma(\beta)^2
    }\, ,
\end{align}
and
\begin{align}
    z^m\log z
    =
    \left.
        \partial_\beta z^\beta
    \right|_{\beta=m}\, .
\end{align}
At the pole $\alpha=\alpha_n^{(0)}$, the alpha transform of
$z^m\log z$ has a double pole only for $1\leq m\leq n+1$.
Applying \eqref{eq:q-one-alpha-gamma-residue-rule-main} therefore
reduces the dynamical contribution to the finite sum
\begin{align}\label{eq:q-one-alpha-dynamical-finite-sum-main}
    \delta\gamma_n
    =
    -\sum_{m=1}^{n+1}
    d_m
    (-1)^{n-m+2}
    \frac{
        (n+m-1)!
    }{
        (n+1-m)![(m-1)!]^2
    }\, .
\end{align}
The factorial coefficient in
\eqref{eq:q-one-alpha-dynamical-finite-sum-main} is naturally
recognized as a coefficient of the Legendre polynomial.  Indeed,
writing $j=m-1$ and using
\begin{align}
\begin{aligned}
P_n(1-2t)
&=
{}_2F_1(-n,n+1;1;t)
\\
&=
\sum_{j=0}^{n}
(-1)^j
\frac{(n+j)!}{(n-j)!(j!)^2}t^j\, ,
\end{aligned}
\end{align}
the finite sum can be evaluated by standard integral representations
of the harmonic numbers, 
\begin{align}\label{eq:q-one-alpha-dynamical-gamma-main}
    \delta\gamma_n
    =
    \frac{16}{\pi^2}
    \left[
        2H_n-H_{\lfloor n/2\rfloor}
        +\frac12
        +\frac{3(-1)^n}{4(2n+1)}
    \right]\, .
\end{align}
Adding the defect-channel identity contribution $2\gamma_1$ gives
\begin{align}\label{eq:q-one-alpha-final-gamma-main}
    \overline\gamma_n
    =
    2\gamma_1+\delta\gamma_n
    =
    \frac{16}{\pi^2}
    \left[
        2H_n-H_{\lfloor n/2\rfloor}
        -\frac74
        +\frac{3(-1)^n}{4(2n+1)}
    \right]
    \, .
\end{align}
This reproduces
\eqref{eq:weighted-gamma-result}.  Alpha-space inversion
therefore recovers the leading adjoint tower and all of its
OPE-weighted anomalous dimensions without the recursive
position-space block subtraction used in
subsection~\ref{subsec:adjoint-SL2-spectrum}.

\bibliographystyle{JHEP}
\bibliography{defect}

\end{document}